\documentclass[%
 reprint,
superscriptaddress,
 amsmath,amssymb,
prx,aps,
showpacs
floatfix,
]{revtex4-2}

\usepackage[english]{babel}
\usepackage{graphicx}
\usepackage{dcolumn}
\usepackage{bm}
\usepackage{mathrsfs}
\usepackage{amsthm}
\usepackage{physics}
\usepackage{gensymb}
\usepackage{siunitx}
\usepackage{textcomp}
\usepackage{upgreek}
\usepackage{romannum}
\usepackage{dsfont}
\usepackage{array}
\usepackage{makecell}
\usepackage{multirow}
\usepackage{soul}
\usepackage{varioref}
\usepackage[table,dvipsnames]{xcolor}
\usepackage{booktabs}
\usepackage{mathtools}
\usepackage{tabularx}

\newcolumntype{L}[1]{>{\raggedright\arraybackslash\hspace{0pt}}p{#1}}
\newcolumntype{C}[1]{>{\centering\arraybackslash}p{#1}}

\definecolor{linkblue}{RGB}{0, 0, 139}       
\definecolor{citegreen}{RGB}{0, 100, 0}      
\definecolor{urlpurple}{RGB}{138, 43, 226}   
\definecolor{sectionred}{RGB}{165, 42, 42}    

\usepackage[breaklinks=true,
unicode = true,
urlcolor = blue,
colorlinks = true,
citecolor = citegreen,
linkcolor = sectionred
]{hyperref}

\newcolumntype{L}[1]{>{\RaggedRight\arraybackslash}p{#1}}
\newcolumntype{C}[1]{>{\Centering\arraybackslash}p{#1}}
\newcolumntype{R}[1]{>{\RaggedLeft\arraybackslash}p{#1}}

\usepackage{float}

\newcounter{extendedfigure}

\newenvironment{extendedfigure}[1][]{%
  \let\oldfigurename\figurename%
  \renewcommand{\figurename}{Extended Data Figure}%
  \refstepcounter{extendedfigure}%
  \begin{figure}[#1]%
}{%
  \end{figure}%
  \let\figurename\oldfigurename%
}

\newenvironment{extendedfigure*}[1][]{%
  \let\oldfigurename\figurename%
  \renewcommand{\figurename}{Extended Data Figure}%
  \refstepcounter{extendedfigure}%
  \begin{figure*}[#1]%
}{%
  \end{figure*}%
  \let\figurename\oldfigurename%
}

\newcounter{extendedtable}

\newenvironment{extendedtable*}[1][]{%
  \let\oldtablename\tablename%
  \renewcommand{\tablename}{Extended Data Table}%
  \refstepcounter{extendedtable}%
  \begin{table*}[#1]%
}{%
  \end{table*}%
  \let\tablename\oldtablename%
}

\newcommand{\eq}{Eq.~}

\newcommand{\rref}{Ref.~}

\newcommand{\ie}{i.e.~}
\newcommand{\eg}{e.g.~}

\DeclareMathOperator{\hc}{H.c.}

\newcommand{\figrefauto}[1]{Fig.~\ref{#1}}
\newcommand{\figref}[1]{\hyperref[#1]{Fig.~\ref{#1}}}
\newcommand{\extfigref}[1]{\hyperref[#1]{Ext Data Fig.~\ref{#1}}}
\newcommand{\extfigrefauto}[1]{Ext. Data Fig.~\ref{#1}}
\newcommand{\tabref}[1]{\hyperref[#1]{Tab.~\ref{#1}}}
\newcommand{\exttabref}[1]{\hyperref[#1]{Ext. Data Tab.~\ref{#1}}}
\newcommand{\eqrefauto}[1]{\hyperref[#1]{\eq\eqref{#1}}}
\newcommand{\methods}[0]{\hyperref[methods]{Methods}}

\newcommand{\subautoref}[2]{\hyperref[#1]{\figrefauto{#1}\textbf{#2}}}

\newcommand{\extsubautoref}[2]{\hyperref[#1]{\extfigrefauto{#1}\textbf{#2}}}

\newcommand{\aref}[1]{\hyperref[#1]{Supplementary Note~\ref*{#1}}}

\begin{document}

\preprint{APS/123-QED}

\title{\texorpdfstring{Superstrong Dynamics and Directional Emission of a Giant Atom in a\\ Structured Bath}
                    {Superstrong Dynamics and Directional Emission of a Giant Atom in a Structured Bath}}
\pagenumbering{arabic}
\author{V. Jouanny}
\affiliation{Hybrid Quantum Circuits Laboratory (HQC), Institute of Physics, \'{E}cole Polytechnique F\'{e}d\'{e}rale de Lausanne (EPFL), 1015, Lausanne, Switzerland}
\affiliation{Center for Quantum Science and Engineering,\\ \ Institute of Physics, \'{E}cole Polytechnique F\'{e}d\'{e}rale de Lausanne (EPFL), 1015, Lausanne, Switzerland}

\author{L. Peyruchat}
\affiliation{Hybrid Quantum Circuits Laboratory (HQC), Institute of Physics, \'{E}cole Polytechnique F\'{e}d\'{e}rale de Lausanne (EPFL), 1015, Lausanne, Switzerland}
\affiliation{Center for Quantum Science and Engineering,\\ \ Institute of Physics, \'{E}cole Polytechnique F\'{e}d\'{e}rale de Lausanne (EPFL), 1015, Lausanne, Switzerland}

\author{M. Scigliuzzo}
\affiliation{Laboratory of Photonics and Quantum Measurements (LPQM), Institute of Physics, \'{E}cole Polytechnique F\'{e}d\'{e}rale de Lausanne (EPFL), 1015, Lausanne, Switzerland}
\affiliation{Center for Quantum Science and Engineering,\\ \ Institute of Physics, \'{E}cole Polytechnique F\'{e}d\'{e}rale de Lausanne (EPFL), 1015, Lausanne, Switzerland}

\author{A. Mercurio}
\affiliation{Laboratory of Theoretical Physics of Nanosystems, Institute of Physics, \'{E}cole Polytechnique F\'{e}d\'{e}rale de Lausanne (EPFL), 1015, Lausanne, Switzerland}
\affiliation{Center for Quantum Science and Engineering,\\ \ Institute of Physics, \'{E}cole Polytechnique F\'{e}d\'{e}rale de Lausanne (EPFL), 1015, Lausanne, Switzerland}

\author{E. Di Benedetto}
\affiliation{Università degli Studi di Palermo, Dipartimento di Fisica e Chimica-Emilio Segrè, Via Archirafi 36, 90123 Palermo, Italy}

\author{D. De Bernardis}
\affiliation{National Institute of Optics (CNR-INO), c/o LENS via Nello Carrara 1, Sesto F.no 500019, Italy}

\author{D. Sbroggi\`o}
\affiliation{Hybrid Quantum Circuits Laboratory (HQC), Institute of Physics, \'{E}cole Polytechnique F\'{e}d\'{e}rale de Lausanne (EPFL), 1015, Lausanne, Switzerland}
\affiliation{Center for Quantum Science and Engineering,\\ \ Institute of Physics, \'{E}cole Polytechnique F\'{e}d\'{e}rale de Lausanne (EPFL), 1015, Lausanne, Switzerland}

\author{S. Frasca}
\affiliation{Hybrid Quantum Circuits Laboratory (HQC), Institute of Physics, \'{E}cole Polytechnique F\'{e}d\'{e}rale de Lausanne (EPFL), 1015, Lausanne, Switzerland}
\affiliation{Center for Quantum Science and Engineering,\\ \ Institute of Physics, \'{E}cole Polytechnique F\'{e}d\'{e}rale de Lausanne (EPFL), 1015, Lausanne, Switzerland}

\author{V. Savona}
\affiliation{Laboratory of Theoretical Physics of Nanosystems, Institute of Physics, \'{E}cole Polytechnique F\'{e}d\'{e}rale de Lausanne (EPFL), 1015, Lausanne, Switzerland}
\affiliation{Center for Quantum Science and Engineering,\\ \ Institute of Physics, \'{E}cole Polytechnique F\'{e}d\'{e}rale de Lausanne (EPFL), 1015, Lausanne, Switzerland}

\author{F. Ciccarello}
\affiliation{Università degli Studi di Palermo, Dipartimento di Fisica e Chimica-Emilio Segrè, Via Archirafi 36, 90123 Palermo, Italy}
\affiliation{NEST, Istituto Nanoscienze-CNR, Piazza S. Silvestro 12, 56127 Pisa, Italy}

\author{P. Scarlino}
\affiliation{Hybrid Quantum Circuits Laboratory (HQC), Institute of Physics, \'{E}cole Polytechnique F\'{e}d\'{e}rale de Lausanne (EPFL), 1015, Lausanne, Switzerland}
\affiliation{Center for Quantum Science and Engineering,\\ \ Institute of Physics, \'{E}cole Polytechnique F\'{e}d\'{e}rale de Lausanne (EPFL), 1015, Lausanne, Switzerland}

\date{\today}

\begin{abstract}
Quantum emitters coupled to waveguides with nonlinear dispersion show rich quantum dynamics with the promise of implementing non-trivial non-Markovian quantum models.
Recent advances in engineered photonic environments now allow the realization of discrete-site waveguides with tailored dispersion, yet most implementations of waveguide QED remain limited to a local qubit-waveguide coupling.
Here, we study a transmon qubit non-locally coupled to a high-impedance coupled cavity array (CCA), thus implementing a \emph{giant atom} in a structured photonic environment. 
The non-local coupling produces interference with the CCA modes, selectively enhancing interaction with even and long-wavelength modes, while suppressing coupling to odd and short-wavelength modes. 
For a subset of symmetric, long-wavelength modes, we reach the superstrong coupling regime.
In this regime, measurements of the atomic participation ratio reveal strongly hybridized eigenmodes on a par with a strongly reduced qubit participation at the frequency of maximum hybridization with the qubit, in agreement with theory. 
Time-domain measurements of the qubit dynamics show clear deviations from the single-mode Jaynes--Cummings model, marked by the emergence of mode--mode interactions. 
By breaking spatial inversion symmetry of the CCA, the qubit seeds dressed eigenmodes confined to either the right or left of the qubit, which we exploit to implement and characterize a directional photon-emission protocol.
These results demonstrate precise control over multimode light--matter interaction in a structured photonic environment.
\end{abstract}

\maketitle

\section{Introduction}

\begin{figure*}
    \centering
    \includegraphics[width = \linewidth]{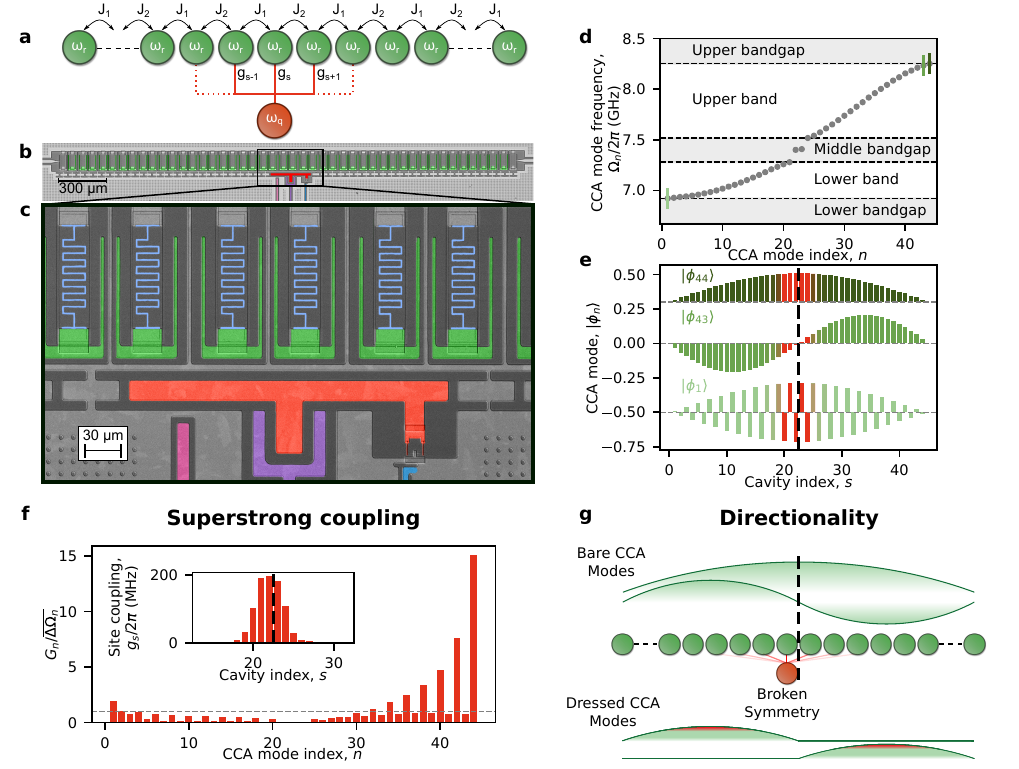}
    \caption{\textbf{Giant atom interacting with a structured photonic bath.}
    \textbf{(a)} Concept of a giant atom coupled non-locally to a dimerized coupled-cavity array (CCA). 
    Resonators of bare frequency $\omega_r/2\pi$ are coupled with staggered rates $J_1/2\pi$ and $J_2/2\pi$. 
    A qubit of frequency $\omega_q/2\pi$ couples to site $s$ with strength $g_s/2\pi$. 
    \textbf{(b)} False-colored optical micrograph of a representative device (aluminum: light gray; silicon: dark gray; cavities: green; qubit: red; drive/readout/flux lines: pink/violet/blue). 
    \textbf{(c)} False-colored scanning electron micrograph. 
    NbN inductors of the CCA resonators are highlighted in light blue. 
    The color code is the same as in the previous panel.
    \textbf{(d)} CCA mode spectrum $\Omega_n/2\pi$ vs mode index $n$ (gray dots). Bandgaps are shaded gray. 
    Long green markers highlight the three modes analyzed in panel \textbf{e}. 
    \textbf{(e)} Spatial profiles of CCA modes$\ket{\phi_{44}}$, $\ket{\phi_{43}}$, and $\ket{\phi_{1}}$ vs site index $s$ (three shades of green).
    Bars at the qubit-coupling sites are overlaid in red. 
    The vertical dashed line marks the CCA center. 
    \textbf{(f)} Superstrong-coupling metric $G_n/\overline{\Delta\Omega_n}$ vs CCA mode index $n$, where $\overline{\Delta\Omega_n}$ is the local average mode spacing around CCA mode $n$ (horizontal dashed line indicates the unity). 
    Inset: spatial profile of the coupling, $g_s/2\pi$, between the qubit and cavity $s$ of the CCA. 
    \textbf{(g)} Schematic illustrating how coupling a qubit away from the CCA center breaks inversion symmetry (middle), leading to hybridization between even- and odd-parity CCA modes (top) and resulting in left- and right-confined dressed modes at specific qubit frequencies (bottom). The vertical dashed line indicates the CCA center.
    }
    \label{fig:fig1}
\end{figure*}

\begin{table}[t!]
  \centering
  \renewcommand{\arraystretch}{1.5}
  \begin{tabular}{l c l}
    \toprule
    \textbf{Parameter Group} & \textbf{Parameter} & \textbf{Description} \\
    \midrule

    \parbox[t]{0.38\linewidth}{\textbf{Indices}} & $s$ & 
      \parbox[t]{0.38\linewidth}{Cavity index} \\
    & $n$ & \parbox[t]{0.38\linewidth}{CCA mode index} \\
    & $m$ & \parbox[t]{0.38\linewidth}{Dressed mode index} \\
    \midrule

    \parbox[t]{0.38\linewidth}{\textbf{CCA sites basis}} & $\omega_r/2\pi$ & 
      \parbox[t]{0.38\linewidth}{Resonant frequency} \\
    & $g_s/2\pi$ & \parbox[t]{0.38\linewidth}{Qubit–cavity coupling rate} \\
    \midrule

    \parbox[t]{0.38\linewidth}{\textbf{CCA modes} $\!\ket{\phi_n}$} & $\Omega_n/2\pi$ & 
      \parbox[t]{0.38\linewidth}{CCA mode frequency} \\
    & $G_n/2\pi$ & \parbox[t]{0.38\linewidth}{Qubit–mode coupling rate} \\
    \midrule

    \parbox[t]{0.38\linewidth}{\textbf{Dressed CCA modes} $\!\ket{\psi_m}$} & $\tilde\omega_m/2\pi$ & 
      \parbox[t]{0.38\linewidth}{Dressed CCA mode frequency} \\
    \midrule

    \parbox[t]{0.38\linewidth}{\textbf{Effective CCA modes}} & $\Omega_n^{\mathrm{eff}}/2\pi$ & 
      \parbox[t]{0.38\linewidth}{Effective CCA mode frequency} \\
    & $G_{n,n'}/2\pi$ & \parbox[t]{0.38\linewidth}{Mode–mode coupling rate} \\
    \midrule

    \parbox[t]{0.38\linewidth}{\textbf{Effective dressed modes}} & $\tilde\omega_m^{\mathrm{eff}}/2\pi$ & 
      \parbox[t]{0.38\linewidth}{Effective dressed mode frequency} \\
    \bottomrule
  \end{tabular}
  \caption{Notation used throughout the manuscript.}
  \label{tab:notation}
\end{table}

Controlling the interaction between emitters and confined photonic modes lies at the heart of cavity quantum electrodynamics (QED), achieving milestones ranging from the demonstration of strong coupling to the development of quantum information processing~\cite{wallraff2004strong, haroche2006exploring, krinner2022realizing}.  
At the other extreme, continuous waveguide QED describes qubits coupled to a 1D continuum of propagating modes, where interference can give rise to collective radiative phenomena and chiral interactions~\cite{chang2007single,sollner2015deterministic,chang2018colloquium,sheremet2023waveguide,kannanOndemandDirectionalMicrowave2023,ciccarello2024waveguide,lodahl2017chiral}. 
Between these limits lies \emph{multimode QED}, where a qubit couples to a discrete set of modes~\cite{sundaresanStrongCouplingMultimode2015a}.  
When the qubit--mode coupling strength exceeds the characteristic mode frequency spacing, the system enters the \textit{superstrong coupling regime
}~\cite{meiserSuperstrongCouplingRegime2006, krimer2014route}: the qubit hybridizes with multiple modes simultaneously, leading not only to a reshaping of their spectrum but also to significant mode--mode hybridization.  
This hybridization drives dynamics beyond the single-mode Jaynes--Cummings paradigm, manifesting in the breakdown of single-mode Rabi oscillations and in strong spatial confinement effects of the eigenmodes~\cite{meiserSuperstrongCouplingRegime2006, krimer2014route, sinha2022radiative}. 
Experiments in the linear-dispersion regime~\cite{sundaresanStrongCouplingMultimode2015a, johnson2019observation, kuzminSuperstrongCouplingCircuit2019a, leger2019observation, mehtaDownconversionSinglePhoton2023a, fraudet2025direct} have revealed rich non-Markovian effects~\cite{lechner2023light}.
However, in such systems the qubit typically remains hybridized across the entire accessible spectrum, hindering single-photon state preparation and limiting time-resolved access to quantum dynamics.

Structured photonic environments with engineered bandgaps overcome these limitations, enabling quantum state preparation inside the gap and subsequent transfer into the band~\cite{castillo2025dynamical}. 
Such environments can be realized with stepped-impedance waveguides~\cite{liu_quantum_2017, sundaresan_interacting_2019}, periodically loaded waveguides~\cite{mirhosseini2018superconducting} or coupled cavity arrays (CCAs)~\cite{indrajeet2020coupling, kim_quantum_2021, scigliuzzoControllingAtomPhotonBound2022a, castillo2025dynamical, jouannyHighKineticInductance2025, Ferreira2020CollapseReservoir,kollar_hyperbolic_2019,naik2017random,owens_chiral_2022}.
At band edges, structured photonic environments support atom–photon bound states~\cite{shi2016bound,calajo_atom-field_2016}, which have been used for analog quantum simulation~\cite{zhang_superconducting_2023}, proposed as alternative routes to quantum-computing architectures~\cite{kim2025fast,gorshkov2025cavity,thorbeck2024high}, and exploited to enable directional interactions~\cite{Bello2019,kim_quantum_2021,pakkiam2024experimental}.
In all of these experiments, artificial atoms have been coupled to a \emph{single site} of the array, fulfilling the electric dipole approximation. 
In this limit, due to parity symmetry, the $n$th CCA mode couples equally to the qubit as the $(N-n)$th mode, where $N$ is the total number of CCA modes.
As a result, strong coupling to the CCA opens large avoided crossings at both band-edges, which hinder non-adiabatic transfer between bandgap and in--band states, effectively ``trapping'' the qubit outside the band and preventing access to the rich dynamics of the superstrong regime.

In this work, we address this challenge by engineering the qubit–mode coupling profile via non-local coupling~\cite{wang2024realizing} of a transmon qubit to a compact, high-kinetic-inductance, gapped CCA [\subautoref{fig:fig1}{\textbf{b–c}}]~\cite{jouannyHighKineticInductance2025}, thereby realizing a \textit{giant-atom} interaction~\cite{gustafsson2014propagating,frisk_kockum_quantum_2021,friskkockumDesigningFrequencydependentRelaxation2014}. 
Such non-local coupling has been studied in continuous waveguide-QED systems~\cite{frisk_kockum_quantum_2021,kannanWaveguideQuantumElectrodynamics2020,vadirajEngineeringLevelStructure2021,joshiResonanceFluorescenceChiral2023,andersson2019non}, but has not yet been realized in structured photonic environments~\cite{soro_interaction_2023,leonforte_quantum_2024,gaoCircuitQEDGiant2024,wang2021tunable,zhang2023quantum}. 
In our device, the qubit strongly couples to more than seven lattice sites [\subautoref{fig:fig1}{\textbf{a}}; inset \subautoref{fig:fig1}{\textbf{f}}].
Destructive interference suppresses coupling to low-frequency band-edge modes, facilitating non-adiabatic transfer into the bands, while constructive interference selectively enhances interactions with a subset of high-frequency modes [see \methods].
This coupling enhancement places the system in the superstrong-coupling regime [\subautoref{fig:fig1}{\textbf{f}}], where we investigate both steady-state and dynamical properties [\autoref{sec:fig2}, \autoref{sec:fig3}]. 
Moreover, the chosen coupling configuration breaks spatial inversion symmetry [inset \subautoref{fig:fig1}{\textbf{f}}], deterministically confining the photonic components of the dressed CCA modes to one side of the array relative to the qubit position [\subautoref{fig:fig1}{\textbf{g}}], which we exploit to demonstrate \textit{directional emission} of a single-photon quantum state [\autoref{sec:fig4}, \autoref{sec:fig5}].

\section{Experiment}
\label{sec:experiment}

The device comprises a Coupled Cavity Array (CCA) with \( N = 44 \) cavities, implemented as a superconducting metamaterial composed of compact (\( 50 \times \SI{125}{\micro\metre\squared} \)) high-kinetic inductance lumped LC resonators with a resonant frequency of \(\omega_r/2\pi = \SI{7.749}{\giga\hertz}\) [\subautoref{fig:fig1}{\textbf{a}--\textbf{c}}]. 
The CCA implements the Su-Schrieffer-Heeger (SSH) model in its symmetry-protected topological phase~\cite{su1980soliton,asbothShortCourseTopological2016}, with dimerized hopping rates of \( J_1/2\pi = \SI{258}{\mega\hertz} \) and \( J_2/2\pi = \SI{370}{\mega\hertz} \) [\subautoref{fig:fig1}{\textbf{a}--\textbf{d}}]. 
Nonetheless, this characteristic will not be involved in what follows. 
A flux-tunable transmon qubit~\cite{kochChargeinsensitiveQubitDesign2007} couples non-locally to several cavities at positions offset from the CCA center [\figref{fig:fig1}].
The device is described by the Hamiltonian (\(\hbar = 1\) in the remainder of the manuscript)
\begin{equation}
    \begin{split}
        H =&\, H_{\rm CCA} + H_{\rm QB} + H_{\rm INT}\\  
        =&\, \omega_r\,\sum_{s=1}^{N}  {a}^\dagger_s {a}_s + J_1\sum_{s=1}^{N/2}\qty( a^\dagger_{2s-1}  a_{2s} + \hc)\\
        &+ J_2\sum_{s=1}^{N/2 - 1}\qty( a^\dagger_{2s}  a_{2s+1} + \hc)\\
        &+ \omega_{q}\, b^\dagger  b - \frac{E_C}{2}\, b^\dagger  b^\dagger  b  b+ \sum_{s=1}^N g_s\,\qty( a^\dagger_s  b + \hc)\,,
    \end{split}
    \label{eq:normalSpaceHamiltonian}
\end{equation}
where $H_{\rm CCA}$, $H_{\rm QB}$ and $H_{\rm INT}$, are the CCA, qubit and interaction Hamiltonian. 
$a_s$ ($b$) annihilates an excitation on cavity $s$ (the transmon qubit), 
$\omega_q/2\pi$ is the tunable transition frequency between $\ket{g}$ and $\ket{e}$ (respectively ground and first-excited states of the qubit), while the qubit anharmonicity is given by $E_C/2\pi = \SI{318}{\mega\hertz}$.
To better model the system, higher order CCA coupling terms are taken into account but not reported in this Hamiltonian [see \aref{app:circ-quant}]. 
Ultimately, these higher-order couplings are responsible for the observed asymmetry of the energy bands [see \subautoref{fig:fig1}{\textbf{d}}].
Finally, $g_s/2\pi$ is the coupling strength between the transmon qubit and cavity $s$, which breaks spatial inversion symmetry around the CCA inversion center [inset \subautoref{fig:fig1}{\textbf{f}}].

Single-photon eigenstates of the bare CCA Hamiltonian $H_{\rm CCA}$ are expressed as $\ket{\phi_n} = \sum_s d_{s,n} \ket{s}$, where $\ket{s} = a_s^\dagger \ket{\mathrm{vac}}$ is the single photon state on cavity $s$ and $d_{s,n}$ is the corresponding probability amplitude.
Analogously, single-excitation eigenstates of the full Hamiltonian $H$ will be written as $\ket{\psi_m} = u_m \ket{e, \mathrm{vac}} + \sum_s c_{s,m} \ket{g, s}$. 
Unlike $\ket{\phi_n}$, these states are dressed by the qubit.
The ground state of the system is $\ket{g, \rm vac}$.
For clarity, \tabref{tab:notation} introduces the notation of the different bases used in this work.

\begin{figure*}[t!]
    \centering
    \includegraphics[width = \linewidth]{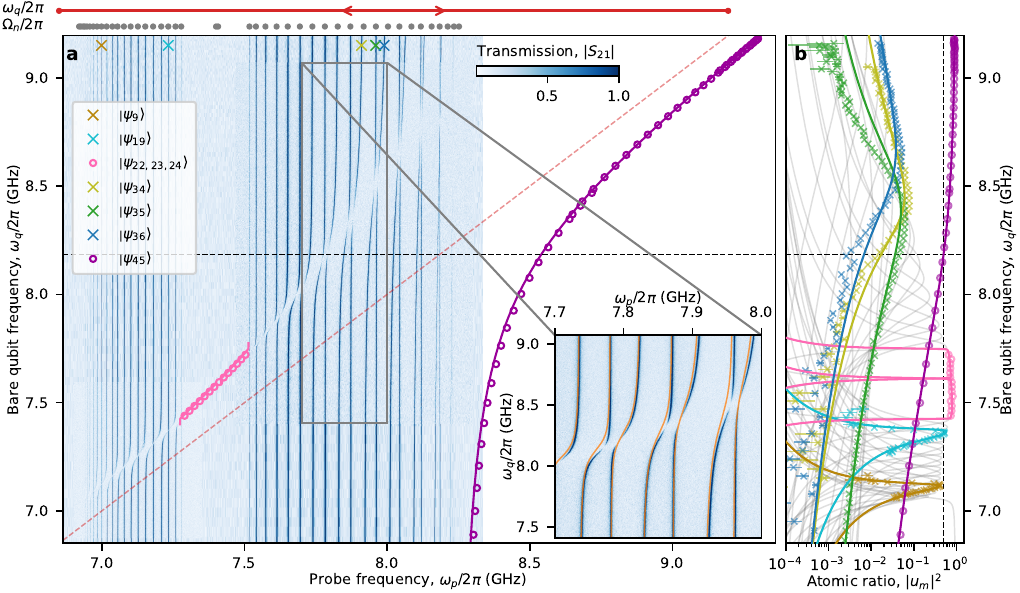}
    \caption{\textbf{Steady-state interaction of a giant atom with a structured photonic bath.} 
    \textbf{(a)} Single-tone transmission of the CCA as a function of bare qubit frequency $\omega_q/2\pi$.  
    The dressed qubit frequency extracted from two-tone spectroscopy is shown above the upper band (purple circles) and inside the middle bandgap (pink circles).  
    Model predictions according to \eqrefauto{eq:normalSpaceHamiltonian} for the dressed qubit are shown as purple and pink lines, and the bare qubit frequency as a red dashed line.  
    Crosses indicate the dressed CCA modes analyzed in panel \textbf{b}. 
    Inset: zoom-in from $\omega_p/2\pi = \SI{7.7}{\giga\hertz}$ to \SI{8}{\giga\hertz}; orange lines show eigenvalues, $\tilde\omega_m/2\pi$, of the fitted disorder-free model according to \eqrefauto{eq:normalSpaceHamiltonian}.  
    \textbf{(b)} Atomic participation ratio of each dressed CCA modes, $|u_m|^2$.  
    Continuous lines: expected atomic participation ratio, $|u_m|^2$, of each dressed CCA modes according to \eqrefauto{eq:normalSpaceHamiltonian}; circles and crosses: values extracted from two-tone and single-tone spectroscopy, respectively (error bars from the standard deviation of fitted dressed-mode frequencies), using $\abs{u_m}^2 = \dd \tilde{\omega}_m / \dd \omega_q$.  
    Vertical black dashed line: reference atomic ratio of $0.5$.  
    Horizontal black dashed line: qubit frequency where the atomic ratio of the APBS, $|u_{45}|^2$, is 0.5 (also marked in panel \textbf{a}). 
    Mode labels are indicated in the legend in panel \textbf{a}.}
    \label{fig:fig2}
\end{figure*}

\section{Superstrong coupling}
\label{sec:fig2}

We first measure the CCA using single-tone spectroscopy in transmission as a function of the qubit frequency at low power to remain in the single-excitation manifold [\subautoref{fig:fig2}{\textbf{a}}].
Due to the dimerization of the CCA, we observe two bands separated by a bandgap of size ${\sim}\left|J_2-J_1\right|$.
The dressed qubit frequency, $\tilde \omega_q/2\pi$, is measured using standard two-tone spectroscopy [see \aref{app:sec:qubit:charac}] and reported in \subautoref{fig:fig2}{\textbf{a}} as purple circles above the upper band and pink circles in the middle bandgap. 
This highlights the presence of an atom-photon bound state (APBS) in the upper bandgap~\cite{liu_quantum_2017,scigliuzzoControllingAtomPhotonBound2022a}. 
The combined single- and two-tone spectroscopy data are globally fitted to the Hamiltonian eigenvalues $\tilde\omega_m$ (dressed CCA modes frequency) in \eqrefauto{eq:normalSpaceHamiltonian}, showing excellent agreement [see inset \subautoref{fig:fig2}{\textbf{a}}].
An alternative fit including resonant frequency disorder allows to extract a more precise estimate of the eigenmode frequencies [see \extfigref{extfig:fig2_1}].

We observe that the qubit interacts strongly with the upper band and weakly with the lower one, as seen from the small and large  CCA modes frequency shifts, respectively [\subautoref{fig:fig2}{\textbf{a}}].
This is a direct consequence of the non-local coupling (giant atom regime) of the transmon qubit to the CCA, which inhomogenously redistributes the coupling strength over the CCA normal modes [see inset \subautoref{fig:fig1}{f} and \extfigref{extfig:fig1_1}]. 
At the upper band edge, the qubit strongly hybridizes with photonic modes, producing a visible frequency shift even when the bare qubit is significantly detuned from the upper band edge [\subautoref{fig:fig2}{\textbf{a}}]. 

When the qubit--mode coupling strength to the $n$th CCA mode, $G_n$ [see \methods], becomes comparable to the mode frequency spacing $\Delta\Omega_n = \Omega_n - \Omega_{n-1}$, the avoided crossing shifts exceed $\Delta\Omega_n$, effectively hybridizing multiple modes simultaneously.
Since there are $N$ qubit--mode coupling strengths but only $N-1$ mode frequency spacings, we define the $n$th \emph{coupling per average mode splitting}, $G_n/\overline{\Delta\Omega_n}$, where each coupling is normalized by the average of the neighboring mode spacings  $\overline{\Delta\Omega_n} = \frac{1}{2}\left(\Delta\Omega_{n+1} + \Delta\Omega_{n}\right)$~[\subautoref{fig:fig1}{\textbf{h}}].
In systems with linear dispersion and uniform coupling, superstrong coupling produces a smooth frequency shift of all modes as a function of the qubit frequency~\cite{kuzminSuperstrongCouplingCircuit2019a}. 
In our case, however, the engineered coupling profile of the qubit to the CCA [see \extfigref{extfig:fig1_1}] leads to an alternating pattern: even modes experience large frequency shifts, while odd modes, weakly coupled at the qubit position, remain nearly unperturbed [see inset \subautoref{fig:fig1}{\textbf{a}} and \aref{app:sec:qubit_CCA}]. 
Consequently, strongly shifted modes overlap and interfere with weakly shifted ones.

\begin{figure*}
    \centering
    \includegraphics[width = \linewidth]{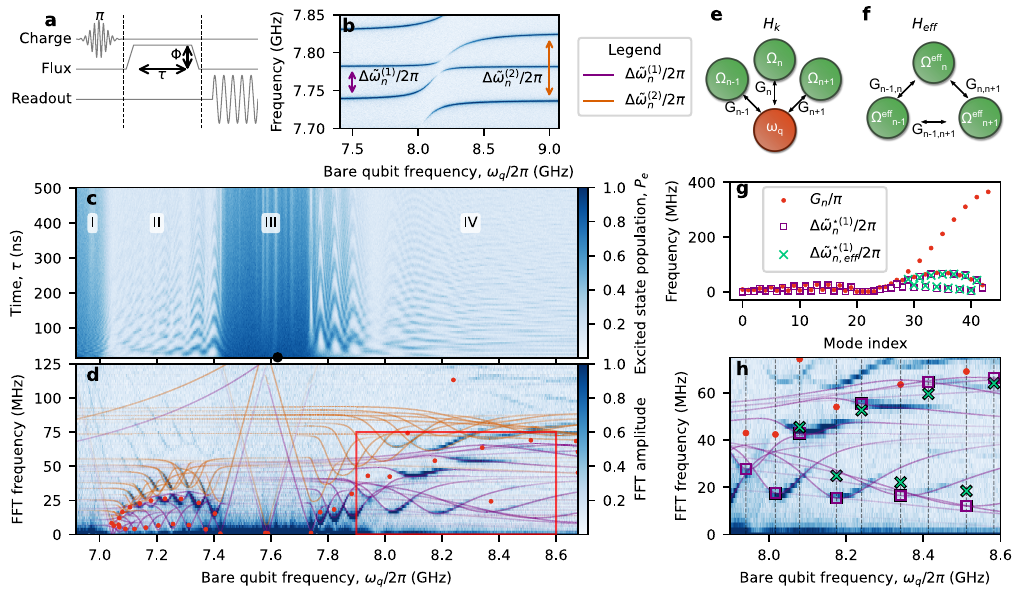}
    \caption{\textbf{Superstrong dynamics.} 
    \textbf{(a)} Pulse sequence: the qubit is excited, non-adiabatically displaced in frequency for a time $\tau$, then returned at its initial frequency for readout.  
    \textbf{(b)} Zoom-in on the spectroscopy of \subautoref{fig:fig2}{\textbf{a}}. 
    Energy spacings between consecutive and next-nearest dressed CCA modes are denoted $\Delta\tilde\omega_n^{(1)}/2\pi$ (purple arrows) and $\Delta\tilde\omega_n^{(2)}/2\pi$ (orange arrows), respectively. 
    \textbf{(c)} Measured qubit excited-state population versus $\tau$ and $\omega_q/2\pi$. 
    The black dot marks the bare qubit frequency at $\tau=0$, corresponding to a dressed frequency $\tilde\omega_q/2\pi = \SI{7.43}{\giga\hertz}$.  
    \textbf{(d)} Fast Fourier transform of the data in panel \textbf{c}. Purple (orange) lines indicate $\Delta\tilde\omega_n^{(1)}/2\pi$ ($\Delta\tilde\omega_n^{(2)}/2\pi$).
    The intensity of the lines is proportional to $|u_i|^2|u_j|^2$. Red dots mark $G_n/\pi$, the expected Rabi rate between the qubit and CCA mode $n$.  
    \textbf{(e)} Hamiltonian $H_k$ in the CCA mode basis [see \methods], with bare eigenmodes of frequency $\Omega_n/2\pi$, coupling rates $G_n/2\pi$, and the bare qubit at $\omega_q/2\pi$. Just three modes are shown.  
    \textbf{(f)} Corresponding effective Hamiltonian $H_{\rm eff}$ from perturbation theory [see \methods], featuring effective eigenmodes of frequency $\Omega_n^{\rm eff}/2\pi$ and induced mode--mode couplings $G_{n,n^\prime}/2\pi$, between modes $n$ and $n^\prime$.  
    \textbf{(g)} Comparison between expected and measured Rabi rates versus mode index. Red dots: $G_n/\pi$. 
    Purple squares: $\Delta\tilde\omega_n^{\star(1)}/2\pi$, the nearest-neighbor mode frequency spacing $\Delta\tilde\omega_n^{(1)}$ at which $|u_n|^2|u_{n+1}|^2$ is maximum.
    Green crosses: $\Delta\tilde\omega_{n, \rm eff}^{\star(1)}/2\pi$, the effective nearest-neighbor mode frequency spacing $\Delta\tilde\omega_{n, \rm eff}^{(1)}/2\pi$ at which $|u_n|^2|u_{n+1}|^2$ is maximum $H_{\rm eff}$.
    \textbf{(h)} Zoom-in of the red frame in panel \textbf{d}, the markers and lines as previously defined in panel \textbf{d} and \textbf{g}. 
    Vertical dashed lines: bare qubit frequencies at which $|u_n|^2|u_{n+1}|^2$ is maximum.  }
    \label{fig:fig3}
\end{figure*}
Due to the high $G_n/\overline{\Delta\Omega_n}$ ratio, the qubit is expected to participate simultaneously in multiple eigenmodes. 
We quantify this with the \emph{atomic participation ratio}~\cite{scigliuzzo2025quantum,minev2021energy}, defined as $\abs{u_m}^2$ for the $m$th dressed eigenstate $\ket{\psi_m}$ [see \autoref{sec:experiment}].
In \subautoref{fig:fig2}{\textbf{b}}, we show the atomic ratios with grey lines, extracted from each of the eigenstates of Hamiltonian \eqrefauto{eq:normalSpaceHamiltonian}, plotted on a logarithmic scale.
Experimentally, we determine $\abs{u_m}^2$ using the Hellmann--Feynman theorem~\cite{scigliuzzo2025quantum,feynmanForcesMolecules1939}, from the measured dressed CCA modes $\tilde\omega_m$, according to $\abs{u_m}^2 = \dd \tilde{\omega}_m / \dd \omega_q$, and report values for several dressed modes in \subautoref{fig:fig2}{\textbf{b}} [additional data in \aref{app:sec:qubit_CCA}].
As required by normalization, $\sum_m \abs{u_m}^2 = 1$. 
The small discrepancies between theory and experiment are attributable to disorder.

For dressed CCA modes with $G_n/\overline{\Delta\Omega_n} \lesssim 1$, the qubit participation ratio exhibits narrow peaks, approaching unity, when the qubit is tuned into resonance with a bare CCA mode.
This behavior is characteristic of a \textit{single-mode coupling} regime, where the qubit primarily interacts with one mode at a time.  
In contrast, when $G_n/\overline{\Delta\Omega_n} > 1$, the participation ratio broadens and its maximum is reduced, marking the onset of the \textit{superstrong coupling} regime, where the qubit interacts simultaneously with multiple CCA modes.  
As $G_n/\overline{\Delta\Omega_n}$ increases, the maxima of the participation ratio decrease smoothly across the upper band.
In this regime, the dressed modes remain mostly photonic, and the strong qubit hybridization induces mode--mode interactions, renormalizing the dynamics of the system~\cite{krimer2014route,lechner2023light}.

\section{Dynamics of the emitter}
\label{sec:fig3}

In \autoref{sec:fig2}, we demonstrated that the qubit interacts simultaneously with several CCA modes by measuring the atomic participation ratio. 
Here, we investigate how the superstrong coupling regime affects the qubit dynamics using the pulse sequence shown in \subautoref{fig:fig3}{\textbf{a}}. 
The qubit is first initialized in the middle bandgap [black dot in \subautoref{fig:fig3}{\textbf{c}}]. 
A constant flux offset, $\Phi$, with a $\SI{2.4}{\nano\second}$ ramp is then applied, shifting the bare qubit frequency for a duration $\tau$. 
Finally, the qubit is returned to its initial frequency and its state measured.

In \subautoref{fig:fig3}{\textbf{c}}, we show the qubit excited-state population as a function of the interaction time $\tau$ and the bare qubit frequency $\omega_q/2\pi$.  
This measurement reveals four distinct interaction regions. 
Within the bandgaps (I, III), minimal interaction with the eigenmodes is observed, as indicated by the absence of Rabi oscillations. 
Nonetheless, enhanced decay rates are noted, attributed to the presence of Two-Level-System fluctuators~\cite{muller2019towards} on the chip and, in the middle bandgap, to weak coupling with the SSH edge modes of the CCA. 
By contrast, when the qubit is tuned within the bands (II, IV), Rabi oscillations appear. 
As the qubit is tuned further into the bands, these oscillations strongly interfere.
The upper bandgap is not resolved in this measurement because the qubit’s strong coupling to the highest-frequency CCA modes suppresses non-adiabatic transfer from the band into the upper APBS; the excitation therefore remains confined to the band and does not probe the upper gap [see \aref{app:sec:qubit_dynamics}].

\begin{figure*}
    \centering
    \includegraphics[width = \linewidth]{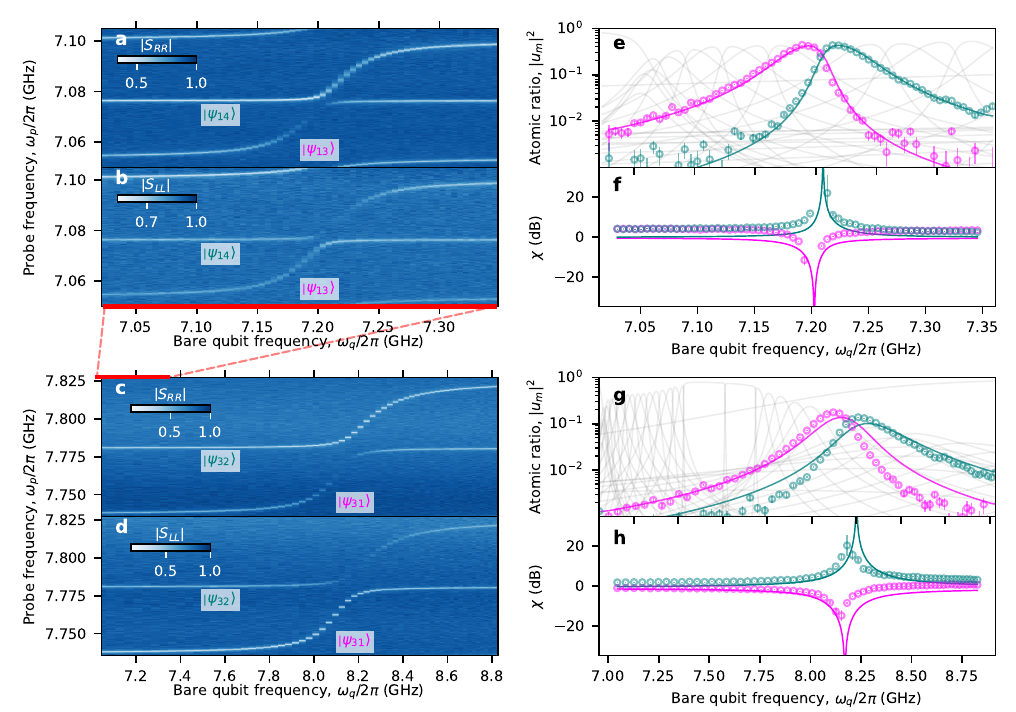}
    \caption{\textbf{Experimental signature of directionality.}
    \textbf{(a)} [\textbf{b}] Magnitude of the reflection spectra from the right [left] port, $|S_{RR}|$ [$|S_{LL}|$], for dressed CCA modes $\ket{\psi_{13}}$ and $\ket{\psi_{14}}$ as a function of the bare qubit frequency.
    \textbf{(c)} [\textbf{d}] Magnitude of the reflection spectra from the right [left] port, $|S_{RR}|$ [$|S_{LL}|$], for dressed CCA modes $\ket{\psi_{31}}$ and $\ket{\psi_{32}}$ as a function of the bare qubit frequency.
    The red lines in panels \textbf{b} and \textbf{c} on the x axes, compares the qubit frequency range spanned in the two cases.
    \textbf{(e)} [\textbf{g}] Extracted atomic ratio $|u_m|^2$ (circles) of dressed CCA modes $\ket{\psi_{13}}$ and $\ket{\psi_{14}}$ [$\ket{\psi_{31}}$ and $\ket{\psi_{32}}$]. 
    Error bars represent the propagated standard deviation of the dressed mode frequencies obtained from fits of the reflection spectra to the input–output response function [see \aref{app:sec:CCA_charac}]. 
    The continuous lines are the expected atomic ratio of each dressed modes extracted from a disorderless fit according to \eqrefauto{eq:normalSpaceHamiltonian}.
    \textbf{(f)} [\textbf{h}] Extracted ratio of external dissipation, $\chi = 10\log_{10}(\gamma_{\rm ext,R}/\gamma_{\rm ext,L})$, reported in dB scale (circles) of dressed CCA modes $\ket{\psi_{13}}$ and $\ket{\psi_{14}}$ [$\ket{\psi_{31}}$ and $\ket{\psi_{32}}$].
    Error bars represent the propagated standard deviation from fits of the reflection spectra to the input–output response function [see \aref{app:sec:CCA_charac}].
    The continuous lines are the expected ratio $\chi$, from a disorder-less non-Hermitian Hamiltonian [see \methods]. 
    }
    \label{fig:fig4}
\end{figure*}

To gain deeper insight, we present in \subautoref{fig:fig3}{\textbf{d}} the Fast Fourier Transform (FFT) of the data shown in \subautoref{fig:fig3}{\textbf{c}}. 
For a qubit coupled to a single mode at frequency $\Omega_n/2\pi$, we expect to observe Rabi oscillations at frequency $\qty(\sqrt{4G_{n}^2 + \Delta_{n}^2}) / 2\pi$, where $\Delta_{n} = \Omega_{n} - \omega_q$ is the CCA mode-qubit frequency detuning.
This picture works in the lower band [see \subautoref{fig:fig3}{\textbf{d}}], where we observe a good match between the expected resonant ($\Delta_n=0$) Rabi frequency and the observed ones. 
This reinforces our claim that the interaction in the lower band mostly remains in the single-mode regime. 
Systematic deviations emerge once the qubit is tuned deep into the upper band, corresponding to a region where the system enters the superstrong coupling regime ($G_n/\overline{\Delta\Omega_n} > 1$). 
Here, the simultaneous participation of the qubit in several CCA modes allows the observation of transitions between (dressed) modes at different (dressed) frequencies $\tilde{\omega}_n$.
This behavior is confirmed by the matching of the transition frequencies (energy differences) $\Delta\tilde\omega_n^{(i)} = \tilde\omega_{n+i} - \tilde\omega_{n}$ for $i=1$ (purple lines) and $i=2$ (orange lines) [\subautoref{fig:fig3}{\textbf{b}}], which the FFT spectrum in \subautoref{fig:fig3}{\textbf{d}} and \extfigref{extfig:dynamics}.

In \subautoref{fig:fig3}{\textbf{g}}, we compare the extracted coupling, $G_n$ (red dots), with the energy difference between neighboring eigenmodes (purple squares) at the qubit frequency of maximum interaction, $\Delta\tilde\omega_n^{\star(1)}$, \ie $\Delta\tilde\omega_n^{(1)}$ at the qubit frequency at which $|u_n|^2|u_{n+1}|^2$ reaches its maximum.
As seen in \subautoref{fig:fig3}{\textbf{d},\textbf{g},\textbf{h}}, these values no longer match once the system enters the superstrong coupling regime [dashed line in \subautoref{fig:fig3}{\textbf{g}}], marking a clear deviation from the single-mode Jaynes-Cummings model.   
To gain further insight into the underlying physics, we derive an effective model that reveals a qubit mediated mode--mode interaction [see \subautoref{fig:fig3}{\textbf{f}} and \methods].  
Although this effective model is strictly valid only when the qubit is detuned from the CCA bands, it exhibits excellent agreement with the energy difference $\Delta \tilde\omega_n^{\star(1)}$ in the upper band [see green crosses vs purple squares in \subautoref{fig:fig3}{\textbf{g},\textbf{h}}]. 
Together, these results demonstrate that our platform enables single-photon dynamics to be tracked across the crossover from the single-mode to the superstrong coupling regime.

\section{Directionality}
\label{sec:fig4}

\begin{figure*}
    \centering
    \includegraphics[width = \linewidth]{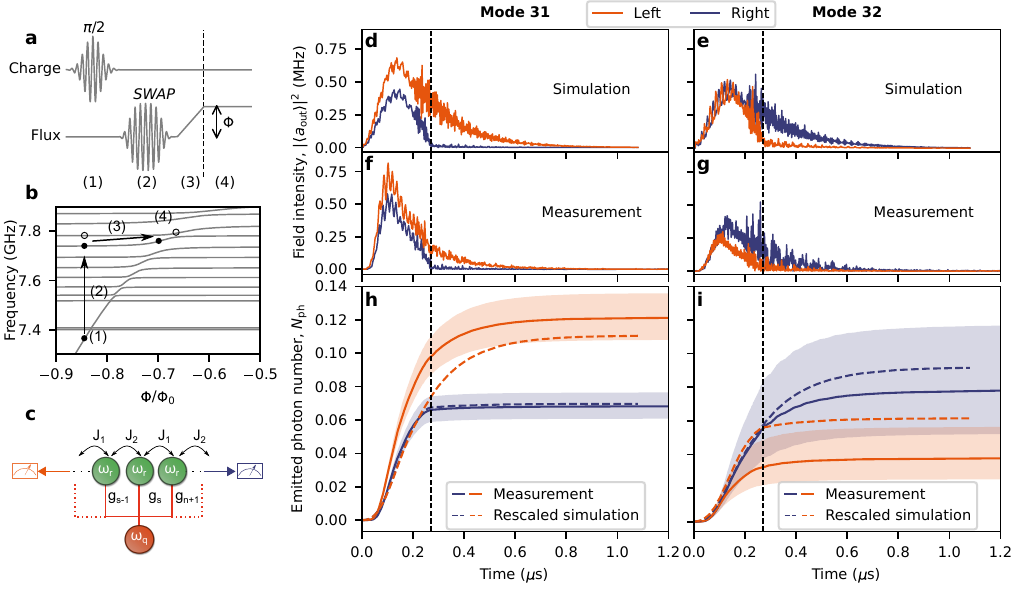}
    \caption{\textbf{Directional emission.} 
    \textbf{(a)} Pulse sequence. The qubit is first excited to the state, \(\qty(\ket{g, \textrm{vac}} + \ket{\psi_{21}})/\sqrt{2}\) (1), it is then transfered to a target dressed CCA mode using a parametric coupling scheme (2). 
    The qubit is adiabatically displaced in frequency (3) to reach the optimal emission point (4). 
    \textbf{(b)} Energy diagram of the system with the pulse sequence. 
    The grey lines represent the dressed CCA modes as a function of the reduced flux, $\Phi/\Phi_0$, threading the SQUID of the transmon. 
    The black dots [circle] show the position of the "excitation" for the steps (1), (2), (3) and (4) as indicated in panel \textbf{a} for mode 31 [32]. 
    \textbf{(c)} Pictorial diagram of the expected emission. 
    The orange (blue) arrows highlight emissions coming from the left (right). 
    This color code is used in the rest of the figure.
    \textbf{(d)} [\textbf{e}] Simulation of the field intensity \(|\langle a_{\rm out}\rangle|^2\) as a function of time [see \methods], following the protocol in panel \textbf{a} for dressed CCA mode 31 [32].
    \(a_{\rm out}\) is the outgoing emitted photon field from the left or right side of the CCA.
    \textbf{(f)} [\textbf{g}] Measured field intensity \(|\langle a_{\rm out}\rangle|^2\), as a function of time, following the protocol in panel \textbf{a} for dressed CCA mode 31 [32].
    \textbf{(h)} [\textbf{i}] Emitted photon number \(N_{\rm ph}\), as a function of time for dressed CCA mode 31 [32]. 
    The continuous lines represent the experimental data, where the shaded area is the uncertainty region due to gain calibration error [see \aref{app:sec:attn_gain_cal}]. 
    The dashed lines are the simulations rescaled to take into account the effect of disorder [see \methods].
    }
    \label{fig:fig5}
\end{figure*}

We focus on the effect of breaking the inversion symmetry of the CCA [inset \subautoref{fig:fig1}{\textbf{f}}], which renders a subset of dressed CCA modes directional, conditional to the bare qubit frequency~\cite{lodahl2017chiral, suarez2025chiral}. 
This effect is evident looking at reflection measurements from both the left and right sides of the CCA, resulting in a different signal contrast near maximum qubit–mode hybridization (\ie maximum atomic ratio) [\subautoref{fig:fig4}{\textbf{a}--\textbf{d}}]. 
More specifically, \subautoref{fig:fig4}{\textbf{a}--\textbf{b}} and \subautoref{fig:fig4}{\textbf{c}--\textbf{d}} show reflection spectra for two representative mode pairs (\((\ket{\psi_{13}}, \ket{\psi_{14}})\) and \((\ket{\psi_{31}}, \ket{\psi_{32}})\)), illustrating the single-mode strong and superstrong coupling regimes, respectively in the lower and upper bands.
For modes $\ket{\psi_{13}}$ ($\ket{\psi_{31}}$), reflection is suppressed on the right and enhanced on the left, whereas the opposite occurs for modes $\ket{\psi_{14}}$ ($\ket{\psi_{32}}$).
As argued in the \methods, this directionality originates from the broken spatial inversion symmetry of the CCA, which enables interference between eigenmodes and confines the dressed wavefunction to one side or the other of the array.

Two main distinctions arise between the two mode pairs: (i) the frequency range over which directionality occurs,  which is broader for the superstrong case [compare red lines on the x-axis in \subautoref{fig:fig4}{\textbf{b},\textbf{c}}], and (ii) the smaller and more widely distributed atomic ratio in the more strongly-coupled modes.

We extract the external dissipation rates to the left and right ports, $\gamma_{\rm ext,L}/2\pi$ and $\gamma_{\rm ext,R}/2\pi$, for the two sets of modes, and report their ratio $\chi = 10\log_{10}(\gamma_{\rm ext,R}/\gamma_{\rm ext,L})$ in dB [see \subautoref{fig:fig4}{\textbf{f},\textbf{h}}].  
The measurements clearly reveal the expected directionality.  
Numerical simulations based on the imaginary parts of the eigenvalues of a disorder-free non-Hermitian Hamiltonian [see \methods] reproduce the main features, but differ in both the background dissipation levels and in the exact qubit frequency at which maximal directionality occurs, which we attribute to disorder [see \aref{app:sec:CCA_charac}].  
At the qubit frequency of maximum directionality, the model predicts ideal localization, with either $\gamma_{\rm ext,R}=0$ and $\gamma_{\rm ext,L}\neq0$ or vice versa.  
Experimentally, however, we do not reach such ideal values: for modes 13 and 14 we extract $\chi = (-11\pm3,\,22\pm9)$~dB, respectively, and for modes 31 and 32, $\chi = (-15\pm3,\,20\pm5)$~dB, comparable to what achieved in \rref~\cite{pakkiam2024experimental} using Rice--Melee lattice.
We attribute the absence of complete suppression not to a fundamental limitation, but to finite signal-to-noise ratio: small external dissipation rates are masked by intrinsic losses and by the noise floor, preventing us from resolving the full directionality predicted by theory.

\section{Directional emission}
\label{sec:fig5}

Directional photon emission has been demonstrated in nanophotonic systems~\cite{sollner2015deterministic, petersen2014chiral}, in superconducting waveguides using interference to generate on-demand directional microwave photons~\cite{kannanOndemandDirectionalMicrowave2023, almanakly2025deterministic,joshiResonanceFluorescenceChiral2023}, and more recently in one-dimensional superconducting CCAs with Rice--Mele topology~\cite{pakkiam2024experimental}.  
Here, building on the directionality of the dressed CCA modes established in the previous section, we exploit this property to selectively suppress qubit emission toward either the left or the right side of the CCA, focusing on dressed modes $\ket{\psi_{31}}$ and $\ket{\psi_{32}}$.  
The pulse sequence and corresponding energy diagram of the system are shown in \subautoref{fig:fig5}{\textbf{a}--\textbf{b}} [see \methods].  
The qubit is first prepared in the middle bandgap in a superposition state 
\(\qty(\ket{g, \textrm{vac}} + \ket{\psi_{21}})/\sqrt{2}\)~\cite{castillo2025dynamical}.  
This state is then transferred into $\ket{\psi_{31}}$ or $\ket{\psi_{32}}$ through a SWAP using a parametric coupling scheme [calibration in \aref{app:sec:cal_SWAP}]~\cite{naik2017random}.  
Finally, the qubit is adiabatically tuned to the point of maximal directional emission via a \SI{120}{\nano\second} flux ramp [see step 3 of \subautoref{fig:fig5}{\textbf{a}--\textbf{b}}].  

In \subautoref{fig:fig5}{\textbf{d}--\textbf{e}}, we show the field intensity, \(|\langle a_{\rm out}\rangle|^2\), to the left and right of the CCA for modes $\ket{\psi_{31}}$ and $\ket{\psi_{32}}$, as obtained from a disorder-free Hamiltonian [see \methods].  
Initially, emission is bidirectional, but after the flux ramp [dashed vertical line in \subautoref{fig:fig5}{\textbf{d}--\textbf{e}}], it becomes predominantly leftward for mode $\ket{\psi_{31}}$ and rightward for mode $\ket{\psi_{32}}$.  
The initial rise of the field intensity reflects the competition between SWAP loading and emission from the mode at rate \(\gamma_m\), producing a wavepacket with finite rise and exponential tail; without the ramp it decays at \(\gamma_m\), while the ramp cancels emission on one side and redirects it to the other side.  
Pronounced oscillations appear at the onset of the ramp, signaling nonadiabatic dynamics, particularly strong for mode $\ket{\psi_{32}}$ whose optimal emission point lies immediately beyond an avoided crossing [hollow circle in \subautoref{fig:fig5}{\textbf{b}}].  

In \subautoref{fig:fig5}{\textbf{f}--\textbf{g}}, we present the measured field intensity for modes $\ket{\psi_{31}}$ and $\ket{\psi_{32}}$ to the left and right of the CCA, averaged over \(75 \times 10^6\) iterations. 
The output gain is directly calibrated at the mode frequencies [see \aref{app:sec:attn_gain_cal}].  
The measurements are in good agreement with the theoretical trends, though they reveal even stronger nonadiabatic behavior for mode $\ket{\psi_{32}}$ than predicted, which we attribute to disorder or distortions in the flux pulses.  

We quantify the directional emission by reporting the emitted photon number \(N_{\rm ph}\), obtained by integrating the measured field intensity as a function of time [solid lines in \subautoref{fig:fig5}{\textbf{h}--\textbf{j}}, for details see \aref{app:sec:emission_measurements}]. 
For comparison, we overlay rescaled theoretical predictions (dashed lines), where the rescaling accounts for disorder-induced changes in mode localization and hence the external dissipation rates [see \methods]. 

The directionality of mode \(m\) is quantified by  
\begin{equation}
    \eta_m = \frac{L - R}{L + R} \, ,
\end{equation}
where \(L\) and \(R\) denote the emitted photon numbers to the left and right ports, respectively. 
By definition, $\eta_m = 1$ ($-1$) corresponds to perfectly left- (right-) directional emission.  
Experimentally, we obtain \(\eta_{31}^{\mathrm{meas}} = 0.279\) and \(\eta_{32}^{\mathrm{meas}} = -0.346\), compared to the rescaled theoretical predictions of \(\eta_{31}^{\mathrm{theory}} = 0.226\) and \(\eta_{32}^{\mathrm{theory}} = -0.196\).  
The discrepancy between theory and measurement is primarily attributed to gain miscalibration of the measurement chain [see \aref{app:sec:attn_gain_cal}].

Three non-idealities limit the emission protocol:
1) the qubit begins emitting immediately after the SWAP gate and becomes directional only after reaching the optimal emission point; 
2) residual nonadiabatic dynamics persist even at the optimal point due to the finite ramp speed;  
3) internal losses in both the qubit and the CCA further reduce the overall emission efficiency. 

Despite these limitations, the experiment demonstrates controlled, mode-selective directional emission of single photons in a superconducting structured environment.

\section{Discussion}

We estimate that directionality could be enhanced by preparing the system in the upper APBS [purple line in \subautoref{fig:fig2}{\textbf{a}}], which allows for direct access of the avoided crossing responsible for directional emission [see \extfigref{ext:fig:ideal_emission}].
In this case, our model predicts nearly ideal performance (\(\eta_{31}^{\mathrm{Theo}} = 0.989\), \(\eta_{32}^{\mathrm{Theo}} = -0.980\)).  
Even better performances are expected from devices optimized for directional emission, with lower internal loss and ports positioned at CCA antinodes.  
In this experiment, this was not implemented due to control-electronics limitations.  

Beyond directionality, several further opportunities emerge. Extending this experiment to two braided qubits would enable decoherence-free interactions in the continuum~\cite{soro_interaction_2023}.  
Generalizing to two-dimensional baths~\cite{gonzalez-tudela_exotic_2018, leonforte_quantum_2024,raaholt2024avoiding,di2025emergent,kollar_hyperbolic_2019,owens_chiral_2022} would provide access to exotic bound states and unconventional transport phenomena.  
Moreover, this platform offers a testbed for nonlinear multimode quantum optics, where the dynamics could be explored beyond the single-photon regime, both inside the bands and in the bandgaps~\cite{longo2010few}.  

Remarkably, our results highlight the role of \textit{coupling engineering}~\cite{wang2024realizing}.  
By tailoring the qubit--CCA connection points, we reshaped the distribution of coupling strengths across the mode spectrum, enhancing interactions with high-frequency symmetric modes while reducing coupling to low-frequency and antisymmetric ones.  
This design enabled selective access to the superstrong regime and already pushes the system toward multimode \textit{ultrastrong} coupling, with \(G_n/\Omega_n \approx 0.05\) at the upper band edge. 
Extending this strategy to qubits with higher anharmonicity—such as fluxoniums~\cite{manucharyan2009fluxonium}, Cooper-pair boxes, or semiconductor quantum dots~\cite{scarlino2022situ}—would open a path to genuinely non-perturbative physics in multimode QED~\cite{ashida2022nonperturbative,sanchez2018resolution}.

\section{Acknowledgements}

The authors thank Carlos Vega, Alejandro Gonzalez-Tudela and Shingo Kono for stimulating discussions. 
P.S. acknowledges support from the Swiss National Science Foundation (SNSF) through the grants Ref. No. 200021\_200418 and Ref. No. 206021\_205335, and from the Swiss State Secretariat for Education, Research and Innovation (SERI) under contract number No MB22.00081. 
P.S. and V.S. acknowledge support from the Swiss State Secretariat for Education, Research and Innovation (SERI) under contract number UeM019-16 and from NCCR SPIN, a National Centre of Competence in Research, funded by the Swiss National Science Foundation (grant number 225153).
V.S. acknowledges support by the Swiss National Science Foundation through Projects No. 200020\_215172, 200021-227992, and 20QU-1\_215928.
D.D.B. acknowledges funding from the European Union - NextGeneration EU, "Integrated infrastructure initiative in Photonic and Quantum Sciences" - I-PHOQS [IR0000016, ID D2B8D520, CUP B53C22001750006].
M.S. acknowledges support from the EPFL Center for Quantum Science and Engineering postdoctoral fellowship. 
E.D.B. and F.C. acknowledge financial support from European Union-Next Generation EU through projects: Eurostart 2022
‘Topological atom-photon interactions for quantum technologies’; PRIN 2022–PNRR No. P202253RLY
‘Harnessing topological phases for quantum technologies’; THENCE–Partenariato Esteso
NQSTI–PE00000023–Spoke 2 ‘Taming and harnessing decoherence in complex networks’.
The device has been fabricated in the Center of Micro-NanoTechnology (CMi) at EPFL.\\

\textbf{Data availability:}
The data used to produce the plots will be available on Zenodo. 
All other data are available from the corresponding authors upon request.\\

\textbf{Code availability:}
The codes used to analyze the data and produce the plots will be available on Zenodo.

\section{Author contributions}

V.J., D.S., M.S., L.P., and P.S. designed the experiment.
D.S., V.J., and S.F. fabricated the devices.
V.J. performed the measurements with guidance from M.S.
V.J. analyzed the data with guidance from L.P.
A.M., E.D.B., V.J., L.P., and M.S. developed the theoretical model; A.M. performed the numerical simulations and derived the effective Hamiltonian, while E.D.B. developed the theory of directionality.
All authors wrote the manuscript.
V.S., F.C., and P.S. supervised the work.

\section{Methods}
\label{methods}

\subsection{Mode-space coupling strength distribution: interference and symmetries}
\label{app:coupling-method}

Given the interaction Hamiltonian in \eqrefauto{eq:normalSpaceHamiltonian}, describing the qubit's interaction with a subset of CCA sites, we would like to compute the qubit's coupling strength to each of the CCA normal modes $\ket{\phi_n}$, \ie the eigenstates of the bare Hamiltonian $H_{\rm CCA}$ with respective eigenstate $\Omega_n$. We then use the form obtained to make some general comments about reaching the \textit{superstrong} coupling regime and the emergence of directional CCA dressed modes.

The interaction Hamiltonian in the site basis reads
\begin{equation}
    H_{\rm INT} = \sum_{s=1}^N g_s \qty(a^\dag_s b+\hc)\,.
\end{equation}
Recalling that $\ket{\phi_n} {=} \sum_{s=1}^N d_{s,n}\ket{s}$, we can express each of the ladder operators $a_s$ as $a_s {=} \sum_n d_{s,n}\phi_n$, where $\ket{\phi_n} {=} \phi_n^\dagger\ket{\rm vac}$. This allows to recast the interaction Hamiltonian in the form 
\begin{equation}
    H_{\rm INT} = \sum_{n} G_n \qty(\phi^\dag_n b+\hc)\,,
\end{equation}
where 
\begin{equation}\label{eq:normal_modes_coupling}
    G_n = \sum_{s=1}^N d^\star_{s,n} \,g_s\,.
\end{equation}

In the case of a standard \textit{small atom}, we have $g_s {=} g\delta_{s,s_0}$, where $s_0$ is the index of the coupling point. Then from \eqrefauto{eq:normal_modes_coupling} we have $G_n{=} g\,d^\star_{s_0,n}$, and the mode-space coupling distribution is only determined by the amplitude of each CCA mode on cavity $s_0$.
However, in the multi-site coupling introduced by the \textit{giant atom}, the coupling distribution is affected by interference between the different eigenmode wavefunctions, that can redistribute the qubit-CCA coupling strength among the CCA eigenmodes in an arbitrary and non-trivial way. In this sense, $G_n$ can be understood as a result of a weighted interference of mode $\ket{\phi_n}$ at different cavities.

For instance, a giant atom configuration allows to couple the qubit strongly only to a subset of CCA bare modes around a certain value of quasi-momentum $k$, corresponding to $G_n$ being peaked around a specific $n$ value. This possibility is intuitively understood given the relation $\Delta k\Delta x {>} 1$ imposed by Fourier analysis, being $\Delta x$ the size of the atom (\eg the number of coupling points). The latter entails that the more the number of coupling points, the smaller number of CCA bare modes will the coupling strength be concentrated to. This is the simple mechanism allowing for selectively reaching the \textit{superstrong} coupling regime for a subset of modes using a giant atom.

In an inversion-symmetric CCA, the CCA bare modes are eigenstates of the inversion symmetry operator $\mathscr{P}$, \ie they have well defined parity around the inversion center. This entails that, if $\mathscr{P}$-symmetry is not broken by the (giant) atom, there cannot exist dressed CCA modes localized only to the left or right of the inversion center (directional CCA modes) and the qubit will couple \textit{only} to even-parity CCA bare modes. 
On the contrary, if the coupling $g_s$ breaks the $\mathscr{P}$-symmetry, such directional CCA modes can emerge at specific qubit frequencies as a result of the mixing of CCA bare modes with even and odd parity.

\subsection{Derivation of the qubit-mediated photon-photon interaction}
\label{app:sec:PMI}

As shown in \autoref{app:coupling-method} the Hamiltonian in \eqrefauto{eq:normalSpaceHamiltonian} can be written in the basis of the eigenmodes of the CCA as
\begin{equation}
   \label{eq:app-total-hamiltonian-in-eigenmodes-basis}
    H =  \omega_{q} {b}^\dagger{b} {-} \frac{E_C}{2} {b}^{\dagger 2} {b}^2{+} \sum_{n=1}^{N} \Omega_n {\phi}_n^\dagger{\phi}_n {+} {G}_n\! \qty( {\phi}_n^\dagger {b} {+} \hc)\,,
\end{equation}
where ${\phi}_n$ is the bosonic annihilation operator of the $n$-th eigenmode of the CCA and $\Omega_n/2\pi$ its frequency. 

We are interested in deriving the effective qubit-mediated photon-photon interaction. 
To do so, we perform a Schrieffer-Wolff transformation~\cite{schriefferRelationAndersonKondo1966,bravyiSchriefferWolffTransformation2011}
\begin{equation}
    \label{eq:app-schrieffer-wolff}
    H_{\rm eff} = e^{{S}} H e^{-{S}} = H + \comm{S} {H} + \frac{1}{2} \comm{S}{\comm{S}{H}}+ \ldots\,,
\end{equation}
where ${S}$ is the generator of the transformation. 
The Hamiltonian in \eqrefauto{eq:app-total-hamiltonian-in-eigenmodes-basis} can be separated into a free part ${H}_0 {=} \sum_{n=1}^{N} \Omega_n {\phi}_n^\dagger{\phi}_n {+} \omega_{q} {b}^\dagger{b}$, a nonlinearity ${H}_\mathrm{nl} {=} {-} \frac{E_C}{2} {b}^{\dagger 2} {b}^2$, and the interaction ${H}_\mathrm{int} {=} \sum_{n=1}^{N} {G}_n \left( {\phi}_n^\dagger {b} + \hc \right)$. 
We choose the generator ${S}$ such that the first-order term in the Schrieffer-Wolff transformation cancels the interaction term, \ie $\comm{S}{H_0}{=} - {H}_\mathrm{int}$. 
This imposes
\begin{equation}
    \label{eq:app-schrieffer-wolff-generator}
    {S} = \sum_{n=1}^{N} \frac{{G}_n}{\Delta_n} \left( {\phi}_n^\dagger {b} - {\phi}_n {b}^\dagger \right) \, ,
\end{equation}
for the generator $S$, where $\Delta_n = \Omega_n - \omega_{q}$ is the detuning between the $n$-th mode and the qubit. This choice is well-posed provided that $\Delta_n{\not=}0$ for each $n$, \ie the qubit is not resonant with any of the CCA bare modes.
Keeping terms in \eqrefauto{eq:app-schrieffer-wolff} up to second order in ${G}_n$, we obtain the effective Hamiltonian
\begin{equation}
    {H}_\mathrm{eff} \simeq {H}_0 + {H}_\mathrm{nl} + \frac{1}{2} [{S}, {H}_\mathrm{int}] + [{S}, {H}_\mathrm{nl}] + \frac{1}{2} [{S}, [{S}, {H}_\mathrm{nl}]] \, .
\end{equation}
Since \eqrefauto{eq:app-total-hamiltonian-in-eigenmodes-basis} conserves the total number of excitations, we can restrict ourselves to a fixed number-of-excitations subspace. 
In particular, in the single-excitation subspace, where ${H}_\mathrm{nl}$ does not appear, the effective Hamiltonian can be written as
\begin{equation}
    \label{eq:app-effective-hamiltonian}
    {H}_\mathrm{eff} {\simeq} \sum_{n=1}^{N} \Omega_n^{\rm eff} {\phi}_n^\dagger {\phi}_n {+} \omega_{q}^{\rm eff} {b}^\dagger {b} {+} \sum_{\substack{n=1 \\ n^\prime \neq n}}^{N} {G}_{n,n^\prime} \!\qty( {\phi}_n^\dagger {\phi}_{n^\prime} {+}\hc)\, ,
\end{equation}
where
\begin{align}
    \label{eq:app-qubit-dispersive-shift}
    \Omega_n^{\rm eff} & = \Omega_n - \frac{{G}_n^2}{\Delta_n} \, , \\
    \omega_{q}^{\rm eff} & = \omega_{q} + \sum_{n=1}^{N} \frac{{G}_n^2}{\Delta_n} \, , \\
    {G}_{n,n^\prime} & = - \frac{{G}_n {G}_{n^\prime} \left(\Delta_n + \Delta_{n^\prime}\right)}{2 \Delta_n \Delta_{n^\prime}} \, .
\end{align}
Thus, an effective photon-photon interaction emerges with a strength ${G}_{n,{n^\prime}}$. 
It is worth noticing that such perturbative approach is valid as long as the ${G}_n / \Delta_n$ ratio is small. 

\extsubautoref{ext:fig:mode-hybridization}{(\textbf{a}-\textbf{b})} compare the eigenfrequencies of the full Hamiltonian in \eqrefauto{eq:app-total-hamiltonian-in-eigenmodes-basis} with the one of effective Hamiltonian in \eqrefauto{eq:app-effective-hamiltonian}.
Despite a small deviation that grows as $\Delta_n{\to}0$, this pertubative approach captures the relevant physics, both in band [\extsubautoref{ext:fig:mode-hybridization}{\textbf{a}}] and out of band [\extsubautoref{ext:fig:mode-hybridization}{\textbf{c}}].
An example at $\omega_q/2\pi = \SI{8.5}{\giga\hertz}$ of the photon-photon coupling matrix ${G}_{n,n'}$ is shown in
\extsubautoref{ext:fig:mode-hybridization}{\textbf{c}}, showing how these mediated interactions are stronger for lower-wavelength modes and, interestingly, mix significantly multiple modes at different frequencies [see \extsubautoref{ext:fig:mode-hybridization}{\textbf{d}-\textbf{f}}].

\subsection{Theory: qubit-induced directionality}
\label{app:localization-method}

We consider in the following a CCA made by $N$ cavities, at a frequency $\omega_r/2\pi$, coupled with nearest-neighbor hopping rate $J/2\pi$ [see \extsubautoref{fig:theory-fig-main}{\textbf{a}}]. 
In the single-excitation sector, the CCA normal modes are $\Omega_n = \omega_r + 2J\cos{\frac{n\pi}{N+1}}$, $n = 1,\dots,N$, and form a band in the energy interval $\comm{\omega_r-2J}{\omega_r+2J}$.
A qubit \textit{locally} coupled to a cavity at position $s_0$ [see \extsubautoref{fig:theory-fig-main}{\textbf{a}}] seeds a set of dressed bound state localized between one of the edges and cavity $s_0$ when the qubit is fine-tuned to a specific frequency~\cite{longhi2007bound}, provided that $s_0$ do not coincide with the CCA inversion center [see \autoref{app:coupling-method}].
Using the Green's function formalism \cite{economou_greens_1979,leonforte_dressed_2021, leonforte_quantum_2024},
in general a dressed bound state having energy $\tilde\omega_{\rm BS}$ can be found in correspondence of a real solution of the equation
\begin{equation}
    \label{eq:pole-equation}
    \tilde\omega_{\rm BS} = \omega_q + g^2 \mel{s_0}{\mathcal G_B(\tilde\omega_{\rm BS})}{s_0}\,,
\end{equation}
where $\omega_q/2\pi$ is the bare qubit frequency, $g/2\pi$ is the qubit's coupling strength to cavity $s_0$ and $\mathcal G_B(z)$ is the CCA's Green's function, defined as $\mathcal G_B(z) = \sum_k \dyad{\phi_n}/(z-\Omega_n)$. 
The wavefunction of the bound state corresponds to the residue of $\mathcal G_B(z)$ at $z=\tilde\omega_{\rm BS}$.
In particular, such state will be localized between one of the edges and cavity $s_0$ if $\mel{s_0}{\mathcal G_B(\tilde\omega_{\rm BS})}{s_0} = 0$, \ie if its wavefunction has a \textit{node} at position $s_0$. From this follows that the energy of this bound state will correspond to the qubit frequency $\omega_q/2\pi$.
In particular, a left-localized state will appear when the bare qubit frequency is resonant with
\begin{align}
    \label{eq:BIC-small}
    \Tilde{\omega}^L_m = \omega_0 + 2J \cos \frac{m\pi}{s_0+1}\,, && m = 1,\dots,s_0,
\end{align}
while for the right-localized dressed states
\begin{align}
    \label{eq:BIC-small2}
    \Tilde{\omega}^R_p = \omega_0 + 2J \cos \frac{p\pi}{N-s_0}\,, && p = 1,\dots,N-s_0-1,
\end{align}
regardless of the coupling strength $g/2\pi$ [see \extsubautoref{fig:theory-fig-main}{\textbf{b}}].
Their localization can be verified using a \textit{directionality quantifier} $Q_m(\omega_q)$ for each dressed mode $\ket{\psi_m}$ defined as
\begin{equation}
    Q_m(\omega_q) = \frac{\sum_{s=0}^{s_0}\abs{\braket{s}{\psi_m}}^2-\sum_{s=s_0+1}^{N}\abs{\braket{s}{\psi_m}}^2}{\sum_{s=0}^N \abs{\braket{s}{\psi_m}}^2}\,,
    \label{eq:directionality_quantifier}
\end{equation}
which take values in $\comm{-1}{1}$, reaching its maximum (minimum) for a mode completely localized to the left (right) of cavity $s_0$. 
In \extsubautoref{fig:theory-fig-main}{\textbf{i-j}}, we plot $Q_m$ for the same pair of modes at different values of $g/J$, confirming that perfect localization happens at the predicted qubit frequencies regardless of $g/J$. Notably, when the qubit coupling is stronger, the directionality quantifier exhibits a broader peak around the predicted frequency [see \extsubautoref{fig:theory-fig-main}{\textbf{m}}] and the APR goes down [see \extsubautoref{fig:theory-fig-main}{\textbf{n}}], signaling that the localized modes become mostly photonic. 
Thus, we predict a better efficiency in the directional emission protocol, introduced in \autoref{sec:fig5}, for strongly coupled qubits.

As shown in the experiment, a \textit{superstrong} coupling can be reached with the aid of a giant atom. We now investigate whether directional localization of the eigenmodes is possible also in this case.
To show this, if $g_s/2\pi$ is the coupling strength of the giant atom to cavity at position $s$, we introduce an effective cavity $\ket{\chi}$ defined as $\ket{\chi} = \Bar{g}^{-1} \sum_s g_s \ket{s}$, where $\Bar{g}^2 = \sum_s g_s^2$, which allows us to express \eqrefauto{eq:pole-equation} for a giant atom as~\cite{leonforte_quantum_2024},
\begin{equation}
   \tilde\omega_{\rm BS} = \omega_q + \Bar{g}^2 \mel{\chi}{\mathcal G_B(\tilde\omega_{\rm BS})}{\chi}
\end{equation}
and the node condition as $\mel{s_0}{\mathcal G_B(\tilde\omega_{\rm BS})}{\chi} \approx 0$. 
For instance, the latter is satisfied at frequencies in \eqrefauto{eq:BIC-small} provided that $G_{m_0}/\delta\omega{\ll} 1$, being $\delta\omega{=}\abs{\Tilde{\omega}_m^L - \Omega_{m_0}}$, where $\Omega_{m_0}/2\pi$ is the frequency of the CCA bare mode quasi-resonant to a given $\Tilde{\omega}_m^L$ and, accordingly, $G_{m_0}/2\pi$ is the coupling to such mode. 
If this is true, maximum localization is obtained for
\begin{equation}
    \omega_q = \tilde\omega_{\rm BS} - \Bar{g}^2 \mel{\chi}{\mathcal G_B(\tilde\omega_{\rm BS})}{\chi}\,,
\end{equation}
now bearing a dependency on the average coupling strength $\Bar{g}$, which introduces a shift between the directional mode and the qubit frequency. This can be estimated as $ \abs{\mel{\chi}{\mathcal G_B(\tilde\omega_{\rm BS})}{\chi}}{\sim} (\delta\omega)^{-1}$, entailing that the stronger the coupling $\Bar{g}$, the bigger the shift.

Condition $G_{m_0}/\delta{\omega}\ll1$ can be achieved easily noticing that the mode $\ket{\phi_{m_0}}$  will be anti--symmetric with respect to cavity $s_0$, meaning that if the coupling $g_s$ is symmetric around cavity $s_0$, $G_{m_0}$ will be highly-suppressed [see discussion in \autoref{app:coupling-method}]. 
Thus, in \extfigref{fig:theory-fig-main} we study a giant atom coupled to $5$ cavities, with $g_s/2\pi$ having a truncated Gaussian shape centered at $s_0$.
The average coupling matches the \textit{small atom} case, \ie  \(\bar g = \sqrt{\sum_s g_s^2} = g_{s_0}\).
Although localization is still observed regardless of $\Bar{g}$ [\extsubautoref{fig:theory-fig-main}{\textbf{d}}], different coupling strength affects significantly both the spectrum [\extsubautoref{fig:theory-fig-main}{\textbf{g-h}}] and the directionality quantifier [\extsubautoref{fig:theory-fig-main}{\textbf{k-l}}]. In particular, if $\Bar{g}/J>1$ changing the qubit frequency induces a slow displacement of the eigenfrequencies and the qubit frequency is strongly shifted from the directional modes frequencies. This is better visible comparing $Q_m$ for a fixed pair of modes at different $\Bar{g}/J$ values [\extsubautoref{fig:theory-fig-main}{\textbf{o}}].
Notice that this coupling configuration highly enhances the coupling to low--wavelength modes, resulting in the peaks in the directionality quantifier for different dressed modes to start overlapping with each other at the same bare qubit frequency. This means that, while localization in the \textit{small atom} system results from fine tuning of the parameter, in a \textit{giant atom} system the qubit can induce localization of a big subset of modes at the same time, removing the need for fine tuning.

Our argument is independent on the detail of the bath provided that the localization problem for a small qubit admits a set of localized solutions. Thus, it explains the results in the present experiment.

\subsection{Simulation of the losses}
\figref{fig:fig4} presents simulations of mode dissipation as a function of the bare qubit frequency, $\omega_q/2\pi$. 
These simulations employ the non-Hermitian Hamiltonian formalism, which is derived similarly to that in Refs.~\cite{scigliuzzoControllingAtomPhotonBound2022a, jouannyHighKineticInductance2025}. 
Here, $j$ and $l$ are indices of the Hamiltonian matrix. 
The non-Hermitian Hamiltonian is given by:
\begin{equation}
\begin{split}
    &{H_{j,l}^\text{Non-Herm}} = \\
    H_{j,l} -&
    \frac{i}{2}\begin{cases}
        \kappa_{\rm int} \text{, if } j = l = \{1,44\}\,,\\
        \kappa_q  \text{, if } j = l = 45\,,\\
        \kappa_{\rm ext,L} \text{, if } j = l = 1\,,\\
        \kappa_{\rm ext,R} \text{, if } j = l = 44\,,\\
        \kappa_{\rm ext,L}^\prime \text{, if } j = l = 2\,,\\
        \kappa_{\rm ext,R}^\prime \text{, if } j = l = 43\,,\\
        2\sqrt{\kappa_{\rm ext,L}\kappa_{\rm ext,L}^\prime} \text{, if } (j,l) = (1,2) \text{ or } (2,1)\,,\\
        2\sqrt{\kappa_{\rm ext,R}\kappa_{\rm ext,R}^\prime} \text{, if }(j,l) = (44,43) \text{ or } (43,44)\,.\\
    \end{cases}
\end{split}
\label{met:eq:nonHermtianHam}
\end{equation}
where $\kappa_{\rm int}/2\pi$ is the internal dissipation rate of the cavity sites, $\kappa_q/2\pi$ is the bare qubit total dissipation rate, $\kappa_{\rm ext,L(R)}/2\pi$ are the external coupling rates of the left (right) cavity to the left (right) measurement port, and $\kappa_{\rm ext,L(R)}^\prime/2\pi$ are the external coupling rates of the second to left (right) cavity to the left (right) measurement port.

The imaginary part of the eigenvalues of the non-Hermitian Hamiltonian provides the dissipation profile of the eigenmodes:
\begin{equation}
    \text{Im}(\text{Eig}({H^\text{Non-Herm}})) = 2\Vec{\gamma}_{\rm Tot}\,,
    \label{met:eq:diss_non_herm}
\end{equation}
where $\Vec{\gamma}_{\rm Tot} = \Vec{\gamma}_{\rm int} + \Vec{\gamma}_{\rm ext,L} + \Vec{\gamma}_{\rm ext,R}$ is the total dressed dissipation rate vector for all eigenmodes.

By setting certain dissipation rates to zero, we can selectively extract the external dissipation rates. 
Specifically, setting $\kappa_{\rm int} = \kappa_q = 0$ and either $\kappa_{\rm ext,L} = \kappa_{\rm ext,L}^\prime = 0$ or $\kappa_{\rm ext,R} = \kappa_{\rm ext,R}^\prime = 0$ allows the extraction of the right or left external dissipation rates, respectively, as formalized in:
\begin{equation}
\begin{split}
    \Im(\text{Eig}({H^\textrm{Non-Herm}})) =\\
    2\begin{cases}
        \Vec{\gamma}_{\rm ext,R} \text{ if }  \kappa_{\rm ext,L} = \kappa_{\rm ext,L}^\prime = \kappa_{\rm int} = \kappa_q = 0\,,\\
        \Vec{\gamma}_{\rm ext,L} \text{ if }  \kappa_{\rm ext,R} = \kappa_{\rm ext,R}^\prime = \kappa_{\rm int} = \kappa_q = 0\,.
    \end{cases}
\end{split}
\label{met:eq:diss_non_herm_spec}
\end{equation}

The extraction of the bare CCA dissipations is described in \aref{app:sec:CCA_charac}.
The external dissipation rates, $\Vec{\gamma}_{\rm ext,R}$ and $\Vec{\gamma}_{\rm ext,L}$, are sensitive to disorder in the resonant frequencies of the CCA sites. 
This sensitivity arises because disorder affects mode localization, thereby altering the field amplitude at the coupling ports. 
Further details are provided in \aref{app:sec:CCA_charac}.

\subsection{Simulation of the open dynamics}

The simulation of the open dynamics is performed by solving the Lindblad master equation
\begin{equation}
    \frac{d}{dt} \rho = - i \comm{H}{\rho} + \kappa_q \mathcal{D} \left[ b \right] \rho + \sum_{s=1}^N \kappa_s \mathcal{D} \left[ a_s \right] \rho \, ,
\end{equation}
where $\mathcal{D} [O] \rho = O \rho O^\dagger - (O^\dagger O \rho + \rho O^\dagger O) / 2$ is the dissipator and $\kappa_s$ is the total dissipation rate of site $s$ [see \eqrefauto{met:eq:nonHermtianHam}]. 
The system preserves a weak $U(1)$ symmetry and we can thus work in the subspace containing only zero and one excitations in total. 
The time evolution is then numerically integrated using the \texttt{QuantumToolbox.jl} package in Julia~\cite{mercurio2025quantumtoolboxjl}.

\subsection{Device parameters}
Device parameters are summarized in \exttabref{ext:tab:params}. 
The extraction of CCA and qubit parameters is described in \aref{app:sec:CCA_charac} and \aref{app:sec:qubit:charac}, respectively.

\subsection{Pulse sequence in Figure 3}
The pulse sequence employed in \figref{fig:fig3} begins with a \SI{40}{\nano\second} DRAG $\pi$-pulse with the qubit initialized at $\omega_q/2\pi = \SI{7.62}{\giga\hertz}$, corresponding to a dressed state frequency of $\tilde\omega_q/2\pi = \SI{7.43}{\giga\hertz}$.
Subsequently, a constant flux offset is applied, with a holding time $\tau$ ranging from 16 to \SI{500}{\nano\second} in \SI{4}{\nano\second} steps.
The ramp time for this flux offset is \SI{2.4}{\nano\second}.
Finally, the qubit is returned to its initial frequency, and its state is read out for \SI{2}{\micro\second} via conventional dispersive readout.

\subsection{Pulse sequence in Figure 5}
The pulse sequence employed in \figref{fig:fig5} begins with a \SI{40}{\nano\second} DRAG $\pi/2$-pulse, with the qubit initialized at $\omega_q/2\pi = \SI{7.56}{\giga\hertz}$, corresponding to a dressed state frequency of $\tilde\omega_q/2\pi = \SI{7.384}{\giga\hertz}$.
The qubit state is transferred to mode 31 (32) using a supergaussian parametric SWAP pulse at \SI{356.5}{\mega\hertz} (\SI{399.5}{\mega\hertz}) for \SI{160}{\nano\second}.
The qubit frequency is then ramped to $\omega_q/2\pi = \SI{8.15}{\giga\hertz}$ for mode 31 and to $\omega_q/2\pi = \SI{8.22}{\giga\hertz}$ for mode 32, with a ramp time of \SI{120}{\nano\second}.
This fast ramp rate causes non-adiabatic transitions between dressed CCA modes 31 and 32, and with other dressed CCA modes [see \aref{app:sec:emission_measurements}].
However, these non-adiabatic transitions can be avoided by omitting the ramp from the pulse sequence, initializing the qubit from the upper APBS as shown in \extfigref{ext:fig:ideal_emission}.

\subsection{Emission measurements}

\paragraph{Rescaling.}
In \figref{fig:fig5}, we compare the experimental data with rescaled theoretical predictions.  
The rescaling compensates for the absence of disorder in the simulations, since implementing the exact disorder realization of our device is not feasible.  
To account for this, the theory is rescaled by the ratio
\[
\Gamma_{L(R)}^{(i)} = \frac{\gamma_{\rm ext,L(R)}^{i,\mathrm{Simu}}}{\gamma_{\rm ext,L(R)}^{i,\mathrm{Meas}}}\,,
\]
which corrects for disorder-induced modifications of eigenmode localization not captured in the disorder-free Hamiltonian.  
Here, \(\gamma_{\rm ext,L(R)}^{i,\mathrm{Simu}}\) and \(\gamma_{\rm ext,L(R)}^{i,\mathrm{Meas}}\) denote the simulated and measured external dissipation rates, respectively, for mode \(i\) from the left (right) port.  
Since the emitted photon number scales linearly with the external dissipation (\(n \propto \gamma_{\rm ext,L(R)}\)), this procedure provides a consistent comparison.  
We find good agreement for mode 31, whereas for mode 32 noticeable discrepancies remain, which we mainly attribute to uncertainties in the gain calibration [see \aref{app:sec:attn_gain_cal}].

\paragraph{Calibration of the gain:} The gain for mode 31 and 32 is calibrated using AC-Stark shift measurements. 
A detailed description of the calibration is in \aref{app:sec:attn_gain_cal}.
The calibration is performed with the bare qubit set to the initial frequency for the directional emission protocol at $\omega_q/2\pi = \SI{7.55}{\giga\hertz}$, corresponding to a dressed qubit frequency of $\tilde \omega_q/2\pi = \SI{7.38}{\giga\hertz}$.
Below we briefly describe the calibration sequence:
\begin{enumerate}
    \item \textbf{Measurement of the modes' dispersive shift:} The frequency response of modes 31 and 32 are measured using single-tone spectroscopy in reflection with the qubit in its ground state, then the qubit in its excited state. This allows for the extraction of the dispersive shift.
    \item \textbf{AC-Stark shift measurement:} The qubit AC-Stark shift is measured as a function of the input power in modes 31 and 32 from both the left and right port of the CCA. This allows for a precise extraction of the attenuation for both modes from both sides of the CCA.
    \item \textbf{Extraction of the gain:} The gain is extracted by measuring modes 31 and 32 using single-tone spectroscopy in reflection from both the left and the right side. Knowing that the baseline of the reflected signal should be equal to 1 for a calibrated system, \ie\(\langle a_{\rm out}\rangle = \langle a_{\rm in}\rangle\), we can find the gain for both modes from both sides of the CCA.  
\end{enumerate}

\clearpage
\section{Extended Data Figure}

\clearpage
\begin{extendedfigure}
    \centering
    \includegraphics[width=\linewidth]{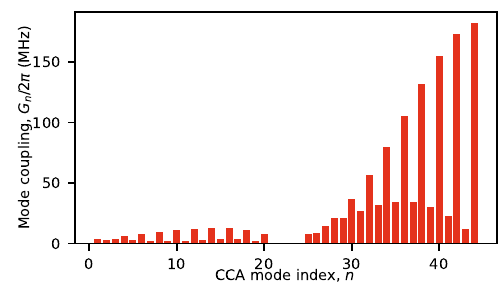}
    \caption{\textbf{Qubit–mode coupling $G_n/2\pi$ versus mode index $n$.}}
    \label{extfig:fig1_1}
\end{extendedfigure}

\begin{extendedfigure*}
    \centering
    \includegraphics[width = \linewidth]{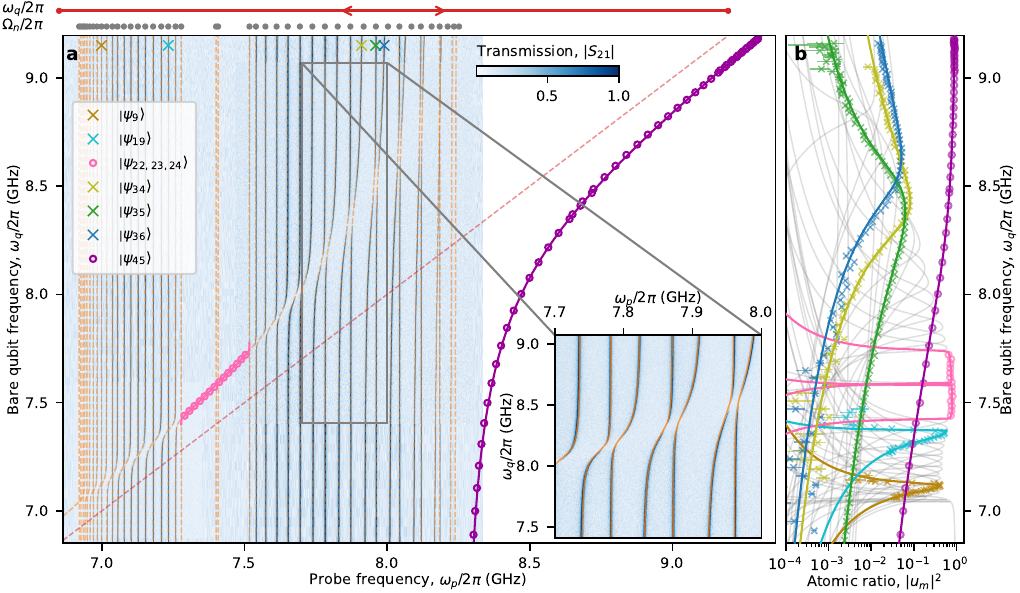}
    \caption{\textbf{Steady-state interaction of a giant atom interacting with a structured photonic bath (comparison to a model including disorder)}
    \textbf{(a)} Single-tone transmission of the CCA as a function of bare qubit frequency $\omega_q/2\pi$.  
    The dressed qubit frequency extracted from two-tone spectroscopy is shown above the upper band (purple circles) and inside the middle bandgap (pink circles).  
    Model predictions according to \eqrefauto{eq:normalSpaceHamiltonian} for the dressed qubit are shown as purple and pink lines, and the bare qubit frequency as a red dashed line.  
    Crosses indicate the dressed CCA modes analyzed in panel \textbf{b}. 
    Inset: zoom-in from $\omega_p/2\pi = \SI{7.7}{\giga\hertz}$ to \SI{8}{\giga\hertz}.
    The orange lines in the inset and orange dashed lines in the main panel show the eigenvalues of the fitted model according to \eqrefauto{eq:normalSpaceHamiltonian}, including resonant frequency disorder with standard deviation $\sigma = \SI{21.8}{\mega\hertz}$.
    \textbf{(b)} Atomic participation ratio of each dressed CCA modes, $|u_m|^2$.  
    Continuous lines: expected atomic participation ratio, $|u_m|^2$, of each dressed CCA modes according to \eqrefauto{eq:normalSpaceHamiltonian}; circles and crosses: values extracted from two-tone and single-tone spectroscopy, respectively (error bars from the standard deviation of fitted dressed-mode frequencies), using $\abs{u_m}^2 = \dd \tilde{\omega}_m / \dd \omega_q$.   
    Mode labels are indicated in the legend in panel \textbf{a}.
    }
    \label{extfig:fig2_1}
\end{extendedfigure*}

\begin{extendedfigure*}
  \centering
    \includegraphics[width=\linewidth]{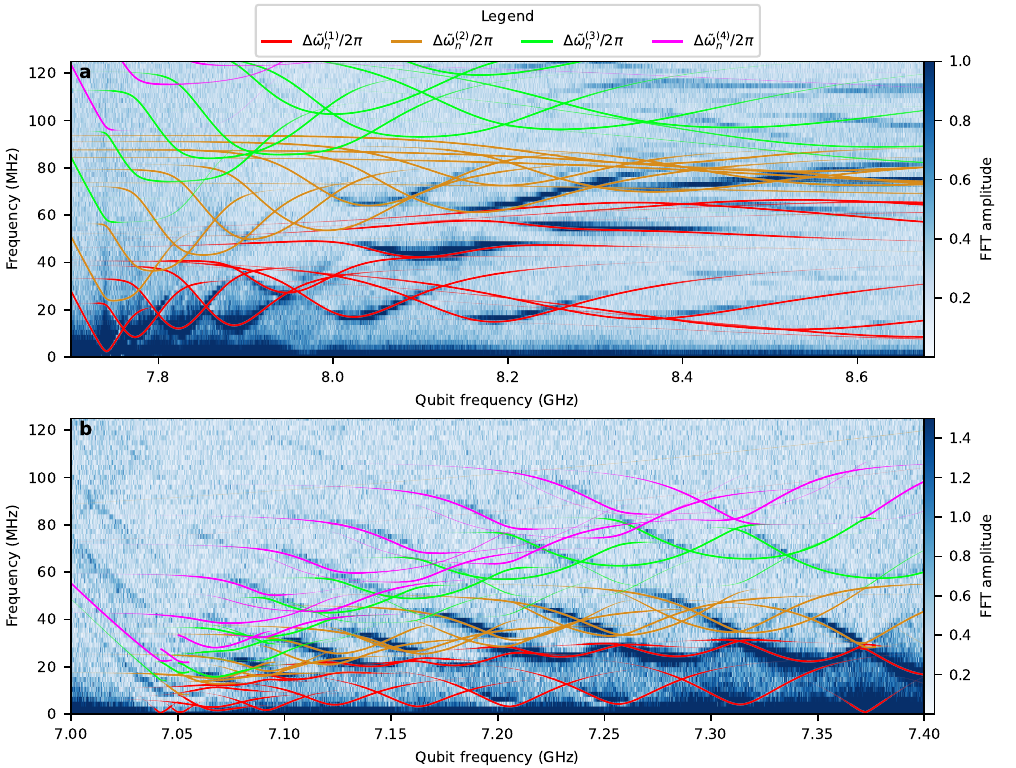}
  \caption{\textbf{Zooms-in on the FFT of the time domain trace of \figref{fig:fig3}.}
  The colormap is saturated to highlight the features.
  \textbf{(a)} Zoom in the upper band.
  \textbf{(b)} Zoom in the lower band.
  The red lines corresponds to $\Delta\tilde\omega_n^{(1)}/2\pi$, the orange lines to $\Delta\tilde\omega_n^{(2)}/2\pi$, the green line to $\Delta\tilde\omega_n^{(3)}/2\pi$ and the purple line to $\Delta\tilde\omega_n^{(4)}/2\pi$.}\label{extfig:dynamics}
\end{extendedfigure*}

\begin{extendedfigure*}
    \centering
    \includegraphics{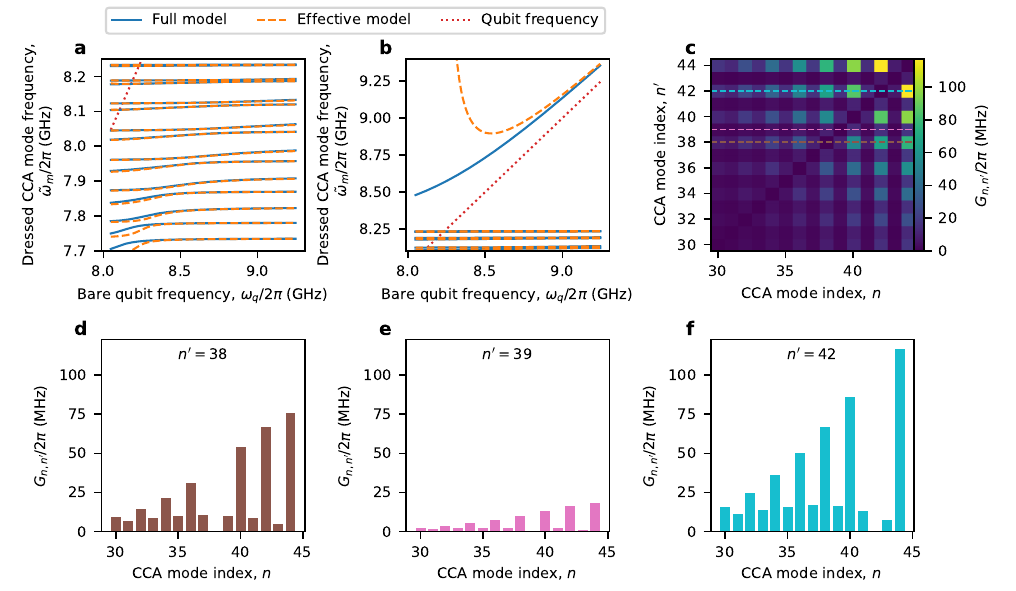}
    \caption{\textbf{Photon–Photon Mode Hybridization.}
    \textbf{(a)} [\textbf{(b)}] Dressed CCA mode frequency comparison between the eigenvalues of the full Hamiltonian, $\tilde\omega_n/2\pi$, (solid blue line) and the effective Hamiltonian, $\tilde\omega_{n, \rm eff}/2\pi$, (dashed orange line), in the upper band [upper bandgap]. The dotted red line corresponds to the bare qubit frequency, $\omega_q/2\pi$. 
    \textbf{(c)} Photon–photon coupling matrix, which increases for higher mode indices.
    \textbf{(d)}, \textbf{(e)} and \textbf{(f)}, line cuts of the photon-photon coupling matrix, for $n^\prime = 37$, 38 and 41, respectively.}
    \label{ext:fig:mode-hybridization}
\end{extendedfigure*}

\begin{extendedfigure*}[h!]
    \centering
    \includegraphics[width=\linewidth]{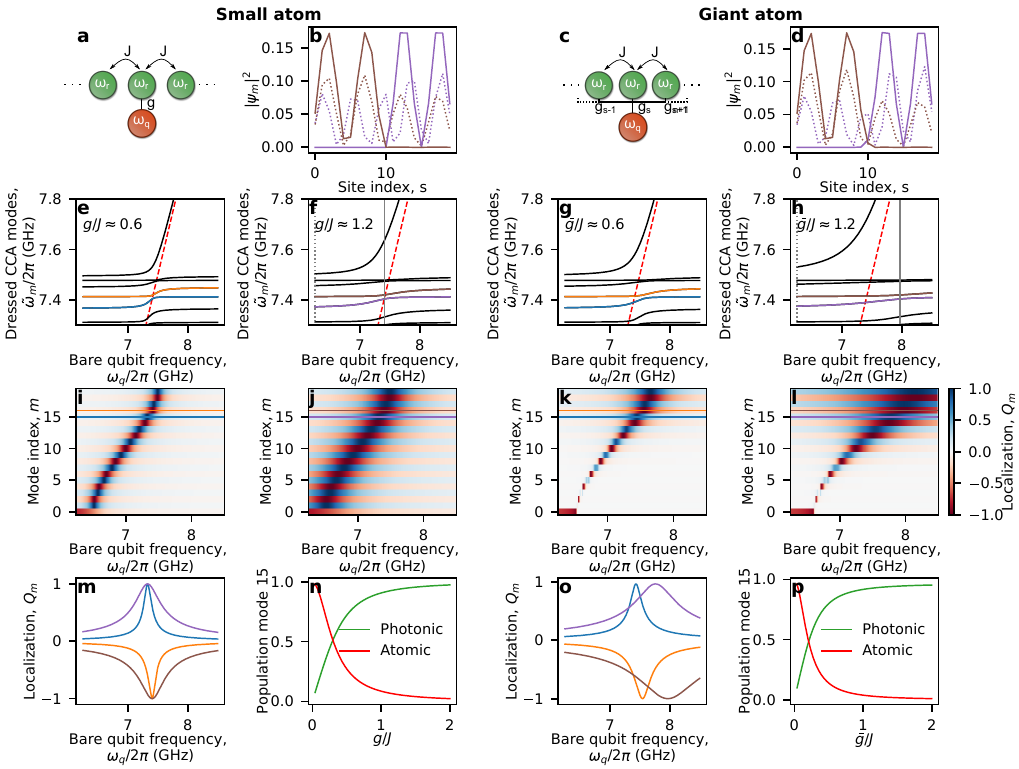}
    \caption{\textbf{Understanding directionality.} 
    (\textbf{a},\textbf{c}) Schematic of a qubit coupled \textit{locally} [\textbf{a}] or \textit{non-locally} [\textbf{c}] to a homogeneous 1D CCA.  
    (\textbf{b},\textbf{d}) Localization of the pair of modes highlighted in purple and brown in panels \textbf{f},\textbf{h}, shown for qubit frequencies where the modes are delocalized (dotted) and localized (solid).  
    (\textbf{e},\textbf{f}) Spectrum as a function of the bare qubit frequency $\omega_q/2\pi$ for a small atom with $g/J = 0.6$ [\textbf{e}] and $g/J = 1.2$ [\textbf{f}].  
    (\textbf{g},\textbf{h}) Spectrum for a giant atom with $\bar g/J = 0.6$ [\textbf{g}] and $\bar g/J = 1.2$ [\textbf{h}], where the qubit couples to five cavities with a truncated Gaussian profile.  
    In all cases, $N=20$, $g=\bar g = \sqrt{\sum_s g_s^2}$, $\omega_r = 28J$, $J/2\pi = \SI{0.25}{\giga\hertz}$, and the qubit couples to cavity 10 (red circle). The red dashed line follows the bare qubit frequency. Blue/orange and purple/brown lines highlight modes 15 and 16 for $g/J=0.6$ and $g/J=1.2$, respectively.  
    (\textbf{i}--\textbf{l}) Localization of the dressed CCA modes $Q_m$ as a function of $\omega_q/2\pi$, corresponding to panels \textbf{e}--\textbf{h}.  
    (\textbf{m},\textbf{o}) Linecuts of $Q_m$ for modes 15 and 16 at $g/J = 0.6$ and $g/J = 1.2$ for the small atom [\textbf{m}] and giant atom [\textbf{o}].  
    (\textbf{n},\textbf{p}) Photonic (green) and atomic (red) fractions of dressed mode 15 as a function of $g/J$, evaluated at the qubit frequency where the mode is maximally localized, for the small atom [\textbf{n}] and giant atom [\textbf{p}].  
    }
    \label{fig:theory-fig-main}
\end{extendedfigure*}

\begin{extendedfigure*}[t!]
  \centering
    \includegraphics[width=\linewidth]{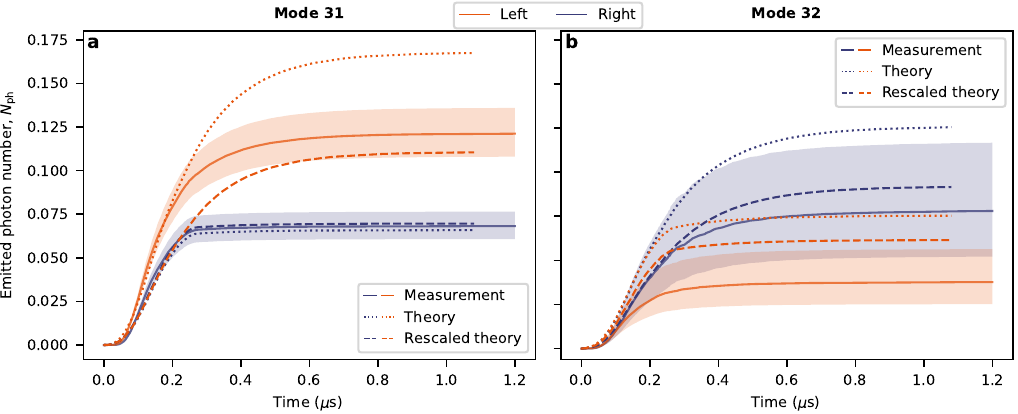}
    \caption{\textbf{Emitted photon number \(N_{\rm ph}\).}
    Same measurement/data as shown in \subautoref{fig:fig5}{\textbf{h}--\textbf{i}}, including non rescaled simulations.
    Emitted photon number, \(N_{\rm ph}\), during the directional emission protocol for mode $\ket{\psi_{31}}$ panel \textbf{a} and mode $\ket{\psi_{32}}$ panel \textbf{b}.
    The continuous lines represent the experimental data, where the shaded area is the uncertainty region due to gain calibration error [see \aref{app:sec:attn_gain_cal}]. 
    The dashed lines are the simulations.
    The dashed lines are the simulations rescaled to take into account the effect of disorder [see \methods].
    }
\end{extendedfigure*}

\begin{extendedfigure}[t!]
    \centering
    \includegraphics[width = \linewidth]{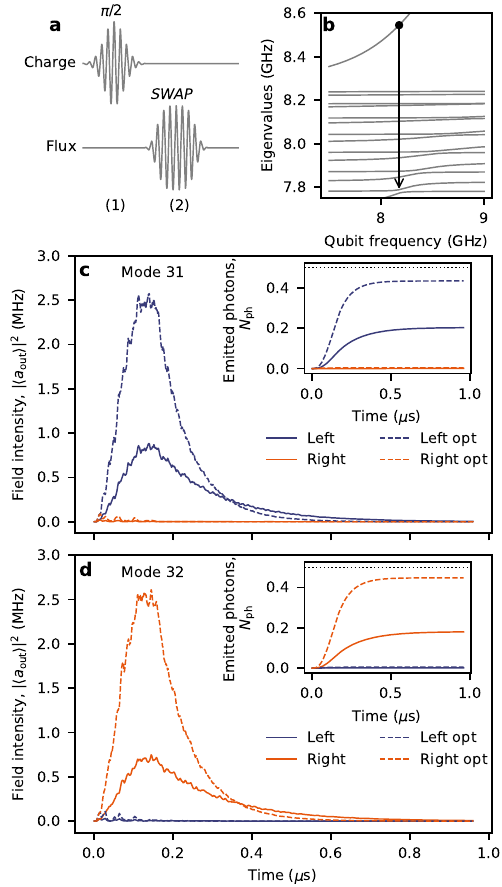}
    \caption{\textbf{Optimized emission simulation.}
    \textbf{(a)} Pulse sequence. Step (1): The APBS is first excited in the state $\qty(\ket{\textrm{vac}, g} + \ket{\psi_{45}})/\sqrt{2}$. Step (2): The state is transferred to the mode of interest with a SWAP gate.
    \textbf{(b)} Energy diagram. The APBS is initialized at the qubit frequency corresponding to the black dot, and then transferred to the mode displaying directionality (arrow).
    \textbf{(c)} and \textbf{(d)} Simulation of the field intensity, \(|\langle  a_{\rm out}\rangle|^2 \), for modes $\ket{\psi_{31}}$ and $\ket{\psi_{32}}$, respectively, as a function of time. 
    The continuous lines show the optimal emission for the device parameters [see \exttabref{ext:tab:params}], while the dashed lines show the emission for a similar device with better loss parameters (\(\kappa_{\rm int}/2\pi = \SI{75}{\kilo\hertz}\), \(\kappa_{\rm ext, L}/2\pi = \kappa_{\rm ext, R}/2\pi = \SI{50}{\mega\hertz}\) and \(\kappa_{q}/2\pi = \SI{16}{\kilo\hertz}\)). 
    The insets shows the emitted photon number, \(N_{\rm ph} \), as a function of time. 
    The dotted black line indicates the theoretical maximum number of emitted photons.}
    \label{ext:fig:ideal_emission}
\end{extendedfigure}

\begin{extendedtable*}[t!]
\centering
\renewcommand{\arraystretch}{1.5}
\begin{tabular}{|c|c|c|}
\hline
\multicolumn{3}{|c|}{\textbf{\large Metamaterial}} \\ \hline
\textbf{Parameter} & \textbf{Value} & \textbf{Method} \\ \hline
$L_g$ & \SI{16.80}{\nano\henry} & Fit of the metamaterial spectrum + microwave simulation. \\ \hline
$C_g$ & \SI{23.04}{\femto\farad} & \multirow{13}{*}{Fit of the metamaterial spectrum} \\ \cline{1-2}
$C_1$ & \SI{1.84}{\femto\farad} &  \\ \cline{1-2}
$C_2$ & \SI{2.72}{\femto\farad} &  \\ \cline{1-2}
$C^\prime$ & \SI{0.38}{\femto\farad} &  \\ \cline{1-2}
$C^{\prime\prime}$ & \SI{0.13}{\femto\farad} &  \\ \cline{1-2}
$Z_r$ & \SI{789}{\ohm} &  \\ \cline{1-2}
$\omega_r/2\pi$ & \SI{7.749}{\giga\hertz} &  \\ \cline{1-2}
$J_1/2\pi$ & \SI{258.8}{\mega\hertz} &  \\ \cline{1-2}
$J_2/2\pi$ & \SI{370.5}{\mega\hertz} &  \\ \cline{1-2}
$J^\prime/2\pi$ & \SI{47.5}{\mega\hertz} &  \\ \cline{1-2}
$J^{\prime\prime}/2\pi$ & \SI{12.7}{\mega\hertz} &  \\ \cline{1-2}
$J^{\prime\prime\prime}/2\pi$ & \SI{5.19}{\mega\hertz} &  \\ \cline{1-2}
$J^{\prime\prime\prime\prime}/2\pi$ & \SI{2.1}{\mega\hertz} &  \\ \hline
$\kappa_\text{ext,R}/2\pi$ & $13.67\pm0.32$ MHz & \multirow{5}{*}{Fit of the metamaterial dissipations.} \\ \cline{1-2}
$\kappa_\text{ext,L}/2\pi$ & $11.12\pm0.14$ MHz &  \\ \cline{1-2}
$\kappa_\text{ext,R}^\prime/2\pi$ & $52.64\pm7.78$ kHz &  \\ \cline{1-2}
$\kappa_\text{ext,L}^\prime/2\pi$ & $28.70\pm2.69$ kHz &  \\ \cline{1-2}
$\kappa_\text{int}/2\pi$ & $590\pm26$ kHz &  \\ \hline

\multicolumn{3}{|c|}{\textbf{\large Qubit}} \\ \hline
\textbf{Parameter} & \textbf{Value} & \textbf{Method} \\ \hline
$\omega_{q,0}/2\pi$ & 9.29 GHz & Fit of the metamaterial spectrum vs flux. \\ \hline
$E_C/2\pi$ & 318 MHz & High-power two-tone spectroscopy measurement. \\ \hline
$E_{J}(0)/2\pi$ & 36.39 GHz & - \\ \hline
$E_{J}(0)/E_C$ & 114 & - \\ \hline
$T_1$ & 871$\pm$ \SI{65}{\nano\second} & Qubit $T_1$ measurements at the sweet spot.\\ \hline
$T_2$ & 1059$\pm$ \SI{213}{\nano\second} & Qubit $T_2$ measurements at the sweet spot. \\ \hline

\multicolumn{3}{|c|}{\textbf{\large Readout resonator}} \\ \hline
\textbf{Parameter} & \textbf{Value} & \textbf{Method} \\ \hline
$\omega_{RO}/2\pi$ & 4.60 GHz & Fit of the readout resonator frequency vs flux. \\ \hline
$g/2\pi$ & 89 MHz & Fit of the readout resonator frequency vs flux. \\ \hline
$\gamma_{\rm RO,ext}/2\pi$ & $1.64\pm0.03$ MHz & Fit of the resonator with the qubit detuned. \\ \hline
$\gamma_{\rm RO,int}/2\pi$ & $330\pm43$ kHz & Fit of the resonator with the qubit detuned. \\ \hline
\end{tabular}
\caption{List of the system parameters.}
\label{ext:tab:params}
\end{extendedtable*}

\clearpage

\renewcommand{\appendixname}{Supplementary Information}

\appendix

\section{Model}
\label{app:circ-quant}
\subsection{Metamaterial}
\label{app:sec:model_metamat}

In this section we derive the Hamiltonian of the bare metamaterial.
The derivation in this section is largely inspired from our previous work~\cite{jouannyHighKineticInductance2025}.
The metamaterial circuit consists of a chain of $N=44$ capacitively coupled LC resonators.
Each resonators has an inductance $L_g$ and capacitance $C_g$ to ground.
The coupling between resonators is dimerized such that the capacitance inside a unit-cell is $C_1$ and the capacitance between unit-cells is $C_2$.
The potential energy in the inductors is expressed as
\begin{equation}
    E_L = \frac{1}{2L_g}\sum_{s=1}^N\varphi_s^2,
    \label{app:eq:ind_energy}
\end{equation}
where $\varphi_s$ is the flux at node $s$. 
The total kinetic energy stored in the chain’s capacitors is given by 
\begin{equation}
\begin{split}
    E_K &= \frac{C_g}{2}\sum_{n=1}^N\dot\varphi_s^2\\
    &+ \frac{C_1}{2}\sum_{s=1}^{N/2}\left(\dot\varphi_{2s-1} - \dot\varphi_{2s}\right)^2\\
    &+ \frac{C_2}{2}\sum_{s=1}^{N/2-1}\left(\dot\varphi_{2s} - \dot\varphi_{2s+1}\right)^2\\
    &+\frac{1}{2}\sum_{\substack{i,j\\2 \leq |i-j| \leq 4 }}C_{i,j} \left(\dot\varphi_{i} - \dot\varphi_{j}\right)^2,
\end{split}
\label{app:eq:cap_energy}
\end{equation}
where $\dot\varphi_s$ is the electric potential at node $s$.
The $C_{i,j}$ terms are taken into account up to order 4 and represent the stray capacitances between higher order neighbors.
The higher order stray capacitances are expressed as $C^\prime$, $C^{\prime\prime}$ and $C^{\prime\prime\prime}$, for order 2,3 and 4, respectively.
Due to the high-impedance of the resonators, the mutual inductance can be neglected.
The Lagrangian of the circuit is,
\begin{equation}
    \mathcal{L} = E_K-E_L.
\end{equation}
We will use its matrix form,
\begin{equation}
    \mathcal{L} = \frac{1}{2}\left(\dot{\bm{\varphi}}^T\left[C\right]\dot{\bm{\varphi}} - {\bm{\varphi}}^T\left[L^{-1}\right]{\bm{\varphi}}\right),
    \label{app:eq:lagrangian_matrix_CCA}
\end{equation}
with the vectors $\dot{\bm{\varphi}}^T = (\dot{\varphi}_1, \dot{\varphi}_2,\hdots, \dot{\varphi}_N)$ and $\bm{\varphi}^T = (\varphi_1, \varphi_2,\hdots ,\varphi_{N})$.
The capacitance matrix is defined as,
\begin{equation}
    \left[ C \right] = 
    \begin{pmatrix}
        C_\Sigma & -C_1 & -C^\prime & -C^{\prime\prime} & \hdots & 0\\
        -C_1 & C_\Sigma & -C_2 &-C^\prime &\ddots& \vdots\\
        -C^\prime & -C_2 & C_\Sigma & -C_1 & \ddots & \vdots\\
        \vdots & \ddots & \ddots & \ddots & \ddots & -C^\prime\\
        \vdots & \ddots & \ddots & \ddots & \ddots & -C_1\\
        0 & \hdots & \ddots &-C^\prime & -C_1 & C_\Sigma
    \end{pmatrix},
    \label{app:eq:capa_matrix}
\end{equation}
with $C_\Sigma$ being the total capacitance of a resonator.
The inductance matrix is defined as,
\begin{equation}
    \left[L^{-1}\right] = \frac{1}{L_g}\mathbf{I}
\end{equation}
where $\mathbf{I}$ is the identity matrix.
The exact Hamiltonian takes the form,
\begin{equation}
    H_{CCA} = \sqrt{\left[C^{-1}\right]\left[L^{-1}\right]}.
    \label{app:eq:exact_ham_CCA}
\end{equation}
The eigenvalues of the Hamiltonian above are used to fit the bare CCA modes' frequencies to extract the circuit parameters of the Hamiltonian according to the method described in~\cite{jouannyHighKineticInductance2025}.

One can obtain an analytical form of the Hamiltonian by neglecting the higher order terms, $C_1C_2, C_1^2,C_2^2$, and higher order stray capacitances, in the inversion of the capacitance matrix.
Doing so, the inverse of the capacitance matrix becomes,
\begin{equation}
    \left[ C^{-1} \right] = 
    L_g\omega_r^2
    \begin{pmatrix}
        1 & \beta_1 & 0 & \hdots & \hdots & 0\\
        \beta_1 & 1 & \beta_2 & \ddots &\ddots& \vdots\\
        0 & \beta_2 & 1 & \beta_1 & \ddots & \vdots\\
        \vdots & \ddots & \ddots & \ddots & \ddots & \vdots\\
        \vdots & \ddots & \ddots & \ddots & \ddots & \beta_1\\
        0 & \hdots & \ddots & \ddots & \beta_1 & 1
    \end{pmatrix},
\end{equation}
with $\omega_r = 1/\sqrt{L_gC_\Sigma}$ and $\beta_i = C_i/C_\Sigma$.

The Hamiltonian can be rewritten using the Legendre transform,
\begin{equation}
    H_{CCA} = \frac{1}{2}\bm{Q}^T[C^{-1}]\bm Q + \frac{1}{2}\bm \varphi^T[L^{-1}]\bm\varphi
\end{equation}
with $\bm Q = \partial\mathcal L/\partial\bm\dot\varphi$ the conjugate charge variable and $\bm Q^T = (Q_1, Q_2, \hdots,Q_N)$.
One gets the Hamiltonian,
\begin{equation}
    \begin{split}
        H_{CCA} = &\frac{1}{2}\sum_{s = 1}^{N}L_g\omega_r^2Q_s^2\\
        &+ \frac{1}{2}\sum_{s = 1}^{N}\frac{1}{L_g}\varphi_s^2\\
        &+\frac{L_g\omega_r^2\beta_1}{2}\sum_{s=1}^{N/2}\left( Q_{2s-1}Q_{2s}+Q_{2s}Q_{2s-1} \right)\\
        &+\frac{L_g\omega_r^2\beta_2}{2}\sum_{s=1}^{N/2-1}\left( Q_{2s}Q_{2s+1}+Q_{2s+1}Q_{2s} \right).
        \label{app:eq:ham_classical_CCA}
    \end{split}
\end{equation}
This Hamiltonian is quantized by introducing the quantized charge, ${Q}_s$, and flux, ${\varphi}_s$, operators, acting on site $s$, satisfying the commutation relation,
\begin{equation}
    \left[{\varphi}_{s},{Q}_{s^\prime}\right] = i\delta_{s,s^\prime}.
\end{equation}
They are defined as,
\begin{align}
    {Q}_s &=\sqrt{\frac{1}{2Z_r}}\left({a}_s - {a}_s^\dagger\right) \label{app:gap:eq:Qq},\\
    {\varphi}_n &=\sqrt{\frac{Z_r}{2}}\left({a}_s + {a}_s^\dagger\right) \label{app:gap:eq:Phiq},
\end{align}
where ${a}_s^\dagger$ (${a}_s$) is the annihilation (creation) operator on the $n$-th site.\\
Inserting \eqrefauto{app:gap:eq:Qq} and \eqrefauto{app:gap:eq:Phiq} into \eqrefauto{app:eq:ham_classical_CCA}, we find the dimerized Hamiltonian,
\begin{equation}
    \begin{split}
        H_{CCA} &= \omega_r\sum_{s=1}^N{a}_s^\dagger{a}_s\\
        &+\underbrace{ J_1\sum_{s=1}^{N/2}\left({a}_{2s-1}^\dagger{a}_{2s} + \hc\right)}_{\text{Intracell coupling}}\\
        &+ \underbrace{ J_2 \sum_{n=1}^{N/2-1}\left({a}_{2s}^\dagger{a}_{2s+1} \hc\right)}_{\text{Intercell coupling}}.
    \end{split}
    \label{app:eq:dimerHamiltonian}
\end{equation}

In our setup, we observe a non-negligible contribution from higher-order neighbor coupling terms.  
Following Ref.~\cite{jouannyHighKineticInductance2025}, we identify two main origins for these terms.  
First, as discussed in the introduction of the capacitance matrix, when the ratio \( C_{1(2)}/C_\Sigma \ll 0.1 \), higher-order couplings naturally emerge upon inverting the capacitance matrix.  
Second, direct stray capacitances can mediate coupling between non-adjacent (next-nearest and beyond) resonators.  
These higher-order neighbor couplings, up to order \( q=4 \), can be written as  
\begin{equation}
    K^{\prime} = \sum_{q=2}^4 \sum_{s=1}^{N-q} J^{(q)} \left( a_s^\dagger a_{s+q} + \mathrm{h.c.} \right),
\end{equation}
where \( J^{(q)}/2\pi \) denotes the coupling strength to the \( q \)-th neighbor.  
In particular, the couplings to the second, third, fourth, and fifth neighbors are denoted \( J^\prime \), \( J^{\prime\prime} \), \( J^{\prime\prime\prime} \), and \( J^{\prime\prime\prime\prime} \), respectively.
The extraction of the Hamiltonian parameters is reported in \aref{app:sec:CCA_charac} and their values are reported in \exttabref{ext:tab:params}.

\subsection{Transmon + Metamaterial}

In this subsection, we add the split-junction transmon qubit~\cite{blais_circuit_2021} to the system.
As we did in subsection \ref{app:sec:model_metamat}, we can derive the Lagrangian of the full circuit to then obtain the full Hamiltonian of the system.
The Lagrangian of the full circuit takes the form~\cite{scigliuzzoControllingAtomPhotonBound2022a},
\begin{equation}
    \mathcal{L}_{Tot} = \frac{1}{2}\left(\dot{\bm{\varphi}}^T\left[C_{Tot}\right]\dot{\bm{\varphi}}\right)- V.
\end{equation}
The total capacitance matrix is defined as,
\begin{equation}
    \left[C_{Tot}\right] =
    \begin{pmatrix}
        \left[C\right] & -\left[C_c\right]\\
         -\left[C_c\right]^T & C_T
    \end{pmatrix},
\end{equation}
\(\left[C_c\right]\) is of dimension $(1,44)$ and represent the coupling capacitances between the transmon qubit and the metamaterial.
$C_T$ is the total capacitance of the transmon qubit and is defined as $C_T = C_{T,g} + \sum_sC_{c,s}$, with $C_{T,g}$ the transmon capacitance to ground and \(C_{c,s}\) the coupling capacitance to site $s$.
The potential energy $V$ is defined as,
\begin{equation}
    V = \frac{1}{2L_g}\sum_{s=1}^N\varphi_s^2 - E_J(\varphi)\,.
\end{equation}
\(E_J(\varphi) = E_J\cos{\left(\Phi/\Phi_0\right)}\) is the Josephson Energy, $\Phi$ is the flux threaded through the SQUID of transmon and $\Phi_0 = h/2e$ is the flux quantum.
Applying the Legendre transformation $H = 1/2\bm{Q}^T[C]^{-1}\bm{Q} + V$~\cite{scigliuzzoControllingAtomPhotonBound2022a}, and applying quantization as done in the previous section one gets the Hamiltonian,
\begin{equation}
    \begin{split}
        H =\,& \omega_r\,\sum_{s=1}^{N}  {a}^\dagger_s {a}_s + J_1\sum_{s=1}^{N/2}\qty( a^\dagger_{2s-1}  a_{2s} + \hc)\\
        &+ J_2\sum_{s=1}^{N/2 - 1}\qty( a^\dagger_{2s}  a_{2s+1} + \hc)\\
        &+ \sum_{q=2}^4\sum_{s=1}^{N-4}J_{s,s+q}\left({a}_s^\dagger{a}_{s+q} + \hc\right)\\
        &+ \omega_{q}\, b^\dagger  b - \frac{E_C}{2}\, b^\dagger  b^\dagger  b  b+ \sum_{s=1}^N g_s\,\qty( a^\dagger_s  b + \hc)\,,
    \end{split}
    \label{app:eq:fullnormalSpaceHamiltonian}
\end{equation}

where \(\omega_q \approx \sqrt{8E_CE_J(\Phi)} - E_C\) and \(g_s/2\pi\) is the qubit site coupling rate.

\section{Metamaterial characterization}
\label{app:sec:CCA_charac}
\subsection{Extraction of the Hamiltonian parameters}

\begin{figure}
    \centering
    \includegraphics[width=\linewidth]{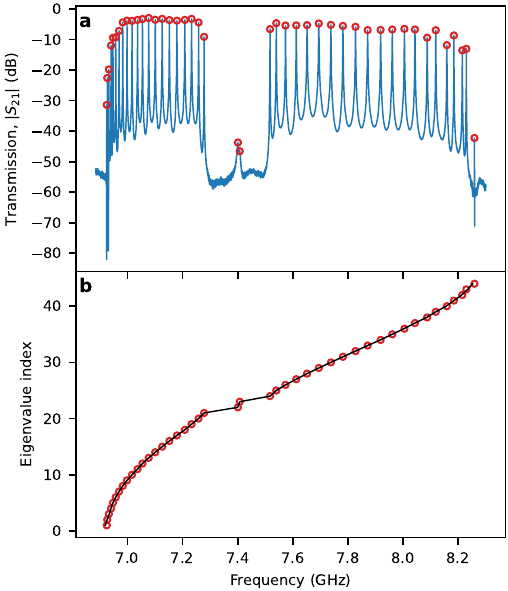}
    \caption{\textbf{CCA characterization: Hamiltonian.}
    \textbf{(a)} Single-tone spectroscopy measurement of the CCA at high photon number with the qubit detuned. 
    The extracted eigenmodes are represented with the red circles.
    \textbf{(b)} Eigenmodes of the CCA as a function of the eigenvalue index (red circle). 
    The black line represent the fit of the eigenmodes of the CCA.}
    \label{app:fig:MM_spectro}
\end{figure}

The circuit parameters of the CCA are extracted by measuring the CCA in transmission via single-tone spectroscopy at high photon number, with the qubit detuned to its lowest frequency (see \subautoref{app:fig:MM_spectro}{\textbf{a}}). 
From this measurement, two distinct bands are observed, each containing 21 modes. 
The two modes located in the middle bandgap are identified as the SSH modes. 
These modes exhibit low amplitudes in transmission due to the minimal overlap of their photonic wavefunctions, which is a consequence of the CCA's effective size. 
Additionally, these modes are non-degenerate due to the presence of disorder that breaks chiral symmetry~\cite{jouannyHighKineticInductance2025}.

A peak-finding algorithm detects the 44 eigenmodes of the CCA, which are shown in \subautoref{app:fig:MM_spectro}{\textbf{b}}. 
The detected mode frequencies are fitted to the eigenvalues of the Hamiltonian in \eqrefauto{app:eq:exact_ham_CCA}, with the fit indicated by a black line in \subautoref{app:fig:MM_spectro}{\textbf{b}}. 
The extracted parameters from this fit are listed in \exttabref{ext:tab:params}.

A strong asymmetry between the lower and upper bands of the CCA spectrum is observed. 
This asymmetry is caused by higher-than-first-neighbor couplings between CCA sites. 
These couplings arise from stray capacitance between higher-neighbor sites and from the high coupling ratio $C_{1(2)}/C_\Sigma$, which introduces additional coupling terms during the inversion of the capacitance matrix (\eqrefauto{app:eq:capa_matrix}).

\subsection{Extraction of the metamaterial dissipation}

\begin{figure}
    \centering
    \includegraphics[width=\linewidth]{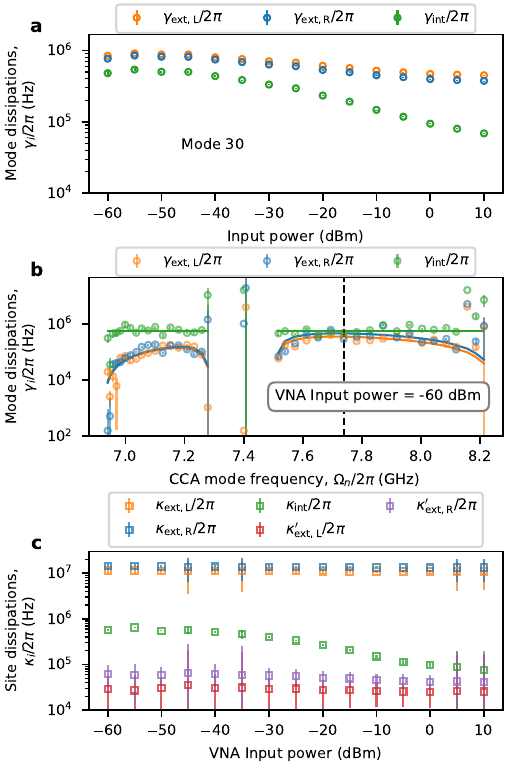}
    \caption{\textbf{CCA characterization: Dissipations.} 
    \textbf{(a)} Mode dissipations, $\gamma$, vs VNA input power for mode 30 of the CCA. The reported dissipation are the external dissipation to the left port (orange), right port (blue) and internal dissipations (green).
    \textbf{(b)} Fit of the mode dissipations, $\gamma$, with the complex part of the eigenvalues of the non-Hermitian Hamiltonian \eq\eqref{app:eq:nonHermtianHam}. The fitted mode in panel \textbf{a} is highlighted by the black dashed line. 
    \textbf{(c)} Extracted site dissipations, $\kappa$, vs VNA input power from the fit in panel \textbf{b}.}
    \label{app:fig:dissipation}
\end{figure}

This subsection focuses on extracting the dissipations of the bare CCA, following the extraction of circuit parameters in the previous subsection.

To extract the bare dissipation rates of the eigenmodes, we first detune the qubit to its lowest frequency. 
Each resonance is then measured via single-tone spectroscopy as a function of the VNA input power, performing reflection measurements from both sides of the CCA.
The reflection scattering coefficients on the left and right port for mode $m$ are defined as,
\begin{equation}
    S_{LL}^{(m)} = A_Le^{-i\alpha_L}\left(1-\frac{\gamma_{\rm ext,L}e^{i\phi_L}}{i(\omega_p-\tilde\omega_m)+\frac{\gamma_{\rm ext,L} + \gamma_{\rm ext,R} + \gamma_{\rm int}}{2}}\right),
    \label{app:eq:SLL_fitting_function}
\end{equation}
and,
\begin{equation}
    S_{RR}^{(m)} = A_Re^{-i\alpha_R}\left(1-\frac{\gamma_{\rm ext,R}e^{i\phi_R}}{i(\omega_p-\tilde\omega_m)+\frac{\gamma_{\rm ext,L} + \gamma_{\rm ext,R} + \gamma_{\rm int}}{2}}\right),
    \label{app:eq:SRR_fitting_function}
\end{equation}
respectively, where the dissipation rates to the left (\(\gamma_{\rm ext,L}\)) and right (\(\gamma_{\rm ext,R}\)) ports are not assumed to be symmetric.
The parameter \(\gamma_{\rm int}/2\pi\) represents the internal dissipation rate of the cavity. 
The parameters \(A_{L(R)}\), \(\alpha_{L(R)}\), and \(\phi_{L(R)}\) represent the baseline amplitude, phase shift, and impedance mismatch, respectively, of the scattering trace measured from the left (right) port. 
Phase delays are corrected during post-processing. 
The mode's resonant frequency is given by \(\tilde\omega_m/2\pi\) and the probe frequency by \(\omega/2\pi\). 
Each mode is fitted in the complex plane using \eqrefauto{app:eq:SLL_fitting_function} and \eqrefauto{app:eq:SRR_fitting_function}. 
Typical fitting results for mode 30 are shown in \subautoref{app:fig:dissipation}{\textbf{a}}.

The procedure above is repeated for all eigenmodes, with results shown in \subautoref{app:fig:dissipation}{\textbf{b}} for a VNA input power of -60~dBm. 
These dissipations can be modeled by transforming the Hamiltonian in \eqrefauto{app:eq:exact_ham_CCA} into a non-Hermitian Hamiltonian that accounts for dissipation effects:
\begin{equation}
\begin{split}
    {H_{\textrm{CCA};k,l}^\textrm{Non-Herm}} &= {H}_{\textrm{CCA};k,l} - i\bm{I}\frac{\kappa_{\rm int}}{2}\\
    &-
    \frac{i}{2}\begin{dcases}
        \kappa_{\rm ext,L} & \text{if } k = l = 0\,,\\
        \kappa_{\rm ext,R} & \text{if } k = l = 44\,,\\
        \kappa_{\rm ext,L}^\prime & \text{if } k = l = 1\,,\\
        \kappa_{\rm ext,R}^\prime & \text{if } k = l = 43\,,\\
        2\sqrt{\kappa_{\rm ext,L}\kappa_{\rm ext,L}^\prime} & \begin{aligned}[t]
            &\text{if } (k,l) = (0,1)\\
            &\quad \text{or } (1,0)\,,
        \end{aligned}\\
        2\sqrt{\kappa_{\rm ext,R}\kappa_{\rm ext,R}^\prime} & \begin{aligned}[t]
            &\text{if } (k,l) = (44,43)\\
            &\quad \text{or } (43,44)\,.
        \end{aligned}
    \end{dcases}
\end{split}
\label{app:eq:nonHermtianHam}
\end{equation}
where the dissipation parameters are defined as:
\begin{itemize}
    \item $\kappa_{\rm int}/2\pi$: Internal cavity dissipation rate (assumed equal for all cavities)
    \item $\kappa_{\rm ext,L}/2\pi$ and $\kappa_{\rm ext,R}/2\pi$: External coupling rates of the leftmost and rightmost cavities to their respective measurement ports
    \item $\kappa_{\rm ext,L}^\prime/2\pi$ and $\kappa_{\rm ext,R}^\prime/2\pi$: External coupling rates of the second-to-leftmost and second-to-rightmost cavities to their respective measurement ports
\end{itemize}
The imaginary part of the eigenvalues of \eqrefauto{app:eq:nonHermtianHam} determines the dissipation profile of the eigenmodes through the relation:
\begin{equation}
    \Im(\text{Eig}({H_{\textrm{CCA}}^\text{Non-Herm}})) = 2\Vec{\gamma}_{Tot}\,,
    \label{app:eq:diss_non_herm}
\end{equation}
where $\Vec{\gamma}_{Tot} = \Vec{\gamma}_{int} + \Vec{\gamma}_{ext,L} + \Vec{\gamma}_{ext,R}$ represents the total dissipation rate vector of length $N = 44$ for all eigenmodes.

By selectively setting dissipation values to zero in the non-Hermitian Hamiltonian (\eqrefauto{app:eq:nonHermtianHam}), we can isolate and extract the different dissipation contributions of the CCA. 
Specifically:
\begin{equation}
\begin{split}
    \Im(\text{Eig}({H_{\textrm{CCA}}^\text{Non-Herm}})) =\\
    2\begin{cases}
        \Vec{\gamma}_{int} & \text{if } \kappa_{ext,R} = \kappa_{ext,L} = \kappa_{ext,R}^\prime = \kappa_{ext,L}^\prime = 0\,,\\
        \Vec{\gamma}_{ext,R} & \text{if } \kappa_{ext,L} = \kappa_{ext,L}^\prime = \kappa_{int} = 0\,,\\
        \Vec{\gamma}_{ext,L} & \text{if } \kappa_{ext,R} = \kappa_{ext,R}^\prime = \kappa_{int} = 0\,.
    \end{cases}
\end{split}
\label{app:eq:diss_non_herm_spec}
\end{equation}

The internal and external dissipations extracted using \eqrefauto{app:eq:SLL_fitting_function} and \eqrefauto{app:eq:SRR_fitting_function} are fitted to \eqrefauto{app:eq:diss_non_herm_spec}, with results shown in \subautoref{app:fig:dissipation}{\textbf{b}} for a VNA input power of -60~dBm. 
As shown in the figure, we observe the expected behavior: external dissipation is larger in the center of the band compared to the edges, and there is a strong asymmetry in external dissipation between the lower and upper bands. 
This asymmetry arises from next-nearest-neighbor interactions and the $\kappa_{ext,L(R)}^\prime$ terms.

The extraction procedure is repeated for multiple input powers, with results shown in \subautoref{app:fig:dissipation}{\textbf{c}}. 
The complete set of extracted dissipation parameters at low power (\(-60 \)dB VNA input power) is reported in \exttabref{ext:tab:params}.
With these results, combined with the previously extracted CCA circuit parameters, we now have complete characterization of both the real and imaginary parts of the non-Hermitian Hamiltonian (\eqrefauto{app:eq:diss_non_herm}) in the single-excitation subspace.

\subsection{Influence of disorder on the dissipations}

\begin{figure}
    \centering
    \includegraphics[width=\linewidth]{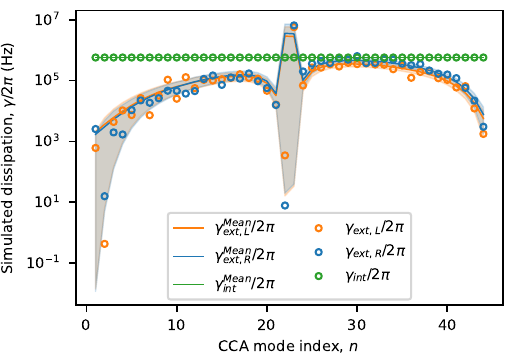}
    \caption{\textbf{Simulation of the CCA mode dissipation, $\gamma$, with disorder.} 
    Dissipations simulated with the eigenvalues of the non-Hermitian Hamiltonian \eqrefauto{app:eq:diss_non_herm_spec} with a standard deviation on the resonant frequency of $\SI{22}{\mega\hertz}$, repeated $5\times10^4$ times.
    The orange (blue) line is the mean external dissipation rates from the left $\gamma_{\rm ext,L}^{\rm Mean}/2\pi$ (right $\gamma_{\rm ext,R}^{\rm Mean}/2\pi$) as a function of the CCA mode index.
    The green line is the mean internal dissipation rate $\gamma_{\rm int}^{\rm Mean}/2\pi$ as a function of the CCA mode index.
    The orange, blue and green shaded areas encompass the $18.57-$th and the $84.13-$th percentile of the dissipation distribution associated to $\gamma_{\rm ext,L}/2\pi$, $\gamma_{\rm ext,R}/2\pi$ and $\gamma_{\rm int}/2\pi$, respectively.
    The orange, blue and green circles corresponds to a single disorder realization associated to $\gamma_{\rm ext,L}/2\pi$, $\gamma_{\rm ext,R}/2\pi$ and $\gamma_{\rm int}/2\pi$, respectively.
    }
    \label{app:fig:disorder_diss}
\end{figure}

Disorder affects mode localization, whether originating from resonant frequency disorder or coupling disorder, which in turn affects the external dissipation rates \(\gamma_{\rm ext, L(R)}/2\pi\). 
This effect is particularly evident in the measurements presented in \autoref{sec:fig4} and \autoref{sec:fig5}. 
To quantitatively study this phenomenon, we perform simulations of the expected dissipations for different CCA modes using \eqrefauto{app:eq:diss_non_herm_spec} over 50{,}000 realizations of resonant frequency disorder (\figref{app:fig:disorder_diss}). 
These simulations highlight the way disorder alters the dissipation profile.

A typical disorder realization is shown in \figref{app:fig:disorder_diss}, where green, blue, and orange circles correspond to \(\gamma_{\rm int}\), \(\gamma_{\rm ext,L}\), and \(\gamma_{\rm ext,R}\), respectively.
Internal dissipation rates are unaffected by frequency disorder since we assume constant losses across all sites, an idealization that differs from experimental conditions. 
Notably, disorder impacts the lower band more significantly than the upper band, due to the lower dispersion, and thus lower kinetic energy, in this band.

\section{Qubit characterization}
\label{app:sec:qubit:charac}
In this section the characterization of the qubit is presented.

\subsection{Two-tone spectroscopy measurements}

\begin{figure}
    \centering
    \includegraphics[width=\linewidth]{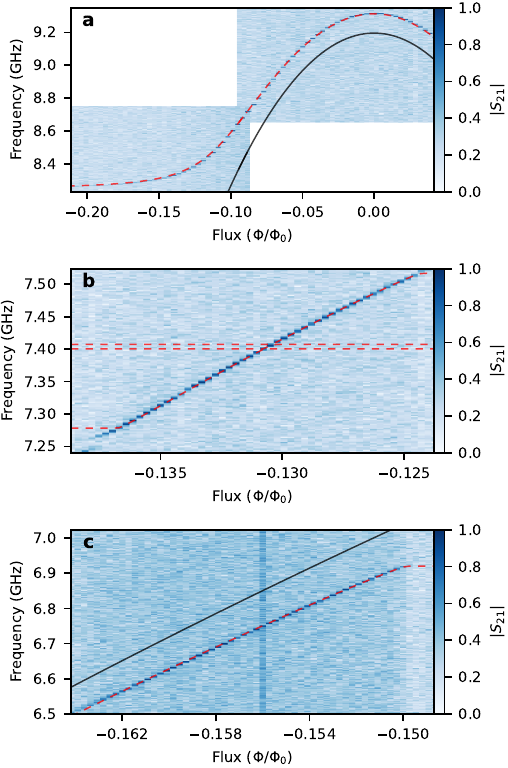}
    \caption{\textbf{Two-tone spectroscopy measurement.}
    \textbf{(a)} Two-tone spectroscopy measurement of the transmon qubit above the upper band as a function of the reduced flux, $\Phi/\Phi_0$, threading the SQUID loop.
    With the qubit initialized at $\omega_q/2\pi = \SI{9.05}{\giga\hertz}$ and $\omega_q/2\pi = \SI{8.46}{\giga\hertz}$.
    The red-dashed line is the theoretical prediction. 
    The black line is the bare qubit frequency ($\omega_q/2\pi$)
    \textbf{(b)} Same as in panel \textbf{a} in the middle bandgap. With the qubit initialized at $\omega_q/2\pi = \SI{7.52}{\giga\hertz}$.
    \textbf{(c)} Same as in panel \textbf{a} and \textbf{b} below the lower band. With the qubit initialized at $\omega_q/2\pi = \SI{6.80}{\giga\hertz}$.}
    \label{app:fig:two_tone_spectro}
\end{figure}

The transmon qubit is characterized using pulsed two-tone spectroscopy at frequencies in the upper, middle, and lower bandgaps.
The measurement protocol consists of applying a fast flux pulse to the qubit via the flux line, followed by a 200~ns \(\pi\)-pulse through the charge line. 
After resetting the flux bias to its initial value, we perform readout.

The transmon qubit's sweet spot is located in the upper bandgap [see \subautoref{app:fig:two_tone_spectro}{\textbf{a}}]. 
We find a good agreement between the theoretical model predictions (red dashed lines) and the experimental measurements shown in \figref{app:fig:two_tone_spectro}.
A strong frequency shift between the bare qubit and the dressed qubit remains evident in the middle bandgap and the lower bandgap.
This is attributed to the strong coupling to the upper band edge modes.

The transmon qubit’s sweet spot is located in the upper bandgap [see \subautoref{app:fig:two_tone_spectro}{\textbf{a}}].  
We find good agreement between the predictions of the theoretical model (red dashed lines) and the experimental measurements shown in \figref{app:fig:two_tone_spectro}.  
A pronounced frequency shift between the bare and dressed qubit remains visible in both the middle and lower bandgaps, which we attribute to the strong coupling to the upper band-edge modes.

\subsection{Losses}
\begin{figure}
    \centering
    \includegraphics[width=\linewidth]{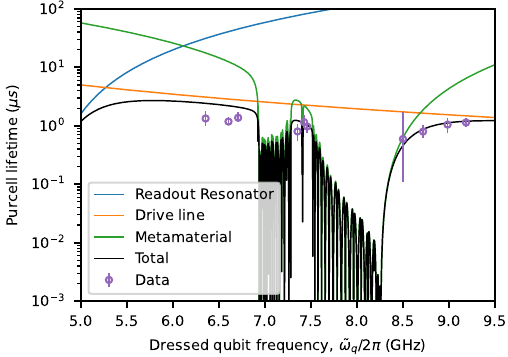}
    \caption{\textbf{Purcell lifetime.} Limit of the qubit lifetime due to the environment as a function of the dressed qubit frequency, $\tilde\omega_q/2\pi$. The blue line is the contribution from the readout resonator. The orange line is the contribution from the drive line. The green line is the contribution from the CCA. The black line is the sum of the contributions. 
    The data are represented with the purple circle. The errorbar corresponds to the minimum and maximum fitted qubit lifetime.
    }
    \label{app:fig:purcell}
\end{figure}
\begin{figure*}
    \centering
    \includegraphics[width=\linewidth]{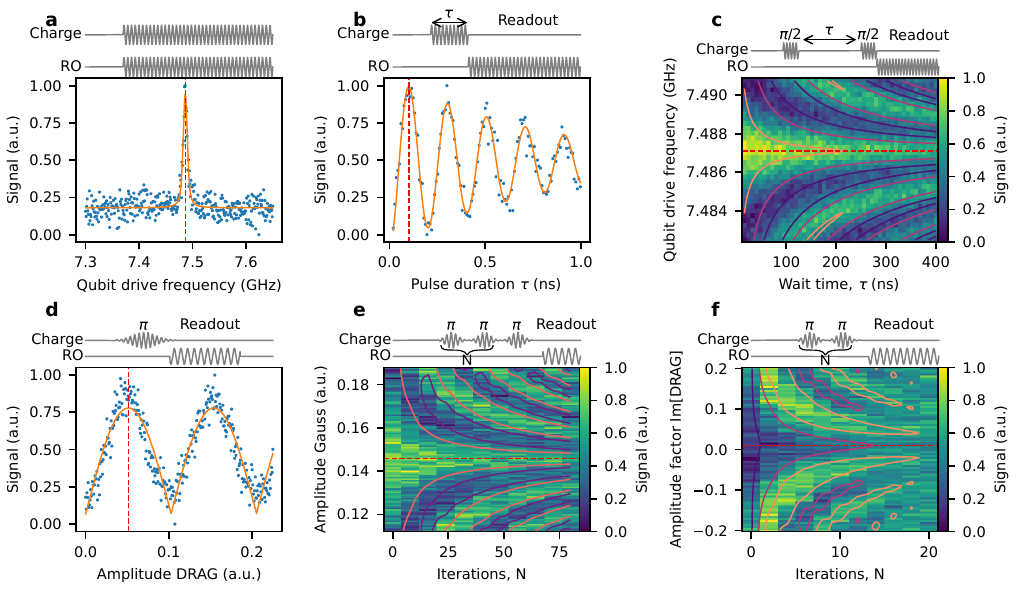}
    \caption{\textbf{$\pi-$pulse calibration sequence.} In all panels the pulse sequence is indicated above and the measurement shown below.
    \textbf{(a)} Two-tone spectroscopy measurement. The data points are indicated with blue dots, the fit is indicated with the orange line. The red dashed line represents the extracted qubit frequency.
    \textbf{(b)} Time-Rabi measurement with the qubit driven at the frequency extracted in the previous measurement. The data points are indicated with blue dots, the fit is indicated with the orange line. The red dashed line represents the pulse time required to perform a $\pi-$pulse.
    \textbf{(c)} Ramsey measurement vs qubit frequency drive. The $\pi/2$ pulse is calibrated from the previous measurement. The data is represented in the colormap and the fit is shown with the contour plot. The red dashed line shows the extracted qubit frequency.
    \textbf{(d)} Power Rabi measurement. The measurement is performed at the frequency of the qubit extracted from the previous measurement. The data points are indicated with blue dots, the fit is indicated with the orange line. The red dashed line shows the amplitude required to perform a $\pi-$pulse.
    \textbf{(e)} Gauss error amplification measurement. The data is represented in the colormap and the fit is shown with the contour plot. The red dashed line shows the optimal amplitude for the $\pi-$pulse.
    \textbf{(f)} DRAG error amplification measurement. The data is represented in the colormap and the fit is shown with the contour plot. The red dashed line shows the optimal amplitude factor on the DRAG asymmetry parameter.}
    \label{app:fig:pi_pulse_calib}
\end{figure*}

In this subsection, we evaluate the Purcell losses experienced by the qubit to identify the factors limiting its lifetime.
The first source of loss we consider is the drive line.
The Purcell decay rate due to the drive line is given by:
\begin{equation}
    \gamma_{\rm Purcell- Drive} = \omega_q^2Z_0\frac{C_c^2}{C_\Sigma},
    \label{app:eq:purcell_drive}
\end{equation}
where \(C_c = \SI{0.5}{\femto\farad}\) is the coupling capacitance between the drive line and the qubit, estimated from finite element simulations, \(C_\Sigma = \SI{60.9}{\femto\farad}\) is the total capacitance of the transmon qubit, and \(Z_0\) is the characteristic impedance of the drive line. 
Given the high ratio \(C_c/C_\Sigma = 0.008\), this Purcell decay mechanism limits the qubit lifetime to \SI{5}{\micro\second} at \SI{5}{\giga\hertz}, decreasing to less than \SI{2}{\micro\second} at \SI{9.5}{\giga\hertz} [orange line in \figref{app:fig:purcell}].

A second source of Purcell decay is due to the readout resonator,
\begin{equation}
    \gamma_{\rm Purcell- Res} = \left(\gamma_{\rm RO, ext} + \gamma_{\rm RO, int}\right)\left(\frac{g}{\Delta}\right)^2.
    \label{app:eq:purcell_res}
\end{equation}
$\gamma_{\rm RO, ext}/2\pi = \SI{1.64}{\mega\hertz}$ is the external coupling of the readout resonator to its feedline, $\gamma_{\rm RO, int}/2\pi = \SI{330}{\kilo\hertz}$, is the intrinsic dissipation rate of the readout resonator, $g/2\pi = \SI{89}{\mega\hertz}$ is the coupling rate of the qubit to the readout resonator, and $\Delta = \omega_q-\omega_{\rm RO}$, is the detuning between the qubit and the readout resonator of resonant frequency $\omega_{\rm RO}/2\pi = \SI{4.6}{\giga\hertz}$.
This source of losses is limiting the qubit lifetime to values greater than \SI{10}{\micro\second} in the measurement range [blue line in \figref{app:fig:purcell}].

The third source of losses is the Purcell decay due to the CCA,
\begin{equation}
    \gamma_{\rm Purcell- CCA} = \sum_{n = 1}^{44}\left(\gamma_{\rm n, int} + \gamma_{\rm n, ext}\right)\left(\frac{G_n}{\Delta_n}\right)^2
    \label{app:eq:purcell_CCA}
\end{equation}
$\gamma_{\rm n, ext}/2\pi$ is the external dissipation rate mode $n$, $\gamma_{\rm n, ext}/2\pi$ is the internal dissipation rate of mode $n$.
$G_n/2\pi$ is the coupling rate from the qubit to mode $n$ and $\Delta_n = (\Omega_n-\omega_q)$ is the detuning between mode $n$ and the qubit.
This source of losses is limiting the qubit lifetime to \SI{10}{\micro\second} below the lower band, around 2-\SI{3}{\micro\second} in the middle bandgap and between 1 and \SI{10}{\micro\second} in the upper bandgap [green line in \figref{app:fig:purcell}].

The combined Purcell effect is represented by the black line in \figref{app:fig:purcell}. 
We compare experimental qubit lifetime measurements at different frequencies [purple circles in \figref{app:fig:purcell}] against the predicted Purcell decay rates. 
Each qubit decay rate measurement is repeated 100 to 2000 times per frequency point over a duration of several hours to ensure statistical reliability.
In the upper bandgap, we observe excellent agreement between experimental lifetimes and Purcell decay predictions. 
Here, the lifetime is limited by the CCA when the qubit is near the band edge and by the drive line when the qubit is far detuned from the band. 
In the middle bandgap, the experimental data show reasonable agreement with theory, where the qubit lifetime is limited by both the drive line and the CCA. 
In the lower bandgap, the qubit lifetime is slightly below the predictions based solely on Purcell decay, suggesting the presence of additional loss mechanisms in this frequency range.

\subsection{Calibration $\pi$-pulse}

For the measurements presented in \autoref{sec:fig3} and \autoref{sec:fig5}, the qubit $\pi$-pulse was calibrated according to the sequence shown in \figref{app:fig:pi_pulse_calib}.
Below we briefly describe each parameter extracted at each steps of the calibration sequence.
The pulse sequences are shown on top of each panels of \figref{app:fig:pi_pulse_calib}.

\begin{enumerate}
    \item \textbf{Two-tone spectroscopy measurement:} The $01$ transition of the qubit is first estimated [red dashed line in \subautoref{app:fig:pi_pulse_calib}{\textbf{a}}] by performing two-tone spectroscopy.
    \item \textbf{Time-Rabi:} The $\pi$-pulse time is then estimated [red dashed line in \subautoref{app:fig:pi_pulse_calib}{\textbf{b}}] by doing time-Rabi measurement with a square pulse shape.
    \item \textbf{Ramsey vs detuning:} The precise qubit frequency is extracted [red dashed line in \subautoref{app:fig:pi_pulse_calib}{\textbf{c}}] by performing a Ramsey measurement as a function of excitation pulse detuning with a square pulse shape.
    \item \textbf{Power Rabi:} Then the time of the pulse is fixed to \SI{40}{\nano\second} and a Power Rabi measurement is done to estimate the optimal $\pi$-pulse amplitude [red dashed line in \subautoref{app:fig:pi_pulse_calib}{\textbf{d}}] with a gaussian pulse shape.
    \item \textbf{Gauss error amplification:}  To refine the $\pi$-pulse amplitude estimate, we perform Gaussian error amplification by applying a train of $2N+1$ $\pi$-pulses as a function of pulse amplitude. The optimal amplitude is determined by fitting the resulting pattern [red dashed line in \subautoref{app:fig:pi_pulse_calib}{\textbf{e}}].
    \item \textbf{DRAG error amplification:} We use DRAG to prevent leakage to higher excited state~\cite{motzoiSimplePulsesElimination2009}. The DRAG amplitude is calibrated by performing a similar type of measurement as done above, but sweeping the imaginary component of DRAG pulse. The extracted parameter is shown with the red dashed line in \subautoref{app:fig:pi_pulse_calib}{\textbf{f}}.
\end{enumerate}

\section{Qubit + metamaterial characterization}
\label{app:sec:qubit_CCA}
\subsection{More data figure 2}

\begin{figure*}
    \centering
    \includegraphics[width=1\linewidth]{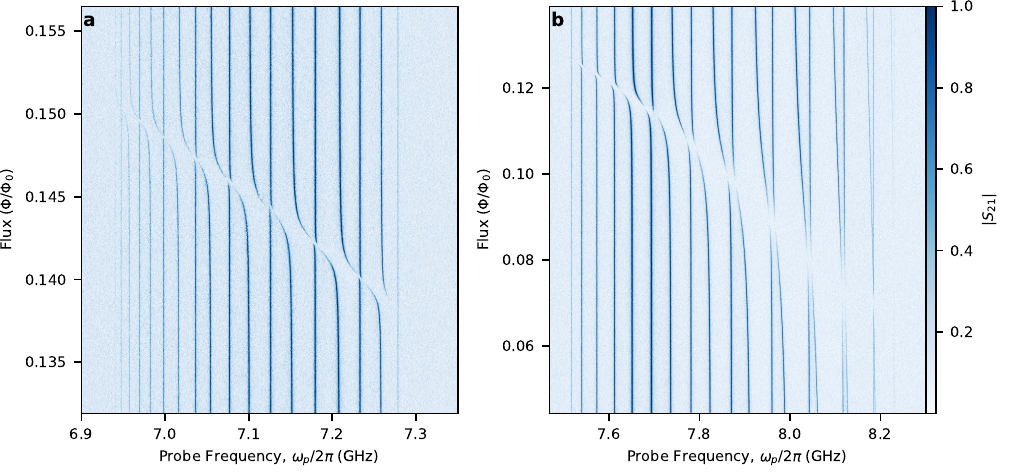}
    \caption{\textbf{Zoom-in on the spectroscopy measurements in figure 2.}
    Single-tone spectroscopy measurements vs coil current in the lower band (\textbf{a}) and the upper band (\textbf{b}). The color bar is shared between the two panels.
    }
    \label{app:fig:supp_spectro_fig2}
\end{figure*}

In \figref{app:fig:supp_spectro_fig2}, we report zoom ins of the single-tone spectroscopy measurement presented in \figref{fig:fig2}.

\subsection{More atomic ratios}
\begin{figure}
    \centering
    \includegraphics[width=\linewidth]{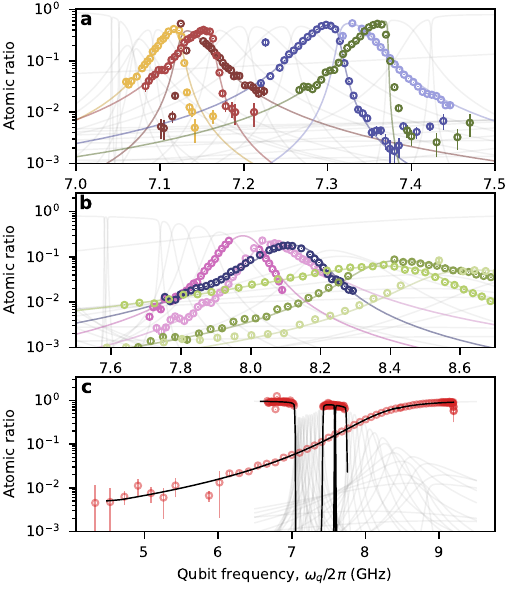}
    \caption{\textbf{More atomic ratio.}
    \textbf{(a)} Atomic ratio in the lower band of modes 9, 10, 11, 17, 18 and 19. Modes 13 and 14 are in the main text.
    \textbf{(b)} Atomic ratio in the upper band of modes 29, 30, 31, 34, 35 and 36. Modes 31 and 32 are also reported in the main text.
    \textbf{(c)} Atomic ratio measured with two-tone spectroscopy. Modes 1, 21, 22, 23 and 45.}
    \label{app:fig:atomic_ratio}
\end{figure}

In \figref{app:fig:atomic_ratio}, we present supplementary atomic ratio measurements performed via single-tone spectroscopy in the lower band [panel~\subautoref{app:fig:atomic_ratio}{\textbf{a}}] and upper band [panel~\subautoref{app:fig:atomic_ratio}{\textbf{b}}], and via two-tone spectroscopy in the bandgaps [panel~\subautoref{app:fig:atomic_ratio}{\textbf{c}}].
These measurements are done using the method introduced in the main text.

\section{Open dynamics of the qubit}
\label{app:sec:qubit_dynamics}

\subsection{Qubit initialized in the middle bandgap}
\begin{figure*}
    \centering
    \includegraphics[width=\linewidth]{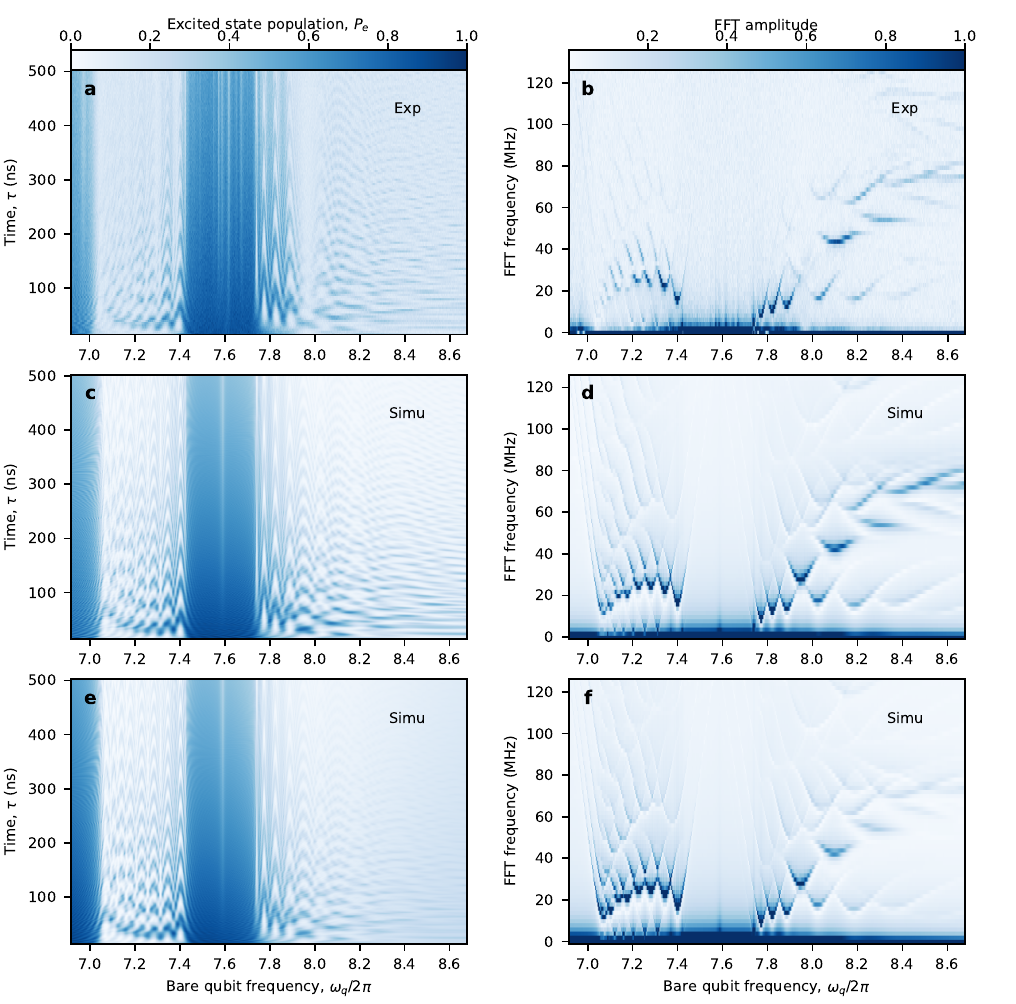}
    \caption{\textbf{Comparing the dynamics of the emitter in the bath with theory.}
    \textbf{(a)} Measurement of the excited state population of the qubit as a function of the time $\tau$ and the bare qubit frequency $\omega_q/2\pi$ with a ramp time of $\SI{2.4}{\nano\second}$. The measurement is done according to the pulse sequence indicated in \figref{fig:fig3}.
    \textbf{(b)} Fourier transform of the data in panel \textbf{a}. The colormap is saturated to better highlight the features.
    \textbf{(c)} Simulation of the time evolution of the qubit as a function of the time $\tau$ and the bare qubit frequency $\omega_q/2\pi$ with a ramp time of $\SI{2.4}{\nano\second}$.
    \textbf{(d)} Fourier transform of the simulated data in panel \textbf{c}. The colormap is saturated to better higlight the features.
    \textbf{(e)} Simulation of the time evolution of the qubit as a function of the time $\tau$ and the bare qubit frequency $\omega_q/2\pi$ with a ramp time of \SI{0.1}{\nano\second}.
    \textbf{(f)} Fourier transform of the simulated data in panel \textbf{e}. The colormap is saturated to better higlight the features.
    }
    \label{app:fig:TD_FFT_fig3_openQ}
\end{figure*}

In the time domain measurements shown in \autoref{sec:fig3}, the qubit is initialized in the middle bandgap. 
Its frequency is then rapidly tuned to a target value, allowing it to interact with the CCA for a defined duration, before returning to the initial frequency for state measurement.
Four distinct frequency regions are identified in \autoref{fig:fig3}:
\begin{itemize}
    \item the lower bandgap (\(\omega_q/2\pi \lesssim \SI{7.015}{\giga\hertz}\)),
    \item the lower band (\(\SI{7.015}{\giga\hertz} \lesssim \omega_q/2\pi \lesssim \SI{7.43}{\giga\hertz}\)),
    \item the middle bandgap (\(\SI{7.43}{\giga\hertz} \lesssim \omega_q/2\pi \lesssim \SI{7.725}{\giga\hertz}\)), and
    \item the upper band (\(\omega_q/2\pi \gtrsim \SI{7.725}{\giga\hertz}\)).
\end{itemize}
However, we do not resolve the upper bandgap due to the strong coupling between the qubit and the last CCA mode.

In \figref{app:fig:TD_FFT_fig3_openQ}, we compare the measured dynamics of the qubit to the simulated dynamics of the qubit using \texttt{QuantumToolbox.jl} [see \methods].
First, we simulate the experimental pulse sequence, where the ramp time is limited to \SI{2.4}{\nano\second}, the simulation is reported in \subautoref{app:fig:TD_FFT_fig3_openQ}{\textbf{c}--\textbf{d}}.
We find excellent agreements between theory and experiment, where the upper bandgap is also not resolvable in the simulation.
Notable differences happens in terms of losses, where in the bandgaps we find enhance losses most likely due to Two-Level-System fluctuators, however we cannot certainly assess this conclusion.

In \figref{app:fig:TD_FFT_fig3_openQ}, we compare the measured dynamics of the qubit to simulations performed using \texttt{QuantumToolbox.jl} [see \methods]. 
Our simulations model the experimental pulse sequence where the ramp time of the flux offset is fixed to \SI{2.4}{\nano\second}, and the results are presented in \subautoref{app:fig:TD_FFT_fig3_openQ}{\textbf{c}--\textbf{d}}. 
We find excellent agreement between theory and experiment, including the inability to resolve the upper bandgap. 
We observe notable differences in the bandgaps, where the measurement shows enhanced losses that we attribute to two-level system (TLS) fluctuators, although we cannot definitively confirm this explanation.

In \subautoref{app:fig:TD_FFT_fig3_openQ}{\textbf{e}--\textbf{f}}, we simulate the qubit dynamics with a ramp time of \SI{0.1}{\nano\second}. 
In this case, we can resolve the upper bandgap, unlike with the ramp time of \(\SI{2.4}{\nano\second}\). 
However, we observe fewer mode interactions than in \subautoref{app:fig:TD_FFT_fig3_openQ}{\textbf{c}--\textbf{d}} in the upper bandgap, which we attribute to population transfer to atom photon bound state above the upper band.

\subsection{Qubit initialized in the upper bandgap}
\begin{figure}
    \centering
    \includegraphics[width=\linewidth]{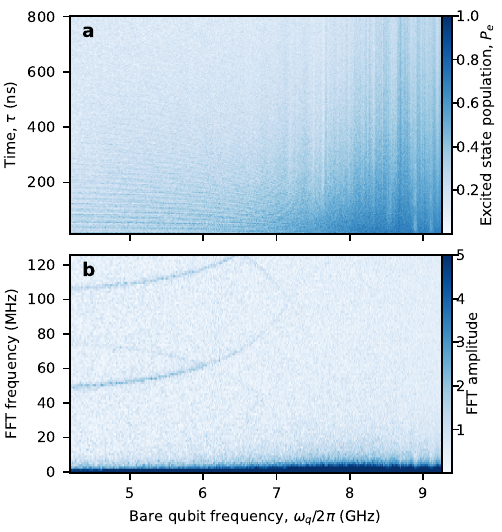}
    \caption{\textbf{Qubit dynamics with the qubit initialized in the upper bandgap.}
    \textbf{(a)} Measurement of the excited state population of the qubit as a function of the time $\tau$ and the bare qubit frequency $\omega_q/2\pi$. 
    The qubit is inialized in the upper bandgap.
    \textbf{(b)} Fourier transform of the experimental data in panel \textbf{a}. The colormap is saturated to highlight its features.}
    \label{app:fig:qubit_TD_UBG}
\end{figure}

In this subsection, we initialize the qubit in the upper bandgap at \(\omega_q/2\pi = \SI{8.69}{\giga\hertz}\) and repeat the same measurement protocol as in the previous subsection.
The time domain measurements are shown in \subautoref{app:fig:qubit_TD_UBG}{\textbf{a}} and the corresponding Fast Fourier transform is presented in \subautoref{app:fig:qubit_TD_UBG}{\textbf{b}}. 
We observe a standard exponential decay for frequencies \(\omega_q/2\pi \gtrsim \SI{7.3}{\giga\hertz}\). 
We observe fewer Rabi oscillations than in \subautoref{fig:fig3}{\textbf{c}}, which we attribute to strong coupling between the qubit and the upper band that prevents leakage into CCA modes. 
This strong coupling can be exploited to implement fast gates on APBS without dynamical hybridization with the band.

\section{Calibration SWAP}
\label{app:sec:cal_SWAP}
\begin{figure}
    \centering
    \includegraphics[width=\linewidth]{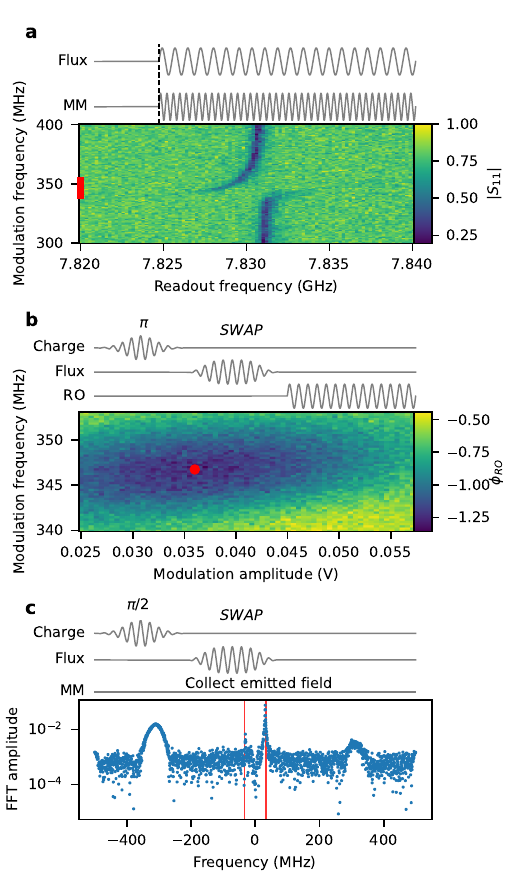}
    \caption{\textbf{Calibration SWAP} 
    \textbf{(a)} Metamaterial spectrosocpy vs qubit modulation frequency.
    The vertical red line highlight the frequency range chosen for the next measurement.
    \textbf{(b)} Evolution of the qubit population as a function of the modulation frequency and amplitude for a fixed modulation length of 120ns. 
    The red point highlight the optimal modulation frequency and amplitude chosen for the emission measurement. 
    \textbf{(c)} Emitted field after calibrating the SWAP pulse. 
    The red dashed lines are the frequency at which we expect the field to be emitted. }
    \label{app:fig:SWAP_pulse_calib}
\end{figure}

This section outlines the typical routine for calibrating the SWAP gate.

The initial step involves obtaining an estimate of the modulation frequency and amplitude required to perform the SWAP. 
This is achieved by applying a flux modulation tone to the qubit and measuring the CCA mode of interest at low power, as shown in \subautoref{app:fig:SWAP_pulse_calib}{\textbf{a}}.

The subsequent step focuses on precisely estimating the SWAP amplitude, duration, and frequency. 
The SWAP duration is fixed at \SI{160}{\nano\second} to limit the amplitude of the SWAP gate, in order to avoid coupling to other CCA modes. 
Typically, such measurements are performed similarly to those in \subautoref{app:fig:pi_pulse_calib}{\textbf{e}}. 
However, due to the high dissipation rate of the eigenmodes, this method proves impractical in our case. 
Instead, the calibration step involves exciting the qubit, applying the frequency modulation as a function of amplitude and modulation frequency for a fixed duration, and then measuring the qubit population to minimize its value. 
The results of these measurements are presented in \subautoref{app:fig:SWAP_pulse_calib}{\textbf{b}}. 
The global minima of this 2D map are determined using a total variation denoising algorithm, which identifies the optimal modulation amplitude and frequency for the fixed SWAP duration set during calibration.

The final step assesses whether neighboring eigenmodes are populated during the SWAP. 
To this end, an emission experiment is conducted. 
The qubit is first excited to a superposition state via a $\pi/2$-pulse, and its population is then transferred to the eigenmode of interest using the calibrated SWAP pulse. 
The emitted field from the CCA is collected throughout the entire pulse sequence and subsequent decay, and is downconverted using a local oscillator at \SI{7.8}{\giga\hertz}.
A fast Fourier transform is applied to the complex signal to identify the modes contributing to the emitted field, as illustrated in \subautoref{app:fig:SWAP_pulse_calib}{\textbf{c}}. 
The frequency range of this diagram spans from -500 MHz to +500 MHz. 
At approximately $\pm30$ MHz, the expected emission frequency is observed. 
At around $\pm300$ MHz, a wide peak is detected, which is attributed to cross-talk from the charge line due to the $\pi/2$-pulse [further details in \aref{app:sec:emission_measurements}].

\section{Calibration of Attenuation and Gain for the emission measurement}
\label{app:sec:attn_gain_cal}

\begin{figure*}
    \centering
    \includegraphics[width=\linewidth]{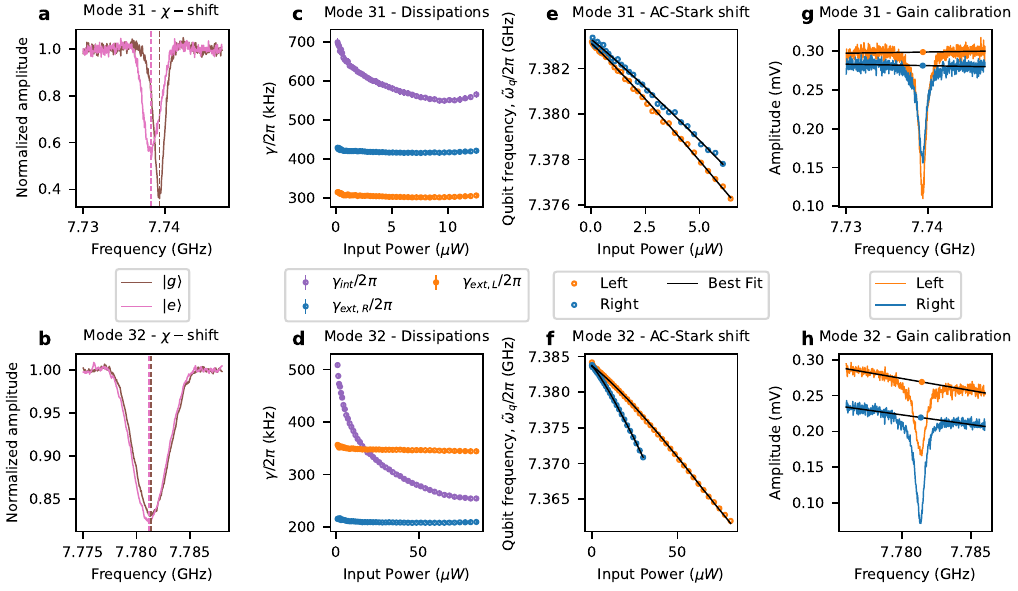}
    \caption{\textbf{Attenuation and gain calibration.}
    \textbf{(a)} [\textbf{b}] Measurement of the dispersive shift $\chi$, of mode 31 (32). 
    The brown color is mode 31 [32] with the qubit in the ground state. 
    The pink color is mode 31 [32] with the qubit in the excited state. 
    The dashed line highlights the resonant frequency of the mode of interest with the qubit in ground (brown) or excited state (pink).
    \textbf{(c)} [\textbf{d}] Measurement of dissipations for mode 31 [32]. 
    The internal dissipations, $\gamma_\text{int}/2\pi$, are reported in purple, the external dissipations to the left, $\gamma_\text{ext,L}/2\pi$, are reported in orange, and to the right, $\gamma_\text{ext,R}/2\pi$, are reported in blue.
    \textbf{(e)} [\textbf{f}] Measurement of the AC-Stark shift for mode 31 [32]. 
    The blue circles are the data acquired from the right and the orange circles the data acquired from the left. 
    The continuous black line is the AC-Stark shift fit according to \eqrefauto{app:eq:stark}.
    \textbf{(g)} [\textbf{h}] Calibration of the gain for mode 31 [32]. 
    Single-tone spectroscopy measurement of the modes to extract the baseline. 
    The data are in orange are acquired from the left and in blue are acquired from the right.
    The dots are the points chosen to calibrate the gain.}
    \label{app:fig:attn_gain_calib}
\end{figure*}

For the measurements presented in \autoref{sec:fig5}, a precise knowledge of the gain was needed.
In this section, we describe the calibration sequence to extract the gain, focusing on modes 31 and 32 studied in \autoref{sec:fig5}.
In order to extract the gain, we first need to calibrate for the attenuation, for this we use a typical AC-stark shift measurement.
We assume a dispersive Hamiltonian of the type,
\begin{equation}
     H_{\chi,{\rm Mode}}^{(m)} = \omega_q\frac{ \sigma_z}{2} + \tilde\omega_{m}\qty(a_m^\dagger a_m + \chi_m\sigma_z)\,,
    \label{app:eq:disp_H}
\end{equation}
with $\sigma_z$, the Pauli $Z$ operator, $\tilde\omega_{m}/2\pi$ the frequency of mode $m$ and $\chi_m$ the dispersive shift associated to mode $m$.
Due to the complexity of the system, we do not analytically estimate $\chi_m$ using the qubit coupling to the mode and detuning.
Indeed, we expect the dispersive shift to be also proportional to the atomic ratio of the eigenmode.
It will only be valid for a specific flux value, hence, for all the measurements described below we are going to work at a fixed flux value with the dressed qubit frequency $\tilde\omega_q/2\pi = \SI{7.38}{\giga\hertz}$, corresponding to the bare qubit frequency $\omega_q/2\pi = \SI{7.55}{\giga\hertz}$.

\subsection{Dispersive shift measurement}

The first step consists of extracting the dispersive shift for both eigenmodes.
To do so, we measure the mode of interest with the qubit in its ground and excited state [\subautoref{app:fig:attn_gain_calib}{\textbf{a}--\textbf{b}}].
From this measurement, we expect to observe the modes to shift in frequency by $2\chi$.
For mode 31, we observe a large dispersive shift, $\chi_{31}/2\pi = -498\pm\SI{12}{\kilo\hertz}$, due to the large qubit coupling to this eigenmode.
However, for mode 32, the dispersive shift is weak, $\chi_{32}/2\pi = -91\pm\SI{29}{\kilo\hertz}$, due to the weak coupling qubit to this eigenmode, this is due to the specific coupling configuration of the qubit to the CCA.
We attribute this low dispersive shift and high relative error to be the main cause of misestimation of attenuation for eigenmode 32.

\subsection{AC-Stark shift measurement}

The second step consists of measuring the AC-stark shift of the qubit, to estimate the attenuation for mode $m$, from the left and right side of the CCA.
To do so we rewrite the Hamiltonian in \eqrefauto{app:eq:disp_H} as,
\begin{equation}
     H_{\chi,{\rm Qubit}}^{(m)} = \frac{\omega_q}{2}\qty( \sigma_z + 2\chi_m a_m^\dagger a_m) + \tilde\omega_{m} a_m^\dagger a_m\,.
    \label{app:eq:ACStark_H}
\end{equation}
Hence, according to the equation above, the dressed qubit is expected to shift as a function of the number of photons in mode $m$, with,
\begin{equation}
    \tilde\omega_{q,n_{m}} = \tilde\omega_{q,0} + 2\chi_m n_m\,,
    \label{app:eq:stark}
\end{equation}
$\omega_{q,0}$ is the qubit frequency at low photon number in the mode and $\chi_m$ is the dispersive shift of eigenmode $m$.
$n_m$ is the number of photons in mode $m$, it will depend from which port it is measured from.
The number of photons in eigenmode $m$ is:
\begin{equation}
\label{app:eq:n_phot_Left}
\begin{split}
    n_m & = {a_m}^\dagger{a_m} \\
    & = \frac{\gamma_{\text{ext},L}}{\frac{1}{4}\left(\gamma_{\text{int}}({a_m}^\dagger{a_m}) + \gamma_{\text{ext},L} + \gamma_{\text{ext},R}\right)^2}{a}^\dagger_{L,m,{\rm in}}{a}_{L,m,{\rm in}}\,,
\end{split}
\end{equation}
when measured from the left port, and,
\begin{equation}
\begin{split}
     n_m & = {a_m}^\dagger{a_m} \\
     & = \frac{\gamma_{\text{ext},R}}{\frac{1}{4}\left(\gamma_{\text{int}}({a_m}^\dagger{a_m}) + \gamma_{\text{ext},L} + \gamma_{\text{ext},R}\right)^2}{a}^\dagger_{R,m,{\rm in}}{a}_{R,m,{\rm in}}\,,
     \label{app:eq:n_phot_Right}
\end{split}
\end{equation}
when measured from the right port.
And finally,
\begin{equation}
    {a}^\dagger_{S,m,{\rm in}}{a}_{S,m,{\rm in}} = \frac{P_{S,m{\rm\, in}}}{\hbar\tilde\omega_{m}} = \frac{\bar P_{S,m{\rm\, in}}10^{-A_{S,i}/10}}{\hbar\tilde\omega_{m}}\,,
\end{equation}
with $P_{S,m{\rm\, in}}$ the input power at the device level for mode $m$, with $S$ taking the value $L$ (Left) or $R$ (Right), depending on which port we are measuring from.
Similarly, $\bar P_{S,m{\rm\, in}}$ is the input power at the output of the digitizer and $A_{S,m}$ is the attenuation of the line from the $S$ port for mode $m$.
$A_{S,m}$ takes into account any amplification and attenuation of the input measurement chain.

Modes 31 and 32, are not overcoupled in the single photon regime, \ie $\gamma_{\rm int} \gtrsim \gamma_{\rm ext, L(R)}$.
Hence, before measuring the AC-Stark shift, we need to measure the dissipation rates of the CCA modes as a function of the input power. 
To be as precise as possible, we use the same input power in the dissipation measurement as the ones for the AC-Stark shift measurements.
This measurement is presented in \subautoref{app:fig:attn_gain_calib}{\textbf{c}--\textbf{d}}.
Mode 31 is more coupled to the right than to the left port [see \subautoref{app:fig:attn_gain_calib}{\textbf{c}}]. 
On the other hand, mode 32 is more coupled to the left than to the right port [see \subautoref{app:fig:attn_gain_calib}{\textbf{d}}].
We attribute this asymmetry to two factors: first to disorder which can affect the localization of the eigenmodes, and second to the qubit already affecting the mode shape of the eigenmodes.
For both cases we are close to the undercoupled limit and we observe a strong dependence of the internal dissipation rate as a function of the input power.

Finally, we proceed to the AC-Stark shift measurement.
For this measurement, the mode of interest is pumped and probed at its resonant frequency, $\tilde\omega_m/2\pi$, and the frequency of the qubit is monitored using standard two-tone spectroscopy measurements.
The measurements are reported in \subautoref{app:fig:attn_gain_calib}{\textbf{e}--\textbf{f}}.
In \subautoref{app:fig:attn_gain_calib}{\textbf{e}-\textbf{f}}, we report the frequency of the dressed qubit, $\tilde\omega_q/2\pi$ extracted from the AC-Stark shift measurement for mode 31 [panel \textbf{e}] and mode 32 [panel \textbf{f}].
We finally fit the qubit frequency shift using \eqrefauto{app:eq:stark}, with the attenuation and initial dressed qubit frequency as the only fitting parameters.
The extracted attenuations for the two modes on both sides of the CCA are reported in \ref{app:tab:attenuation_gain_lines}.
The error on the gain is estimated from the convergence of the fit of the AC-Stark shift as a function of the number points in the fit and the $\chi$-shift error.
The nonlinear trend of the AC-stark shifts is due to the varying $\gamma_\textrm{int}$.

\begin{table}[]
    \centering
    \renewcommand{\arraystretch}{1.5}
    \begin{tabular}{c|c|c}
        \textbf{Line} & \textbf{Attenuation} (dB) & \textbf{Gain} (dB)\\
        \hline\hline
        Left - Mode 31 & $104.59^{-1}_{+1}$ & $86.15^{-1}_{+1}$\\
        Right - Mode 31 & $103.80^{-1}_{+1}$ & $84.83^{-1}_{+1}$\\
        \hline
        Left - Mode 32 & $103.95_{-1.67}^{+1.5}$ & $84.58_{-1.67}^{+1.5}$\\
        Right - Mode 32 & $103.83_{-1.67}^{+1.5}$ & $82.69_{-1.67}^{+1.5}$
    \end{tabular}
    \caption{Attenuation and gain extracted for the modes at which the emission measurement is done.}
    \label{app:tab:attenuation_gain_lines}
\end{table}

\subsection{Extraction of the gain}
The last step consists of extracting the gain for each modes from the two sides of the CCA.
According to input/output theory, we expect the single-tone spectroscopy response of modes 31 and 32 to follow \eqrefauto{app:eq:SLL_fitting_function} and \eqrefauto{app:eq:SRR_fitting_function}, from the left and right side,  respectively.
Specifically, the absolute value of the scattering coefficient is equal to 1 out of resonance, \ie $\hat a_{\rm out} = \hat a_{\rm in}$.
Thus, we can write, at the baseline, 
\begin{equation}
    \bar P_{S,m{\rm\, out}}^{\rm dB} = \bar P_{S,m{\rm\, in}}^{\rm dB} - A_{S,m} + G_{S,m}\,,
    \label{app:eq:gain_extract}
\end{equation}
where $\bar P_{S,m{\rm\, out}}^{\rm dB} = P_{S,m{\rm\, out}}^{\rm dB} + G_{S,m}$ is the output power at the input of the analog to digital converter and $G_{S,m}$ is the gain in dB.
We then perform single-tone spectroscopy measurements of modes 31 and 32 in reflection [see \subautoref{app:fig:attn_gain_calib}{\textbf{g}--\textbf{h}}], fit the baseline of the norm of the signal [black lines in panels \subautoref{app:fig:attn_gain_calib}{\textbf{g}--\textbf{h}}] and extract the amplitude of the expected baseline at the resonant frequency of the modes, at which the attenuation is calibrated and at which the field is emitted.
We then find the gain for the two modes from both sides of the CCA by applying \eqrefauto{app:eq:gain_extract}, the values of the gain are reported \tabref{app:tab:attenuation_gain_lines}.

\section{Emission measurements}
\label{app:sec:emission_measurements}
\subsection{Data cleanup}

\begin{figure}
    \centering
    \includegraphics[width=\linewidth]{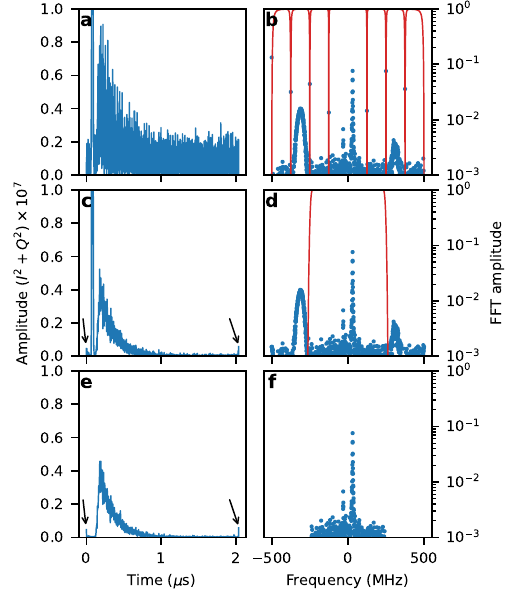}
    \caption{\textbf{Emission data cleanup.}
    \textbf{(a)} Initial time domain trace of the emitted field, where the DC offset has been removed.
    \textbf{(b)} Fourier transform of the time domain trace presented in panel \textbf{a} [blue dots]. The red lines are the notch filter applied to remove the spurious tones due to the measurement setup at $[\pm125, \pm250,\pm375, \pm500] \SI{}{\mega\hertz}$.
    \textbf{(c)} Time domain trace after application of the notch filters in panel \textbf{b}. The two black arrows highlight lobes appearing due to the filtering of the data.
    \textbf{(d)} Fourier transform of the time domain trace presented in panel \textbf{c} [blue dots]. The red line is the FIR filter with a Hamming window applied to the data to remove the $\pi$-pulse crosstalk.
    \textbf{(e)} Time domain trace after application of the FIR filter in panel \textbf{d}.
    \textbf{(f)} Fourier transorm of the time domain trace presented in panel \textbf{f}.}
    \label{app:fig:cleaning_emission}
\end{figure}

The initial emission data acquired from the CCA are very noisy and do not seem to converge to 0 as a function of time [see \subautoref{app:fig:cleaning_emission}{\textbf{a}}].
We attribute this main source of noise to spurious tones arising from the measurement setup.
These can actually be seen in the Fourier transform of the time trace where sharp peaks at $[\pm125, \pm250,\pm375, \pm500] \SI{}{\mega\hertz}$ are visible [see \subautoref{app:fig:cleaning_emission}{\textbf{b}}].
These spurious peaks are also present when unplugging the digitizer from the cryostat.
To minimize the effect of those spurious tones we place the local oscillator such that the expected emitted field is in the range $[10,100] \SI{}{\mega\hertz}$ of the digitizer to limit the influence of those spurious tones.
Then we digitally apply sharp notch filters at the frequencies of these spurious tones [red line in \subautoref{app:fig:cleaning_emission}{\textbf{b}}].

After applying the notch filters, the signal is cleaner [see \subautoref{app:fig:cleaning_emission}{\textbf{c}}] and converge to 0 as a function of time.
However, we can notice the appearance of two lobes near $t = \SI{0}{\micro\second}$ and $t = \SI{2}{\micro\second}$, these appear due to the application of the numerical filters.
We can still observe at $\SI{80}{\nano\second}$ the presence of a sharp high amplitude tone [also present in \subautoref{app:fig:cleaning_emission}{\textbf{a}}].
It is due to cross-talk between the charge line and the CCA ports.
This signal is stronger when measuring the emission from the right than from the left side of the CCA, due to the proximity of the charge line pad to the right pad of the of the CCA input port.
We can also notice that this tone does not decay, and is $\approx\SI{40}{\nano\second}$ wide corresponding to the length of the $\pi$-pulse.
We filter out this tone by applying a low pass Finite Impulse Response filter with a cut-off at \SI{250}{\mega\hertz} with Hamming window [red line in \subautoref{app:fig:cleaning_emission}{\textbf{d}}].
The resulting time trace is shown in \subautoref{app:fig:cleaning_emission}{\textbf{e}} and its Fourier transform shown in \subautoref{app:fig:cleaning_emission}{\textbf{f}}.

The last step consists of converting the signal to field intensity and photon number.
The field intensity is converted using,
\begin{equation}
    |\langle a_{\rm out}\rangle|^2 = \frac{I^2 + Q^2}{G}\frac{1}{Z_0\hbar\omega}\,,
\end{equation}
where $G$ is the gain, \(a_{\rm out}\) is the outgoing emitted photon field from the left or right side of the CCA, $I$ and $Q$ are the in-phase and quadrature phase of the signal, $Z_0 = \SI{50}{\ohm}$ is the waveguide impedance and $\omega/2\pi$ is the frequency of the emitted signal.
Finally, we find the emitted photon number at time $t$ using, 
\begin{equation}
    N_{\rm ph} = \int_0^t \frac{I^2 + Q^2}{G}\frac{1}{Z_0\hbar\omega} d\tau\,.
\end{equation}

\subsection{Spectrograms}
\begin{figure*}
    \centering
    \includegraphics[width=\linewidth]{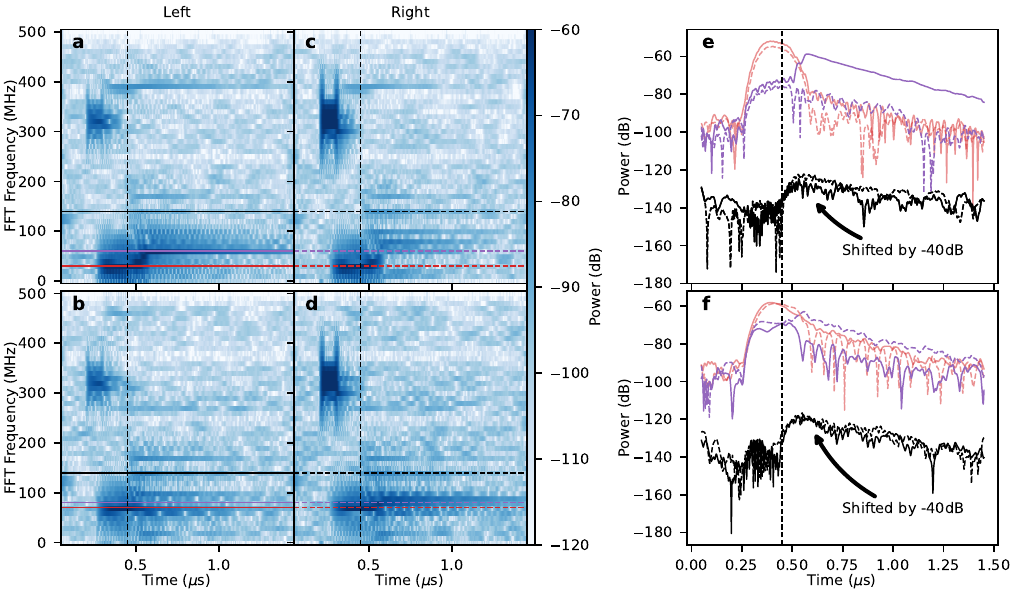}
    \caption{\textbf{Spectrograms of the emission measurements.} 
    \textbf{(a)} [\textbf{b}] Spectrogram of the emission measurement acquired from the left for mode 31 [32].
    \textbf{(c)} [\textbf{d}] Spectrogram of the emission measurement acquired from the right for mode 31 [32].
    The horizontal red and purple lines correspond to the initial frequency of emission and the final frequency of emission after the ramp, respectively. 
    The horizontal black lines highlight spurious tones arising during the ramp. 
    The vertical dashed line highlight the point from which the spurious tones start arising.
    \textbf{(e)} [\textbf{f}] Line cuts of the spectrogram of the emission measurement acquired from the left (continuous lines) and right (dashed lines) for mode 31 [32]. 
    The black lines are shifted down by -40dB for clarity.}
    \label{app:fig:window_FFT}
\end{figure*}

In \subautoref{app:fig:window_FFT}{\textbf{a}--\textbf{d}}, we present spectrograms of the directional emission data from \figref{fig:fig5}, computed using 200~ns windows with a 1~ns time step. 
This analysis allows us to trace the temporal evolution of the various frequency tones involved in the emitted data.

The $\pi$-pulse is visible in all four spectrograms near \SI{320}{\MHz}, appearing most prominently on the right side due to the proximity of the charge line to the right CCA port. 
During the SWAP operation, we observe population transfer from the qubit to the mode of interest at the frequencies highlighted by red lines in \subautoref{app:fig:window_FFT}{\textbf{a}--\textbf{d}}. 
Line cuts at these frequencies are shown in \subautoref{app:fig:window_FFT}{\textbf{e}--\textbf{f}}.

During the DC ramp, we observe the emitted field frequency shifting to the value indicated by the purple line, corresponding to the expected frequency for directional emission. 
The cancellation of emission to one side is excellent for mode 31 but poor for mode 32 [see \subautoref{app:fig:window_FFT}{\textbf{e}--\textbf{f}}], consistent with the measurements shown in \figref{fig:fig5}.

We also observe spurious tones arising during the DC ramp, as indicated by the black lines in the spectrograms. 
These tones only appear during and after the DC ramp and not during the SWAP operation (\subautoref{app:fig:window_FFT}{\textbf{e}--\textbf{f}}), suggesting they originate from non-adiabatic transitions to other eigenmodes. 
While these spurious tones could be avoided using the measurement protocol shown in \extfigref{ext:fig:ideal_emission}, our current experimental setup did not permit implementation of this protocol.

\section{Fabrication}
\label{app:sec:fab}

\begin{enumerate}
    \item \textbf{Wafer cleaning}\\
    The device is fabricated on a high resistivity (\(\geq\SI{10}{\kilo\ohm\centi\metre}\)) 4-inch intrinsic silicon wafer with a thickness of \SI{525}{\micro\metre}. The wafer is first cleaned in a Piranha solution (sulfuric acid/hydrogen peroxide mixture) to remove organic contaminants, followed by a hydrofluoric acid (HF) bath to remove oxide.

    \item \textbf{Ground plane evaporation}\\
    Immediately after cleaning, we deposit a \SI{150}{\nano\metre} aluminum ground plane using electron-beam evaporation (Leybold Optics LAB 600H) at a rate of \SI{0.2}{\nano\metre\per\second}.

    \item \textbf{Markers patterning}\\
    Alignment markers are patterned using double-layer photolithography (LOR-5A and AZ1512-HS resists) with a laser-writer (Heidelberg Instruments MLA-150). The pattern is developed in AZ 726 MIF developer for \SI{70}{\second}, followed by deposition of \SI{10}{\nano\metre} Ti and \SI{50}{\nano\metre} Pt using electron-beam evaporation. The lift-off process is then performed to complete marker fabrication.

    \item \textbf{Ground plane etching}\\
    The wafer is coated with ECI 3007 photoresist, exposed with the laser-writer, and automatically developed. The aluminum ground plane is etched using Aluetch for \SI{2.25}{\minute} at \SI{26}{\celsius}, followed by two water rinses.

    \item \textbf{NbN sputtering and lift-off}\\
    The NbN deposition areas are patterned using electron-beam lithography (Raith EBPG5000+) on a bilayer resist stack (MMA EL9/PMMA 495K A8). After development (\SI{60}{\second} in 3:1 MiBK:IPA followed by \SI{60}{\second} in IPA), NbN is sputtered using a Kenosistec RF sputtering system following established protocols \rref\cite{frasca2023nbn}. The standard lift-off procedure completes the patterning.

    \item \textbf{Inductor patterning and etching}\\
    The metamaterial inductors are patterned on CSAR C04 electron-beam resist using electron-beam lithography. After development (\SI{60}{\second} in n-Amyl acetate followed by \SI{60}{\second} in MiBK/IPA), the inductors are dry-etched in an Oxford Plasmalab 80+ using a CF4/Ar gas mixture with a stepped recipe \rref\cite{frasca2023nbn}.

    \item \textbf{Inductor patching}\\
    A double layer of electron-beam resist (MMA EL9/PMMA 495K A8) is applied and patterned using electron-beam lithography. After development, the areas to be patched are milled and aluminum is evaporated in a Plassys MEB550SL3 system. The standard lift-off procedure completes the process.

    \item \textbf{Josephson junction patterning and evaporation}\\
    The Josephson junctions are designed for Manhattan-style evaporation. After coating with a double layer of MMA EL9/PMMA 495K A8 resist, the junction pattern is exposed using electron-beam lithography. The sample undergoes a \SI{15}{\second} low-power oxygen plasma descum before aluminum evaporation in an angled evaporator (Plassys MEB550SL3). The standard lift-off procedure is then applied.

    \item \textbf{Josephson junctions patch}\\
    Following the same process as inductor patching, this step repairs any defects in the junction fabrication. The double-layer resist is exposed and developed as before, and aluminum is evaporated and lifted off to complete the junctions.

    \item \textbf{Packaging}\\
    Completed devices are diced using a DISCO DAD321 dicer, mounted onto copper holders using MMA EL9, and wire-bonded to PCBs using a semi-automatic bonder (Bondtec).
\end{enumerate}

The standard lift-off procedure consists of sequential solvent baths with sonication: 12 hours in 1165 remover at \SI{70}{\celsius}, followed by fresh 1165 remover, methanol, and isopropanol, each with 5 minutes of mild sonication.

\section{Measurement setup}
\label{app:sec:meas_setup}

\begin{figure*}
    \centering
    \includegraphics[width = \linewidth]{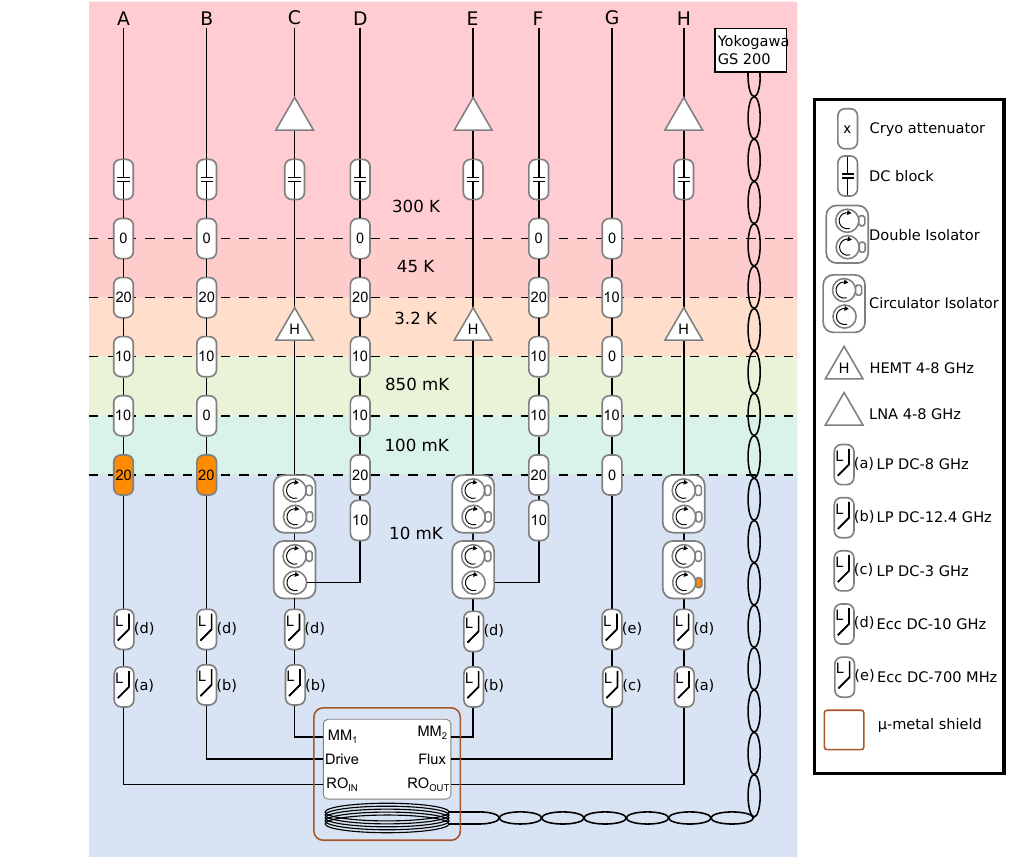}
    \caption{\textbf{Schematic of the cryogenic setup.}}
    \label{app:fig:measSetup_cryo}
\end{figure*}

A schematic of the measurement setup is depicted in \figref{app:fig:measSetup_cryo}. 
The device is thermally anchored to the mixing chamber plate of a commercial dry dilution cryostat (Bluefors-LD) at a temperature of $\SI{10}{\milli\kelvin}$ enclosed with an Al and Cu shield, and placed inside a $\mu-$metal shield. 
The device has 5 input lines: Readout input [\textbf{A}], Drive [\textbf{B}], CCA left [\textbf{D}], CCA right [\textbf{F}] and flux [\textbf{G}].
The input lines are attenuated with cryogenic attenuators at different stages of the cryostat, as reported in \ref{app:fig:measSetup_cryo}.
The device has 3 output lines: Readout output [\textbf{H}], CCA left [\textbf{C}] and CCA right [\textbf{E}].
All the output lines consist of two double circulator-isolator brackets (LNF-CIIS48A and LNF-ISIS48A), input lines \textbf{D} and \textbf{F} are connected to the circulator of lines \textbf{C} and \textbf{E}, respectively, to be able to measure the device in reflection.
A \SI{50}{\ohm} copper termination is placed on the circulator of the readout output.
The signal is then amplified with a 4-8 GHz HEMT (LNF-LNC4\_8C) thermalized on the 4K stage of the cryostat, it is then re-amplified at room temperature with a 4-8 GHz low noise amplifier (Agile AMT-A0284).
All lines are filtered at the mixing chamber stage first with an Eccosorb filter (QMC-CRYOIRF-003MF-S [filter (d) in \figref{app:fig:measSetup_cryo}] and QMC-CRYOIRF-001MF-S [filter (e) in \figref{app:fig:measSetup_cryo}]) and with cryogenic low pass filters (RLC-F-30-8000 [filter (a) in \figref{app:fig:measSetup_cryo}], RLC-F-30-12.4 [filter (b) in \figref{app:fig:measSetup_cryo}], RLC-F-10-3000 [filter (c) in \figref{app:fig:measSetup_cryo}]).
The coil bias is controlled with a current source Yokogawa GS200.

\begin{figure}
    \centering
    \includegraphics[width = \linewidth]{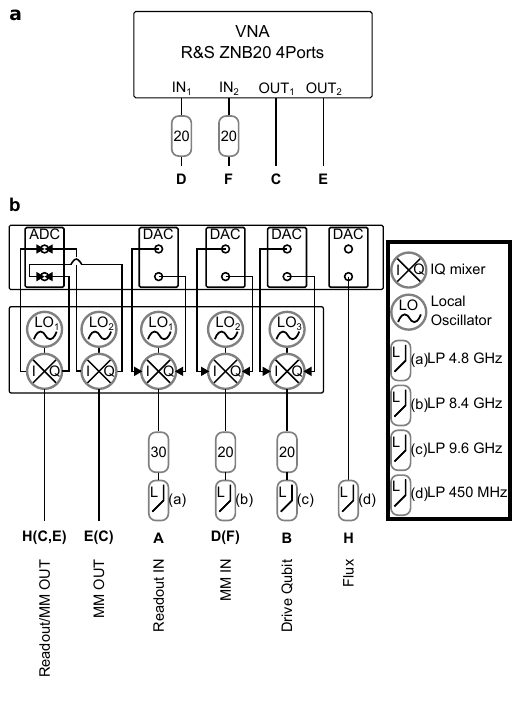}
    \caption{\textbf{Room temperature setup.}
    \textbf{(a)} Schematic of the VNA setup.
    \textbf{(b)} Schematic of the time-domain setup.}
    \label{app:fig:measSetup_RT}
\end{figure}

The spectroscopy measurements used in \figref{fig:fig2}, \figref{fig:fig3} and \figref{app:fig:dissipation} are performed using a 4 ports R\&S ZNB20 vector network analyzer (VNA) [\subautoref{app:fig:measSetup_RT}{\textbf{a}}] with \SI{20}{\decibel} input attenuators.
The rest of the measurements are performed using an OPX+ coupled to an Octave from from Quantum Machine, implementing standard heterodyne measurement techniques.
The readout input line [\textbf{A}] is attenuated with a \SI{30}{\decibel} attenuator and filtered with a low-pass filter (ZLSS-4R8G-S+ [filter (a) in \subautoref{app:fig:measSetup_RT}{\textbf{b}}]).
The CCA input line is attenuated with a \SI{20}{\decibel} attenuator and filtered with a low-pass filter (VLF-8400+ [filter (b) in \subautoref{app:fig:measSetup_RT}{\textbf{b}}]).
The Drive line is attenuated with a \SI{20}{\decibel} attenuator and filtered with a low-pass filter (FLP0960 [filter (c) in \subautoref{app:fig:measSetup_RT}{\textbf{b}}]).

\end{document}